\documentclass[lettersize,journal]{IEEEtran}
\usepackage{amsmath,amsfonts}
\usepackage{algorithmic}
\usepackage{algorithm}
\usepackage{array}
\usepackage{textcomp}
\usepackage{stfloats}
\usepackage{url}
\usepackage{verbatim}
\usepackage{graphicx}
\usepackage{cite}
\usepackage{booktabs}
\usepackage{multirow}
\usepackage[table]{xcolor}
\usepackage{caption}
\usepackage{subcaption}
\usepackage{calc}
\newlength{\mywidthA}
\newlength{\mywidthB}
\newlength{\mywidthC}

\begin{document}

\setlength{\mywidthA}{0.33\columnwidth} 
\setlength{\mywidthB}{0.66\columnwidth} 
\setlength{\mywidthC}{0.86\columnwidth} 

\title{Feature-based Inversion of 2.5D Controlled Source Electromagnetic Data using Generative Priors}

\author{Hongyu Zhou,~\IEEEmembership{Graduate Student Member,~IEEE},
	Haoran Sun,~\IEEEmembership{Graduate Student Member,~IEEE},
	Rui Guo,~\IEEEmembership{Member,~IEEE,}
	Maokun~Li,~\IEEEmembership{Fellow,~IEEE,}
	Fan~Yang,~\IEEEmembership{Fellow,~IEEE,}
	Shenheng~Xu,~\IEEEmembership{Member,~IEEE,}
	\thanks{This work was funded by the National Key Research and Development Program of China (2023YFB3905003), 
			National Natural Science Foundation of China (624B2085), and
		the Science and Technology Project of CNPC (grant no. 08-01-2024).}
	\thanks{Hongyu Zhou, Haoran Sun, Maokun Li, Fan Yang and Shengheng Xu are with the
		Department of Electronic Engineering, Beijing National Research Center for
		Information Science and Technology (BNRist), Tsinghua University, Beijing
		100084, China (e-mail: maokunli@tsinghua.edu.cn).}
	\thanks{Rui Guo is with Faculty of Math and Computer Science, Weizmann Institute of Science, Rehovot, Israel.}
}


\maketitle

\begin{abstract}
\color{black}
In this study, we investigate feature-based 2.5D controlled source marine electromagnetic (mCSEM) data inversion using generative priors. Two-and-half dimensional modeling using finite difference method (FDM) is adopted to compute the response of horizontal electric dipole (HED) excitation. Rather than using a neural network to approximate the entire inverse mapping in a black-box manner, we adopt a plug-and-play strategy in which a variational autoencoder (VAE) is used solely to learn prior information on conductivity distributions. 
During the inversion process, the conductivity model is iteratively updated using the Gauss--Newton method, while the model space is constrained by projections onto the learned VAE decoder. This framework preserves explicit control over data misfit and enables flexible adaptation to different survey configurations. Numerical and field experiments demonstrate that the proposed approach effectively incorporates prior information, improves reconstruction accuracy, and exhibits good generalization performance.
\color{black}
\end{abstract}

\begin{IEEEkeywords}
2.5D electromagnetic modeling, electromagnetic data inversion, controlled source electromagnetics, generative models, generative priors, machine learning, deep learning, constrained inversion, joint inversion
\end{IEEEkeywords}

\section{Introduction}
Electromagnetic (EM) methods have been widely used in geophysical prospecting. Among them, the controlled source electromagnetic (CSEM) method has been successfully used to evaluate \color{black} targets with significant resistivity contrasts, including ore deposits, water bodies, and hydrocarbon reservoirs, etc \cite{constable2006marine, gribenko2007rigorous, abubakar20082, commer2008new, key20091d, LiYuguo, 2p5dCSEM_parallel, brown2012resolution, kang2015mcsem, key2016mare2dem,  zhou2017spectral, oh2020cooperative, ohwoghere-asuma_use_2020, 9164913, wu_fine_2023, farzamian_landscape-scale_2023, qin20243, liang2024physics, lu_feasibility_2024, gao2025two}\color{black}. This method involves stimulating the low-frequency EM field by \color{black} artificial sources, \color{black} measuring the EM field at selected locations, and inverting and interpreting the measurements with data inversion algorithms.

Accurate CSEM modeling is essential for reliable data inversion and interpretation. Due to the lack of translational symmetry in the horizontal electric dipole (HED) source, the EM field in CSEM surveys cannot be accurately simulated in a purely 2D space. A direct solution is to perform full 3D modeling \cite{commer2008new, gribenko2007rigorous}, but this approach is computationally expensive and resource-intensive. Alternatively, if the geological structure is assumed to vary only in two dimensions and remains constant along one direction, 2.5D modeling provides an efficient compromise \cite{abubakar20082, 2p5dCSEM_parallel, zhou2017spectral}. It enables accurate representation of the EM field while significantly reducing computational cost, making it suitable for practical inversion and interpretation tasks.

EM data inversion is inherently ill-posed and nonlinear, often leading to non-unique or unstable solutions. To mitigate this, prior knowledge is typically incorporated to constrain the inversion. A common strategy is to add regularization terms, such as the $L_1$ norm or Tikhonov regularization \cite{tikhonov1963regularization, benning2018modern, liu2024robust}. However, simple regularizations may struggle to capture more complex prior information.
An alternative approach involves reparameterizing the model in a compact transformed space, which can be constructed using various techniques. 
For instance, model-based inversion uses horizons and polygons to define regions with assumed uniform properties. The unknowns include the physical parameters within each region and the coordinates of boundary control points \cite{Marcuello2D1992, auken2004layered, li2010inversion, chen2012stochastic}, which are typically optimized using gradient-based methods.
While this reduces the number of unknowns, it is often sensitive to the initial model and prone to local minima.

Another strategy involves reparameterizing the model with a linear combination of global basis functions, such as Fourier or wavelets. This method has been applied in EM inversion \cite{nittinger2016inversion, liu2018wavelet, abubakar2012model, lochbuhler2014probabilistic} and seismic tomography \cite{chiao2001multiscale, chiao2003multiresolution, loris2007tomographic, hung2010first}. 
However, selecting appropriate basis functions is often heuristic and lacks the flexibility to encode task-specific prior knowledge.
The basis functions can also be constructed with machine learning methods. In hydrogeologic ERT inversion, for example, complex model patterns are described using a training dataset, and singular value decomposition (SVD) is applied to learn representative image bases \cite{oware2019basis}.

Deep learning provides a powerful framework to distill the accumulation of human experience and historical data into neural models and conduct effective inference on similar tasks \cite{gribenko2007rigorous, hung2010first}. Generative modeling \cite{zhao2023generative} allows for learning the underlying data distribution and generating new samples consistent with prior knowledge.
Unlike supervised learning, generative models are typically trained solely in the model space, independent of physical simulations. As a result, paired data and model samples are not required for constructing the training set.
In recent years, generative models have been explored for inversion tasks across various domains, including biomedical imaging \cite{song2021solving, chung2023solving}, electromagnetic inverse scattering \cite{guo2022nonlinear, guo2023three}, image processing \cite{asim2020blind, rout2023solving}, and geophysical applications \cite{wang2022dual, zhou2024feature}.
Among the commonly used generative models, variational autoencoders (VAEs) and generative adversarial networks (GANs) learn nonlinear mappings from the original data space to a lower-dimensional latent space, providing a compact representation that serves as a generative prior in the inversion process \cite{zhao2023generative}.

The generative prior can be integrated into the inversion process by directly cascading the generative model with the physical forward operator in the objective function \cite{bora2017compressed, guo2022nonlinear, guo2023three, zhou2024feature}. In this framework, gradients are directly backpropagated from the objective function to the latent variables. However, the nonlinearity of the generative model can lead the optimization process trapped in local minima \cite{shah2018solving}. This challenge becomes even more pronounced when the physical forward operator is also nonlinear, further complicating the optimization landscape. 
To address this, in some gradient-based optimization with projections (e.g., projected gradient descent, PGD), the updated model is iteratively projected back onto the prior space, which can be defined by the range of a VAE decoder or GAN generator \cite{shah2018solving, raj2019gan, li2019using, chung2023solving, rout2023solving}. However, these methods are primarily designed for linear inverse problems. When the forward operator is nonlinear and the solution space is 2.5D or 3D, the computational cost of first-order methods may become prohibitive.

\color{black}
In this work, we develop a feature-based 2.5D marine CSEM inversion framework regularized by a generative prior. Unlike recent end-to-end or latent-space inversion approaches that use neural networks to directly infer subsurface models from data, the proposed method retains a physics-driven Gauss–Newton inversion scheme and introduces a variational autoencoder (VAE) solely to constrain the admissible conductivity model space. The VAE is trained to learn the prior distribution of conductivity models and is incorporated through a series of projection steps, in which intermediate Gauss–Newton updates are periodically projected onto the decoder output space. This procedure enforces feature-level geological consistency without altering the forward operator or the data misfit formulation. The resulting plug-and-play strategy decouples prior learning from physical inversion and differs fundamentally from conventional generative-prior regularization approaches.
The main contributions of this work are threefold:
\begin{enumerate}
	\item We propose a plug-and-play deep learning framework for 2.5D marine CSEM inversion, in which neural networks are used only to learn conductivity-domain prior information while the physics-based Gauss–Newton inversion scheme is fully preserved.
	\item We construct a feature-level generative prior using a variational autoencoder trained on synthetic conductivity models to encode geological prior information without modifying the physical forward modeling.
	\item We develop a projected Gauss–Newton inversion scheme for marine CSEM, where intermediate model updates are periodically projected onto the decoder space to integrate generative priors into a conventional Hessian-based inversion framework.
\end{enumerate}
Numerical and field experiments demonstrate that the proposed method effectively incorporates \emph{a priori} knowledge into the inversion, leading to improved reconstruction quality and robust performance across different survey configurations.
\color{black}


\section{Method}
\subsection{2.5D forward modeling and Jacobian computation}\label{IIA}
Starting from the frequency domain Maxwell's equations, and neglecting the magnetic source, and using the $j$-notation, we have
\begin{subequations}
	\begin{align}
		\nabla \times \bar{\mathbf{E}} &= -j \omega \mu \bar{\mathbf{H}}, \label{eq:1a} \\
		\nabla \times \bar{\mathbf{H}} &= j \omega \epsilon^\prime \bar{\mathbf{E}} + \bar{\mathbf{J}}, \label{eq:1b}
	\end{align}
\end{subequations}
where $\bar{\mathbf{E}}$, $\bar{\mathbf{H}}$ and $\bar{\mathbf{J}}$ are respectively the electric field, magnetic field and electric source in the spatial domain. $\omega$ is the angular frequency, $\mu$ is the permeability, and $\epsilon^\prime=\epsilon-j\frac{\sigma}{\omega}$ is the complex permittivity in which $\sigma$ is the conductivity and $\epsilon$ is the real permittivity \cite{nabighian1988electromagnetic}. 

In a typical 2.5D survey, the EM field has to be modelled in the 3D space (e.g., the response of the dipole excitation), while the distribution of physical parameter is only variant in the 2D plane but invariant along the third dimension. Without loss of generality, we assume the model extends along the $x$ direction, and the EM field in the $yoz$ plane is of interest. Solving 2.5D problem generally involves transforming the field and source with Fourier transform along the spatially invariant dimension ($x$-direction), that is,
\begin{subequations}
	\begin{align}
		\hat{\mathbf{F}}(k_x, y, z) &= \int_{-\infty}^{+\infty} \bar{\mathbf{F}}(x, y, z) e^{-j k_x x} dx,  \label{eq:2a} \\
		\bar{\mathbf{F}}(x, y, z) &= \frac{1}{2 \pi}\int_{-\infty}^{+\infty} \hat{\mathbf{F}}(k_x, y, z) e^{j x k_x} d(k_x), \label{eq:2b} 
	\end{align}
\end{subequations}
where $\hat{\mathbf{F}}$ denotes the field or source in the wave-number domain, and $\mathbf{F}$ can be either $\mathbf{E}$, $\mathbf{H}$ or $\mathbf{J}$. Rewriting (\ref{eq:1a}) and (\ref{eq:1b}) in the wave-number domain, and neglecting the $z$ component of $\mathbf{J}$, we have
\begin{subequations}
	\begin{align}
		\frac{\partial \hat{E}_z}{\partial y} - \frac{\partial \hat{E}_y}{\partial z} &= -j \omega \mu \hat{H}_x, \label{ea1} \\
		\frac{\partial \hat{E}_x}{\partial z} -j k_x \hat{E}_z &= -j \omega \mu \hat{H}_y, \label{ea2} \\
		j k_x \hat{E}_y - \frac{\partial \hat{E}_x}{\partial y} &= -j \omega \mu \hat{H}_z, \label{ea3} \\
		\frac{\partial \hat{H}_z}{\partial y} - \frac{\partial \hat{H}_y}{\partial z} &= (j \omega \epsilon + \sigma) \hat{E}_x + \hat{J}_x, \label{ea4} \\
		\frac{\partial \hat{H}_x}{\partial z} - j k_x \hat{H}_z &= (j \omega \epsilon + \sigma) \hat{E}_y + \hat{J}_y, \label{ea5} \\
		j k_x \hat{H}_y - \frac{\partial \hat{H}_x}{\partial y} &= (j \omega \epsilon + \sigma) \hat{E}_z. \label{ea6}
	\end{align}
\end{subequations}
Further organizing (\ref{ea1})-(\ref{ea6}), we have 
\begin{subequations}
	\begin{align}
		&-\frac{\partial}{\partial y}(\frac{y_0}{k_e^2} \frac{\partial \hat{E}_x}{\partial y})-\frac{\partial}{\partial z}(\frac{y_0}{k_e^2}\frac{\partial \hat{E}_x}{\partial z}) + jk_x\frac{\partial }{\partial y}(\frac{1}{k_e^2}) \frac{\partial \hat{H}_x}{\partial z} \nonumber \\
		&- jk_x\frac{\partial}{\partial z}(\frac{1}{k_e^2}) \frac{\partial \hat{H}_x}{\partial y} + y_0 \hat{E}_x = -\hat{J}_x + \frac{\partial}{\partial y}(\frac{j k_x}{k_e^2}\hat{J}_y) , \label{et1} \\
		&-\frac{\partial}{\partial y}(\frac{z_0}{k_e^2} \frac{\partial \hat{H}_x}{\partial y})-\frac{\partial}{\partial z}(\frac{z_0}{k_e^2}\frac{\partial \hat{H}_x}{\partial z}) + jk_x \frac{\partial}{\partial z}(\frac{1}{k_e^2})\frac{\partial \hat{E}_x}{\partial y} \nonumber \\
		&- jk_x\frac{\partial}{\partial y}(\frac{1}{k_e^2})\frac{\partial \hat{E}_x}{\partial z} + z_0 \hat{H}_x
		= - \frac{\partial}{\partial z}(\frac{z_0}{k_e^2}\hat{J}_y), \label{et2}
	\end{align}
\end{subequations}
where $k_e^2 = k_x^2-k^2$, $k^2=-z_0 y_0$, $z_0=j \omega \mu$, $y_0=j \omega \epsilon + \sigma$. Discretizing (\ref{et1}) and (\ref{et2}) with finite difference method (FDM), a set of linear equations can be derived and solved with respect to the $x$ component electric and magnetic fields in the wave-number domain ($\hat{E}_x$ and $\hat{H}_x$). Further resorting to (\ref{ea1})-(\ref{ea6}), the $y$ and $z$ components of $\hat{\mathbf{E}}$ and $\hat{\mathbf{H}}$ can be computed, and the spatial domain fields $\bar{\mathbf{E}}$ and $\bar{\mathbf{H}}$ can be solved using the inverse Fourier transform in (\ref{eq:2b}). In this work, we focus on marine CSEM surveys, in which inline electric measurements ($\bar{E}_y$) are collected along the single survey line (extending along the $y$ direction). Both transmitters and receivers are HEDs, whose orientations are along the survey line direction. We use the function $\mathcal{F}(.)$ to represent the relationship between the conductivity $\sigma$ and the measured $\bar{E}_y$ at the receivers.

\begin{figure*}[!t]
	\centering
	\includegraphics[width=0.7\linewidth]{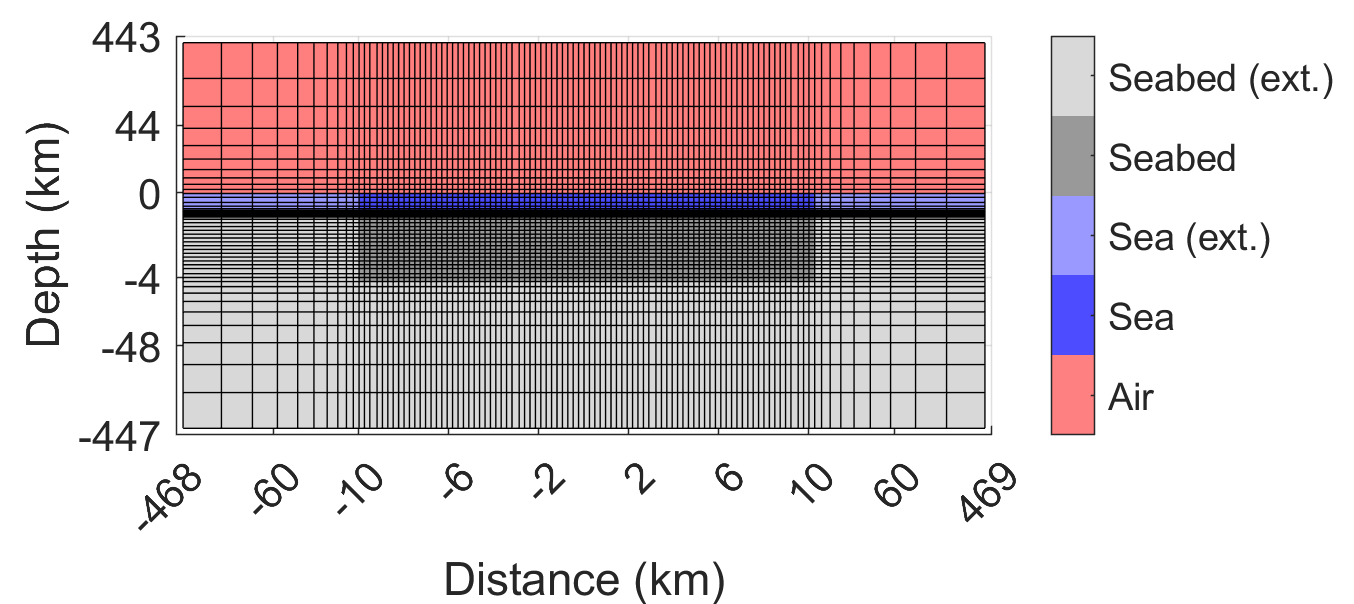}
	\caption{\color{black} Schematic of the finite difference mesh for 2.5D mCSEM forward modeling. "Ext." denotes the extended region. \color{black}}
	\label{mesh}
\end{figure*}

For the numerical simulation, the computational domain is discretized into square cells, and both the electric and magnetic fields in wave-number and spatial domains are defined at the center of each cell. \color{black}

A schematic of the adopted FDM mesh is shown in Fig. \ref{mesh}. For clarity, the coordinates shown in the figure are not scaled to the actual physical distances, and the number of grid points is proportionally reduced to avoid an overly dense mesh illustration. Fig. \ref{mesh} shows the entire computational domain, within which the seabed region of interest (i.e., domain of investigation, DoI) is indicated in dark gray. Above the DoI is the seawater layer, shown in dark blue, and the interface between the seawater and the DoI can be either flat or undulating. The mesh is refined near the seawater–DoI interface and gradually transitions to a coarser mesh away from the interface. Both the seawater layer and the DoI are extended in the horizontal direction. The DoI is also extended downward, and an air layer is located above the seawater. The extended region and the air layer are both discretized using nonuniform meshes, in which the grid spacing increases logarithmically away from the central region. Thirty cells are used in each direction, and the thickness of the outermost cells is set to 100 km. In Fig. \ref{mesh}, the extended cells and air cells are colored light blue, light gray, and red, respectively. \color{black}

The Jacobian matrix can be computed with adjoint method \cite{mcgillivray1994calculation, abubakar20082}. The relationship between the adjoint field, adjoint source and the sensitivity can be formulated with \cite{mcgillivray1994calculation}
\begin{equation}\label{eq5}
	\int_D (\bar{\mathbf{J}}_m^{+} \cdot \frac{\partial \bar{\mathbf{H}}}{\partial \sigma_k}+\bar{\mathbf{J}}_e^{+} \cdot \frac{\partial \bar{\mathbf{E}}}{\partial \sigma_k})d\nu = \int_D \bar{\mathbf{E}}^{+} \cdot \bar{\mathbf{E}} \phi_k d\nu,
\end{equation}
where $\phi_k$ is the basis function at the $k$-th pixel, $D$ is the effective domain of $\phi_k$, and $\bar{\mathbf{J}}_m^{+}$, $\bar{\mathbf{J}}_e^{+}$ and $\bar{\mathbf{E}}^{+}$ are respectively adjoint magnetic source, electric source and electric field. The sensitivity of $\bar{E}_y$ to conductivity $\sigma$ can be written with
\begin{equation}\label{gb}
	\frac{\partial \bar{E}_y}{\partial \sigma} = \int_{-\infty}^{+\infty}\int_{D}\bar{\mathbf{E}}^{+}_{J_y} \cdot \bar{\mathbf{E}}dsdy.
\end{equation}
where $ds$ denotes the infinitesimal size inside effective domain $D$, $dy$ denotes the infinitesimal length along the $y$ direction, and $\bar{\mathbf{E}}^{+}_{J_y}$ is the adjoint electric field excited with $y$-direction electric source. (\ref{gb}) represents the integration of the product of the spatial electric fields within a rectangular prism with a base $D$ and infinitely extending along $y$ direction. Transforming both $\bar{\mathbf{E}}^{+}_{J_y}$ and $\bar{\mathbf{E}}$ to wave-number domain with (\ref{eq:2a}), (\ref{gb}) can be expressed with
\begin{equation}\label{meh}
	\frac{\partial \bar{E}_y}{\partial \sigma} = \frac{1}{2\pi} \int_{-\infty}^{+\infty}\int_{D}\hat{\mathbf{E}}^{+}_{J_y}(-k_x, y, z) \cdot \hat{\mathbf{E}}(k_x, y, z)dsdk_x.
\end{equation}
Noticing that $\bar{E}_x$ is an odd function of $k_x$ while $\bar{E}_y$ and $\bar{E}_z$ are even functions of $k_x$, (\ref{meh}) can finally expressed with
\begin{equation}\label{bew}
	\frac{\partial \bar{E}_y}{\partial \sigma} = \frac{1}{\pi} \int_{0}^{+\infty}\int_{D}(-\hat{E}^{+}_{{J_y},x}\hat{E}_{x}+\hat{E}^{+}_{{J_y},y}\hat{E}_{y}+\hat{E}^{+}_{{J_y},z}\hat{E}_{z})dsdk_x.
\end{equation}

\subsection{Variational autoencoder (VAE)}
The variational autoencoder (VAE) is a typical generative neural network structure, which is able to learn the distribution from the training dataset and generate new realizations conforming to the prior distribution \cite{kingma2013auto, Kingma_2019}. Generally, VAE is composed of an encoder $\mathcal{E}$ and a decoder $\mathcal{D}$. 
\color{black} In this framework, the encoder-decoder structure provides a probabilistic mapping between the high-dimensional conductivity model space and a compact latent space, enabling the extraction of representative geological features from training samples. 
\color{black} 
The encoder input is the data sampled from the prior distribution, and the encoder outputs respectively describe the mean and variance of a high-dimensional Gaussian distribution. \color{black} This stochastic encoding allows different realizations sharing similar structural features to be represented by nearby latent variables, which is essential for enforcing feature-level consistency during inversion. \color{black} Denoting the encoder input with $\mathbf{m}$, the output can be expressed with
\begin{equation}
	\mathcal{E}(\mathbf{m})=(\mathcal{E}_1(\mathbf{m}), \mathcal{E}_2(\mathbf{m}))=(\boldsymbol{\mu}_\mathcal{E}, \rm{log} \, {\boldsymbol{\sigma}}^2_\mathcal{E}),
\end{equation}
where $\boldsymbol{\mu}_\mathcal{E}$ and ${\boldsymbol{\sigma}}^2_\mathcal{E}$ are respectively the mean and variance of a Gaussian distribution $\mathcal{N}(\boldsymbol{\mu}_\mathcal{E}, {\boldsymbol{\sigma}}^2_\mathcal{E})$, from which the latent variable $\mathbf{v}$ is sampled:
\begin{equation}
	\mathbf{v} \sim \mathcal{N}(\boldsymbol{\mu}_\mathcal{E}, {\boldsymbol{\sigma}}^2_\mathcal{E}).
\end{equation}
$\mathbf{v}$ is then input into the decoder $\mathcal{D}$, yielding the reconstruction of $\mathbf{m}$:
\begin{equation}
	\tilde{\mathbf{m}} = \mathcal{D}(\mathbf{v}).
\end{equation}
The training objectives of the VAE are two-fold: making the reconstruction loss of VAE as small as possible, and simultaneously make the latent space approximate a standard normal distribution as close as possible. \color{black} After training, the decoder defines an implicit model manifold, onto which intermediate inversion results can be projected to constrain the solution space without altering the physical forward operator. \color{black} The objective function in VAE training can be expressed with 
\begin{equation}\label{VAEobjectfunc}
	\mathcal{L} = \mathcal{L}_{\text{rec}} + \lambda\sum_{q=1}^{Q}({\boldsymbol{\mu}}^2_{\mathcal{E},q}+{\boldsymbol{\sigma}}^2_{\mathcal{E},q}-\rm{log} \, {\boldsymbol{\sigma}}^2_{\mathcal{E},q}-1).
\end{equation}
Here, $\mathcal{L}_{\text{rec}}$ is the reconstruction loss, $Q$ is the dimension of the latent space, and $\lambda$ is the coefficient adjusting the relative weight of the reconstruction loss and Kullback-Leibler divergence loss. For a more detailed derivation of the objective function please refer to \cite{kingma2013auto}. After the training is finished, the decoder $\mathcal{D}$ can be regarded as a data generator that characterizes the prior distribution and is able to generate new realizations conforming to the prior distribution. 

\begin{figure}[!t]
	\centering
	\includegraphics[width=1\linewidth]{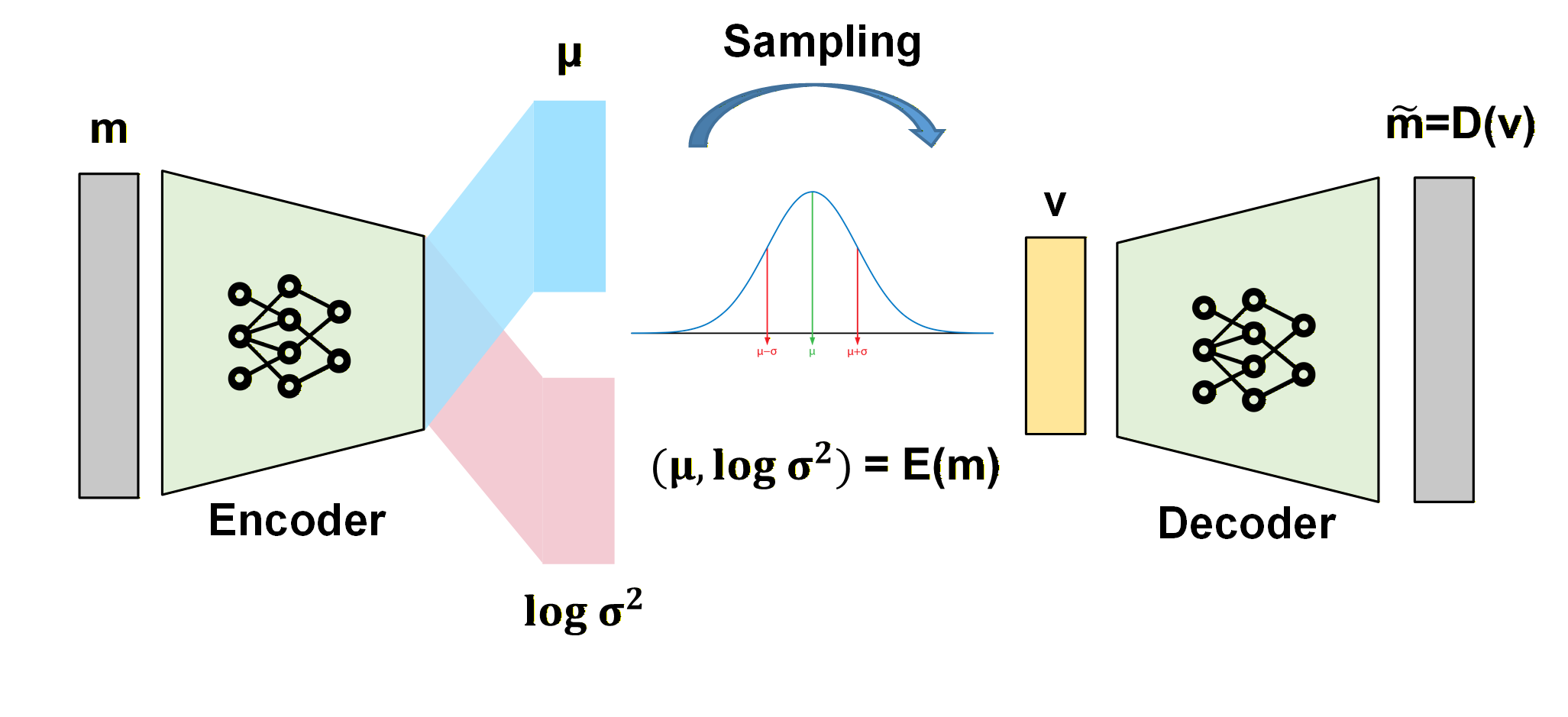}
	\caption{\color{black} Schematic of the variational autoencoder used in this work. \color{black}}
	\label{VAE}
\end{figure}

\color{black} The specific implementation of the VAE is not unique. 
In our implementation, the VAE is fine-tuned from the $\mathtt{kl-8f}$ VAE introduced in the noted latent diffusion work \cite{rombach2022high}. A schematic showing the structure is illustrated in Fig. \ref{VAE}, and the detailed structure of the VAE is included in Fig. S1 of the Supplementary Materials. The VAE have 86 million trainable parameters and 17 million non-trainable parameters in total. The three channel input and output are respectively merged into one channel by averaging parameter values along the channel direction. In our implementation, the reconstruction loss $\mathcal{L}_{\text{rec}}$ as included in (\ref{VAEobjectfunc}) is composed of the $L_2$ loss both in the pixel domain and feature domain, i.e., 
 \begin{equation}\label{Lrec}
 	\mathcal{L}_{\text{rec}} = \|\mathbf{m}-\tilde{\mathbf{m}}\|^2_2 + \gamma\|\mathcal{G}(\mathbf{m})-\mathcal{G}(\tilde{\mathbf{m}})\|^2_2,
 \end{equation}
 where $\mathcal{G}(.)$ denotes a pre-trained and frozen image processing neural network such as AlexNet \cite{krizhevsky2012imagenet}, and $\gamma$ is the hyper-parameter adjusting the relative weights of the two types of loss. The inclusion of the feature domain loss aims to better capture the detailed features of small-sized layers.

\color{black}

\subsection{Feature-based 2.5D CSEM inversion}
In the deterministic EM inversion framework, the 2.5D CSEM inversion can be formulated as optimizing the unknown model $\mathbf{m}$ by minimizing a loss function composed of both the data misfit term and the Tikhonov regularization \cite{habashy2004general}, i.e.,
\begin{equation}
    \mathcal{L}(\mathbf{m}) = \|\mathbf{W}_d(\mathbf{d}_\text{obs}-\mathcal{F}(\mathbf{m}))\|^2+\alpha \|\mathbf{W}_m\mathbf{m}\|^2, 
\end{equation}
and
\begin{equation}
	\hat{\mathbf{m}} = \arg \min_{\mathbf{m} \in \mathcal{S}} \mathcal{L}(\mathbf{m}).
\end{equation}
Here, $\mathbf{m}$ is the 2D conductivity distribution in the seabed, $\mathbf{d}_{\text{obs}}$ is the measured complex value $y$ component electric field as introduced in Section \ref{IIA}, and $\mathcal{S}\subseteq\mathbb{R}^{N_\mathbf{m}}$ is a set that captures the distribution that $\mathbf{m}$ is \emph{a priori} required to obey (e.g., non-negativity). In this work, we adopt a more informative generative prior described by the range of VAE decoder ($\mathcal{R}(\mathcal{D})$), i.e.,
\begin{equation} \label{chs}
	\hat{\mathbf{m}} = \arg \min_{\mathbf{m} \in \mathcal{R}(\mathcal{D})} \mathcal{L}(\mathbf{m}).
\end{equation}
As suggested by existing works \cite{bora2017compressed, guo2022nonlinear, guo2023three, zhou2024feature}, one possible approach to solve (\ref{chs}) is to reformulate the optimization problem in the latent space $\mathbb{R}^Q$, i.e., 
\begin{equation} \label{mqf}
	\hat{\mathbf{v}}=\arg\min_{\mathbf{v} \in \mathbb{R}^Q}\|\mathbf{W}_d(\mathbf{d}_\text{obs}-\mathcal{F}(\mathcal{D}(\mathbf{v})))\|^2+\alpha \|\mathbf{W}_m\mathcal{D}(\mathbf{v})\|^2,
\end{equation}
and obtain the optimized model $\hat{\mathbf{m}}$ from the optimized latent variable $\hat{\mathbf{v}}$, i.e.,
\begin{equation}\label{eq17}
	\hat{\mathbf{m}}=\mathcal{D}(\hat{\mathbf{v}}).
\end{equation}
However, we find (from both reports in \cite{shah2018solving, raj2019gan} and our numerical experiments) that (\ref{mqf}) can be easily stuck in local minima, since the cascade of two nonlinear operators $\mathcal{F}$ and $\mathbf{D}$ exaggerates the ill-posedness of the problem. 
References \cite{shah2018solving,raj2019gan} solve (\ref{chs}) with projected gradient-descent (PGD), that is, in each iteration $\mathbf{m}$ is first updated with gradient descent method and then projected to the $\mathcal{R}(\mathcal{D})$ set. 
However, first-order optimizations can take hundreds or even thousands of iterations to converge. Since in a 2.5D EM survey, each forward modeling and gradient computation requires solving dozens of linear systems with tens of thousands of unknowns, higher-order optimizations (e.g., Gauss-Newton) are preferred to accelerate the convergence.

\begin{figure*}[!t]
	\centering
	\includegraphics[width=1\linewidth]{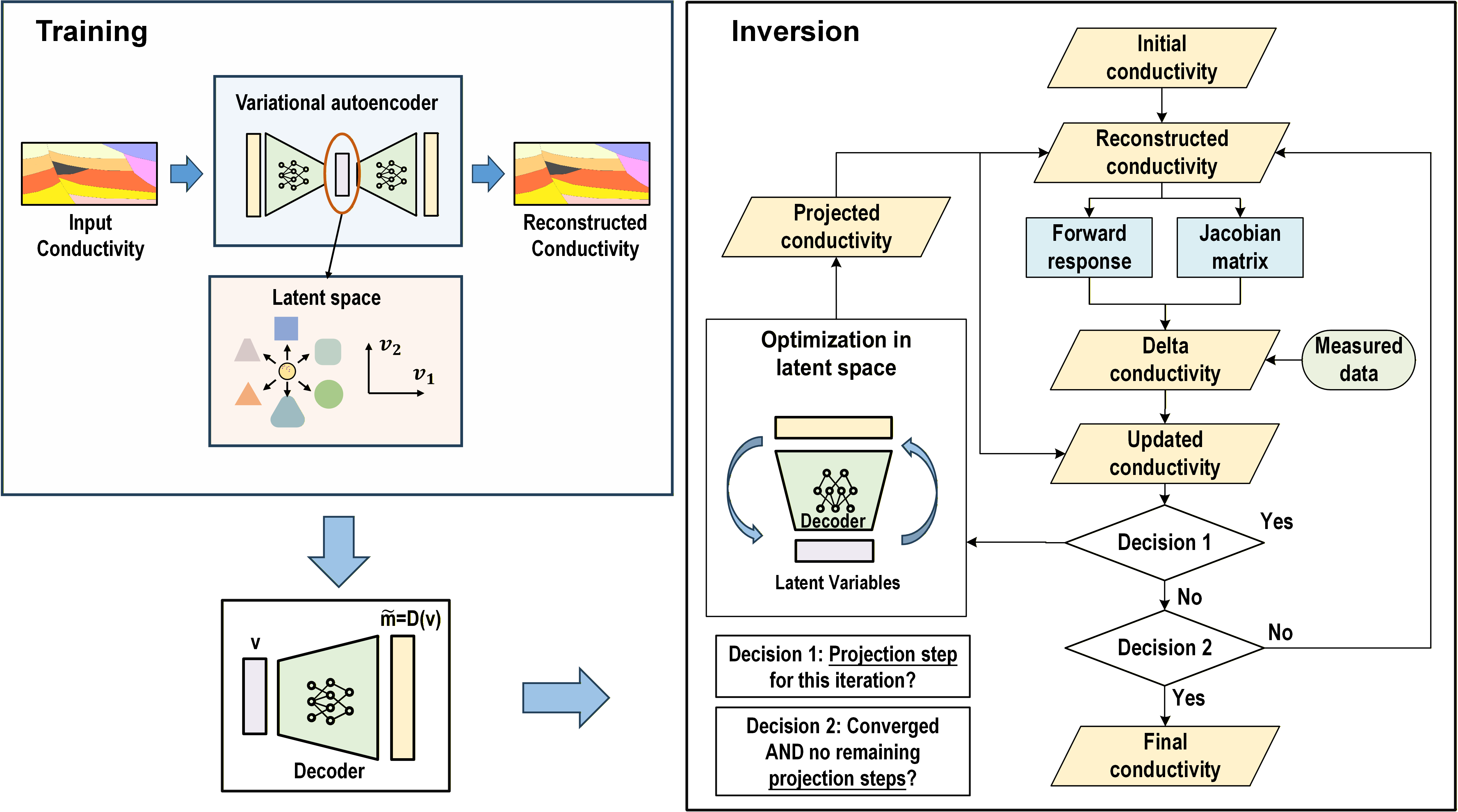}
	\caption{
\color{black}
The workflow of the proposed method consists of two stages: training and inversion. In the training stage, a variational autoencoder (VAE) is used to learn prior patterns in conductivity models. After training, the decoder is extracted and incorporated into the inversion stage, where iteratively updated conductivity models are projected onto the output space of the decoder. This approach differs from supervised end-to-end schemes, which typically require paired (data, model) examples for training.
\color{black}
	}
	\label{workflow}
\end{figure*}

We solve (\ref{mqf}) with a projected Gauss-Newton (PGN) method \cite{gonccalves2020gauss, nabou2023modified}. Similar to PGD, the PGN can also be separated into two stages. At the $k$-th iteration, $\mathbf{m}$ is first updated with Gauss-Newton method, that is
\begin{equation}
	\mathbf{m}_{k+1} = \mathbf{m}_k - \mathbf{H}^{-1}_k\mathbf{g}_k,
\end{equation}
where $\mathbf{H}_k$ and $\mathbf{g}_k$ are respectively Hessian matrix and gradient of $\mathcal{L}(\mathbf{m})$ with respect to $\mathbf{m}$ at the $k$-th iteration:
\begin{subequations}
	\begin{align}
		\mathbf{H}_k &= \mathbf{J}^T_k\mathbf{W}_d^T\mathbf{W}_d\mathbf{J}_k+\alpha\mathbf{W}^T_m\mathbf{W}_m,\\
		\mathbf{g}_k &= -\mathbf{J}^T_k\mathbf{W}_d^T\mathbf{W}_d(\mathbf{d}_\text{obs}-\mathcal{F}(\mathbf{m}_k))+\alpha\mathbf{W}_m^T\mathbf{W}_m\mathbf{m}_k,
	\end{align}
\end{subequations}
where $\mathbf{J}_k$ is the Jacobian matrix calculated by (\ref{eq5})-(\ref{bew}). In the second stage, the updated $\mathbf{m}_k$ is projected to $\mathcal{R}(\mathcal{D})$ by solving the following minimization problem: 
\begin{equation}\label{eq20}
	\hat{\mathbf{v}}_{k+1} = \arg \min_{\mathbf{v}}\|\mathbf{m}_{k+1}-\mathcal{D}(\mathbf{v})\|^2_2,
\end{equation}
and compute the $\hat{\mathbf{m}}_{k+1}$ with (\ref{eq17}).
The inversion terminates when the data misfit reaches a predefined level or exceeds the maximum iterations. The workflow of the proposed inversion is shown in Fig. \ref{workflow}. 
We would like to point out that in the inversion, the projection is only conducted at several \emph{projection steps} (the total number is generally smaller than 5 and with intervals between each other) instead of each iteration. The aim is to minimize the  generalization risk of VAE therefore reducing the possibility that the inversion is trapped in the local minima. In addition, because the total number of projections is relatively small, and the time to solve (\ref{eq20}) is not a computational bottleneck compared to the time for forward modeling and gradient computation, we directly use first-order gradient descent (e.g., adaptive moment estimation (ADAM, \cite{kingma2017adam})).

\section{Numerical Experiments}
\color{black}
\subsection{VAE Algorithm Test} \label{IIIA}
\color{black}

\begin{figure}[!t]
	\centering
	\includegraphics[width=1\linewidth]{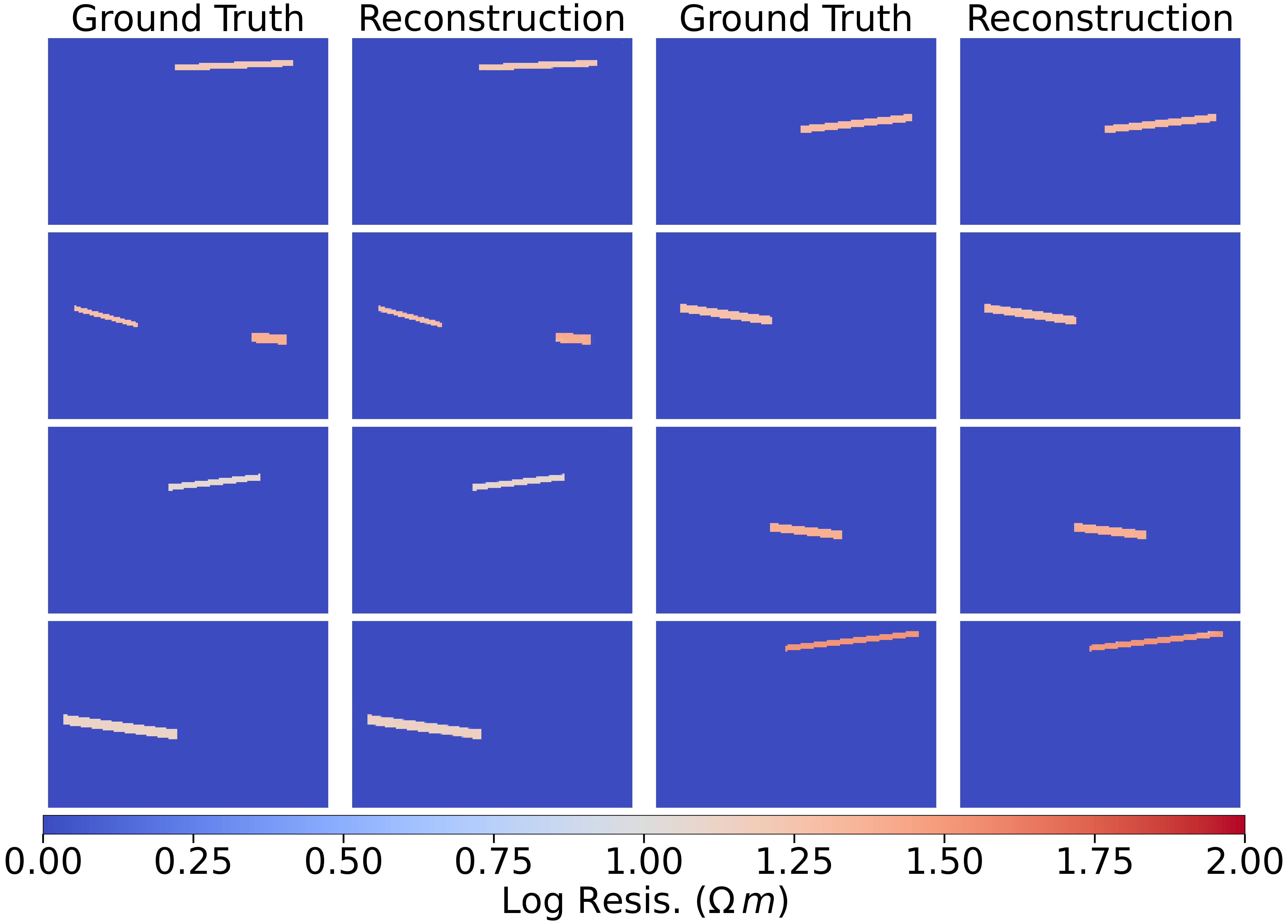}
	\caption{Visualization of some example images in the training dataset and corresponding reconstructions.}
	\label{datasets}
\end{figure}

\begin{figure}[!t]
	\centering
	\includegraphics[width=1\linewidth]{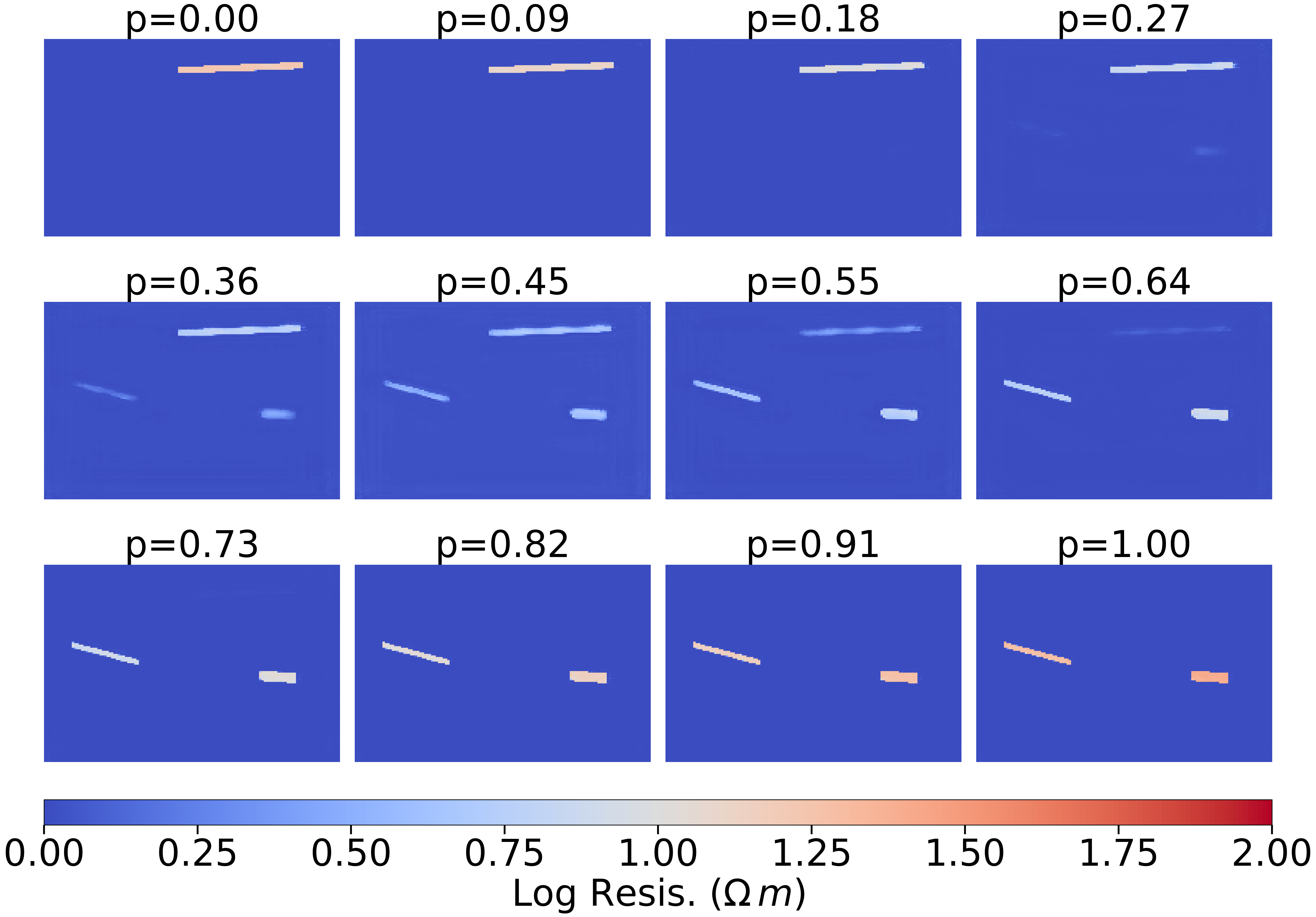}
	\caption{Visualization of intermediate models transferring between two ground truth models in Fig. \ref{datasets}.}
	\label{9s}
\end{figure}

\begin{figure*}[tb]
	\centering
	\includegraphics[width=1.5\columnwidth]{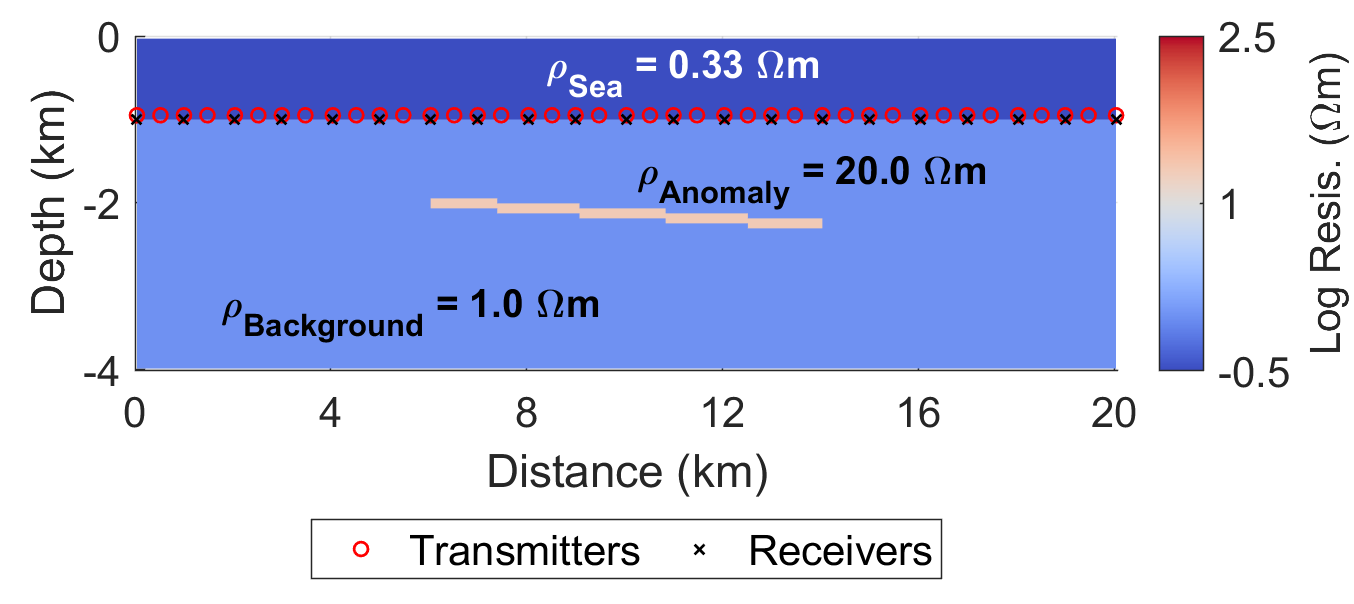}
	\caption{\color{black}A visualization of test model A. Forty-one transmitters and twenty-one receivers are respectively positioned at the depth of 950 m and 1000 m. The resistivities of the seawater, seabed background, and resistive anomaly are clearly labeled in the image. \color{black}}
	\label{model7}
\end{figure*}


\color{black}
A fine-tune dataset containing 50000 images is simulated, 90\% used for training and 10\% for validation. Each image has a size of 128 $\times$ 128, and has a pattern of one or two thin resistive layers exist in the uniform conductive background. This corresponds to the \emph{a priori} knowledge that, the DoI contains a thin and layer-shaped resistive hydrocarbon reservoir residing in the uniform conductive seabed, but the location, shape and size of the layer is to be determined in the inversion. The attribute value is set to 0 for the background and sampled from uniform distribution $\mathcal{U}(1.3, 2)$ for the layers. 
Some examples in the training dataset is shown in Fig. \ref{datasets}. The fine-tuning converges in 10 steps.  

After the training finished, the performance of the fine-tuned VAE is tested. \color{black} First, the ground truth models $\mathbf{m}$ are compared to the corresponding VAE reconstructions $\mathbf{m}_\text{rec}=\mathcal{D}(\mathcal{E}_1(\mathbf{m}))$. 
Second, the continuity of the latent space is checked: given two models $\mathbf{m}_1$ and $\mathbf{m}_2$ and their corresponding latent variables $\mathbf{v}_1=\mathcal{E}_1(\mathbf{m}_1)$ and $\mathbf{v}_2=\mathcal{E}_1(\mathbf{m}_2)$, the intermediate models $\mathbf{m}_{1,2,p}=\mathcal{D}((1-p)\mathbf{v}_1+p\mathbf{v}_2)$ are calculated. 
As shown in Figs. \ref{datasets} and \ref{9s}, the reconstructions well resemble the ground truths, and the reconstructed models can transform smoothly between each other. The two experiments together verify the efficacy of the VAE.


\color{black}
\subsection{Simple Model Test}\label{Experiment1}
We set up the same survey configuration as adopted in \cite{abubakar20082} to verify the correctness of the proposed algorithm. The DoI ranges from 0 to 20 km horizontally and from --4 to --1 km vertically, with a 1-km-thick seawater layer above it. Forty-one inline electric dipoles are uniformly distributed at the depth of 950 m, excited one at a time, and twenty-one receivers are uniformly distributed on the seafloor. We adopt a nonuniform mesh that is refined in the transmitter–receiver region and uniform elsewhere, with 251 grids in the horizontal direction and 156 grids in the vertical direction. Fig. \ref{model7} shows the survey geometry and transmitter–receiver configuration introduced in this section. The transmitters are denoted by red circles, and the receivers are denoted by black crosses. 
\color{black}
\color{black}

\begin{figure*}[!t]
	\centering
	\begin{minipage}{\textwidth}
		
		\centering
		
		\subcaptionbox{\label{fig10_a}}{\includegraphics[width=\mywidthB]{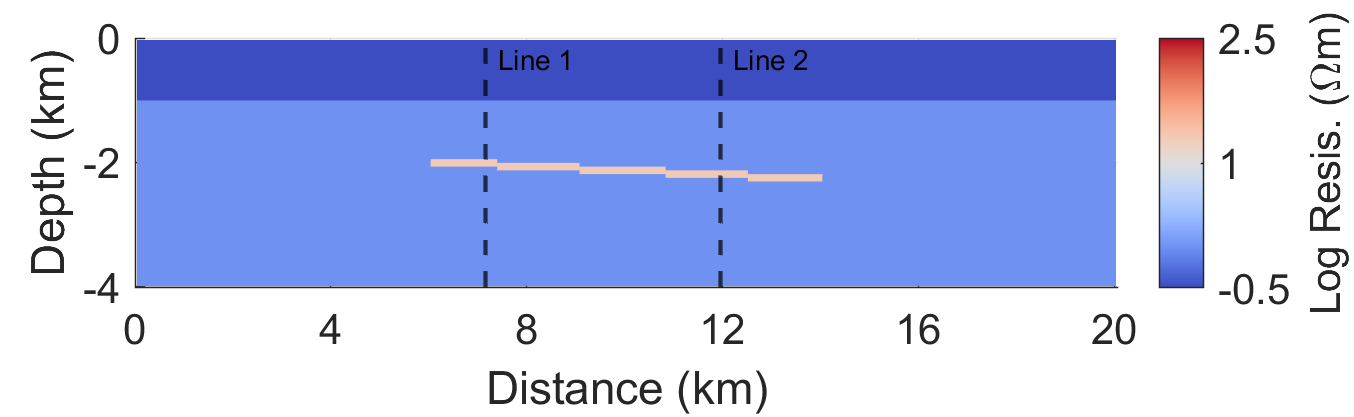}}\hfil
		\subcaptionbox{\label{fig10_b}}{\includegraphics[width=\mywidthB]{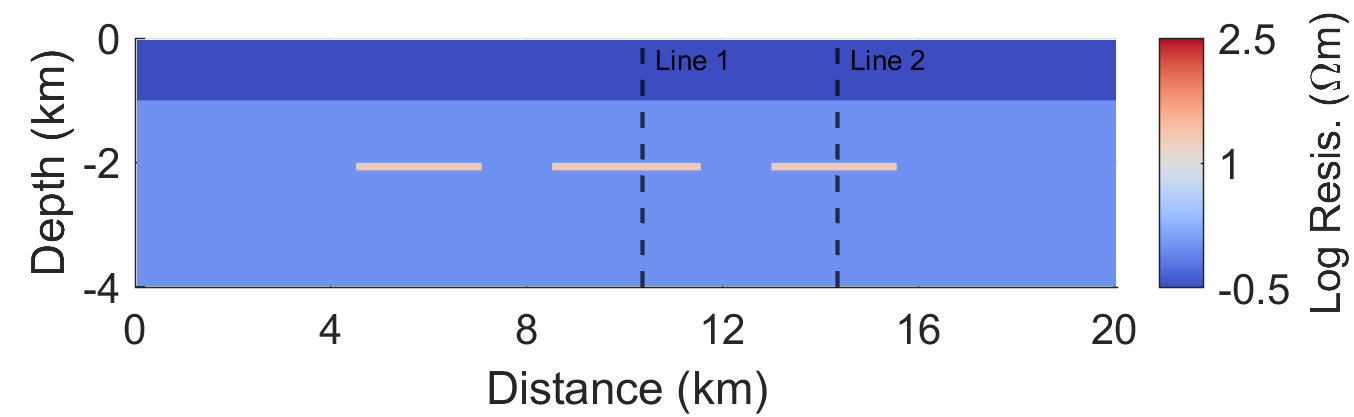}}\hfil
		\subcaptionbox{\label{fig10_c}}{\includegraphics[width=\mywidthB]{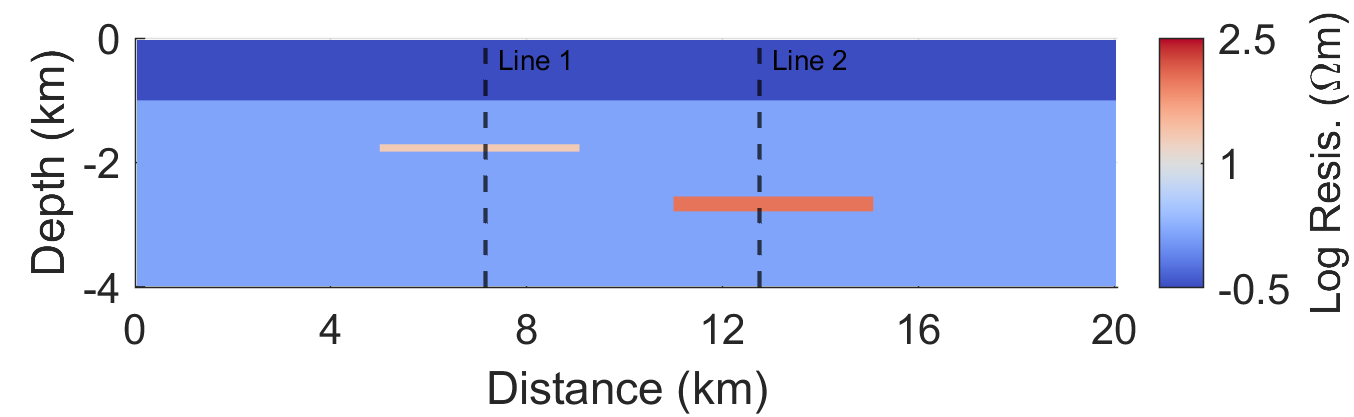}}
		
		\subcaptionbox{\label{fig10_d}}{\includegraphics[width=\mywidthB]{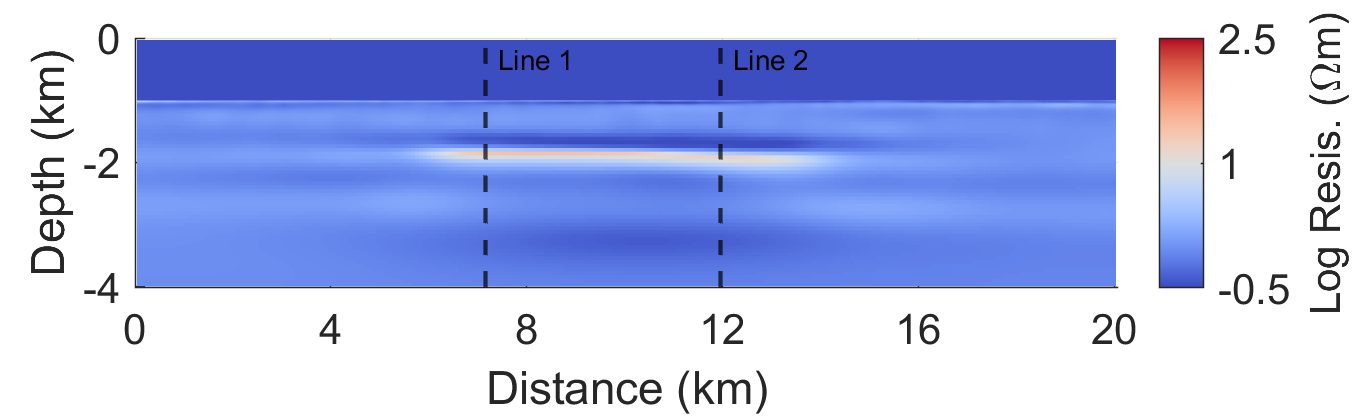}}\hfil
		\subcaptionbox{\label{fig10_e}}{\includegraphics[width=\mywidthB]{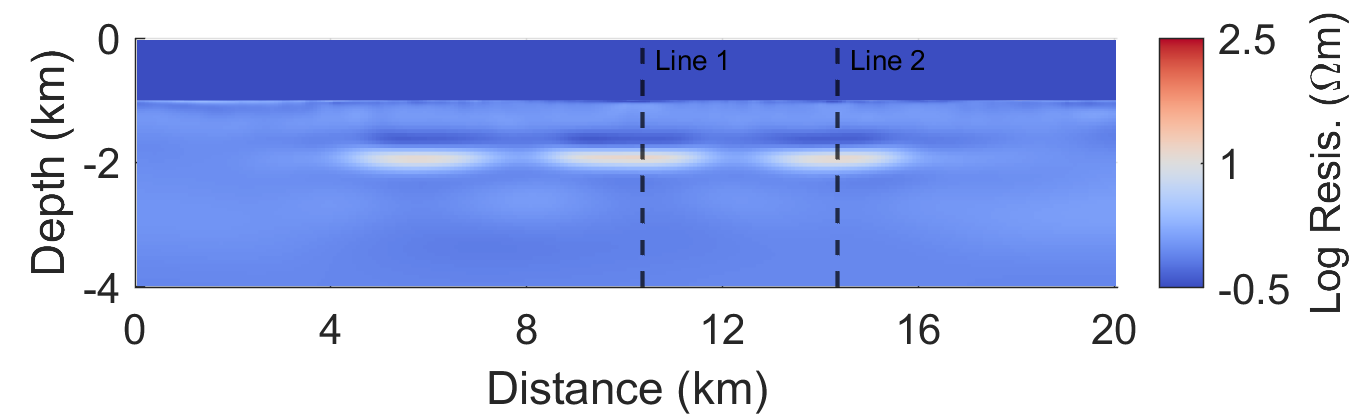}}\hfil
		\subcaptionbox{\label{fig10_f}}{\includegraphics[width=\mywidthB]{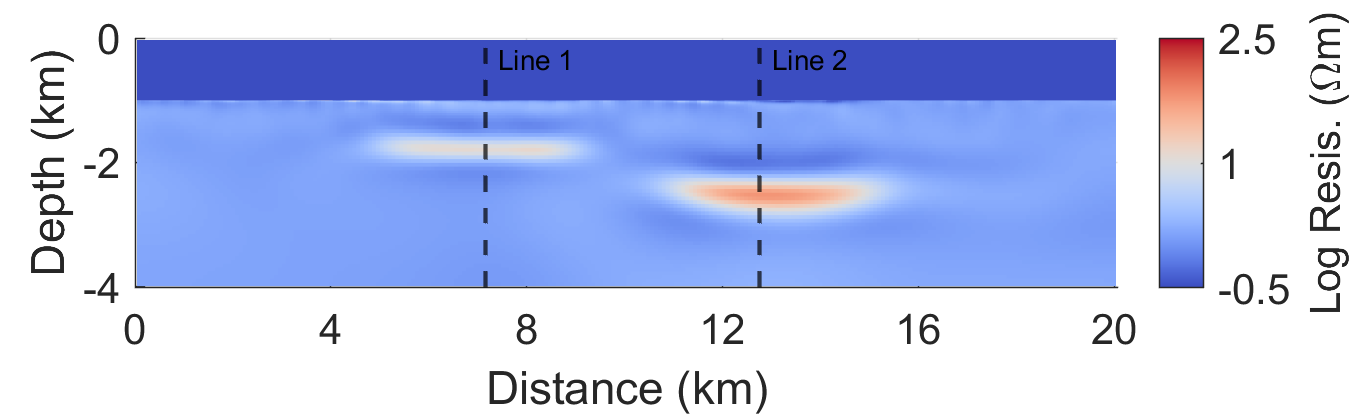}}
		
		\subcaptionbox{\label{fig10_g}}{\includegraphics[width=\mywidthB]{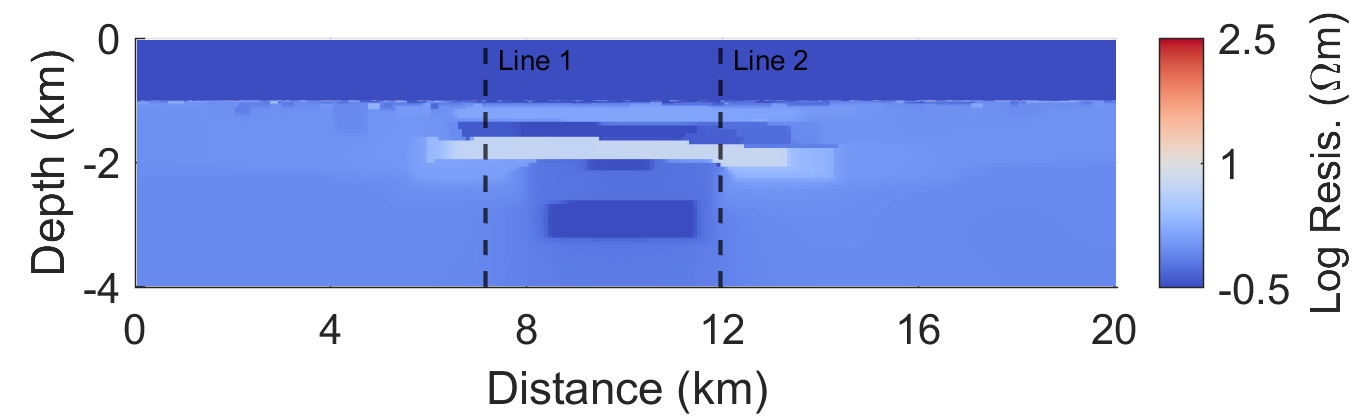}}\hfil
		\subcaptionbox{\label{fig10_h}}{\includegraphics[width=\mywidthB]{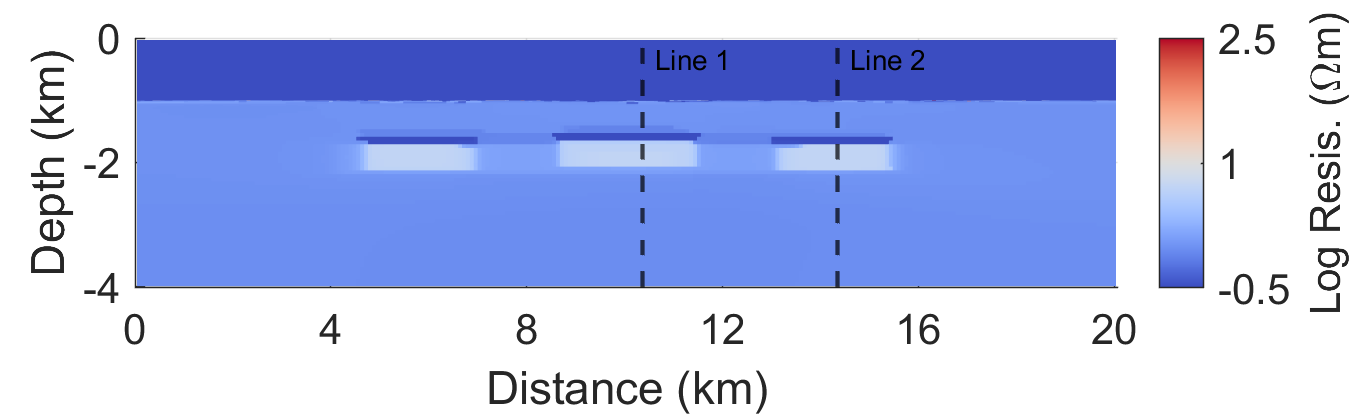}}\hfil
		\subcaptionbox{\label{fig10_i}}{\includegraphics[width=\mywidthB]{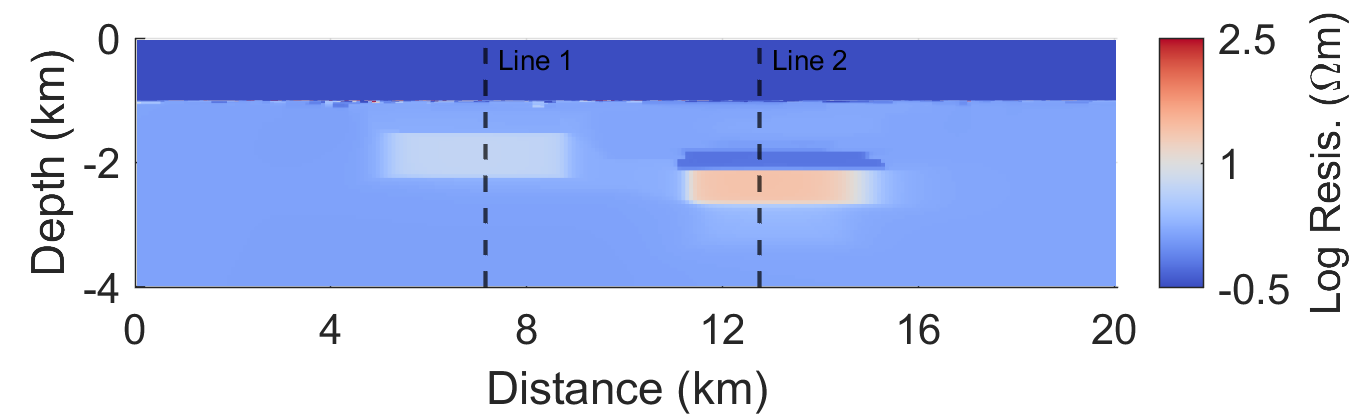}}
		
		\subcaptionbox{\label{fig10_j}}{\includegraphics[width=\mywidthB]{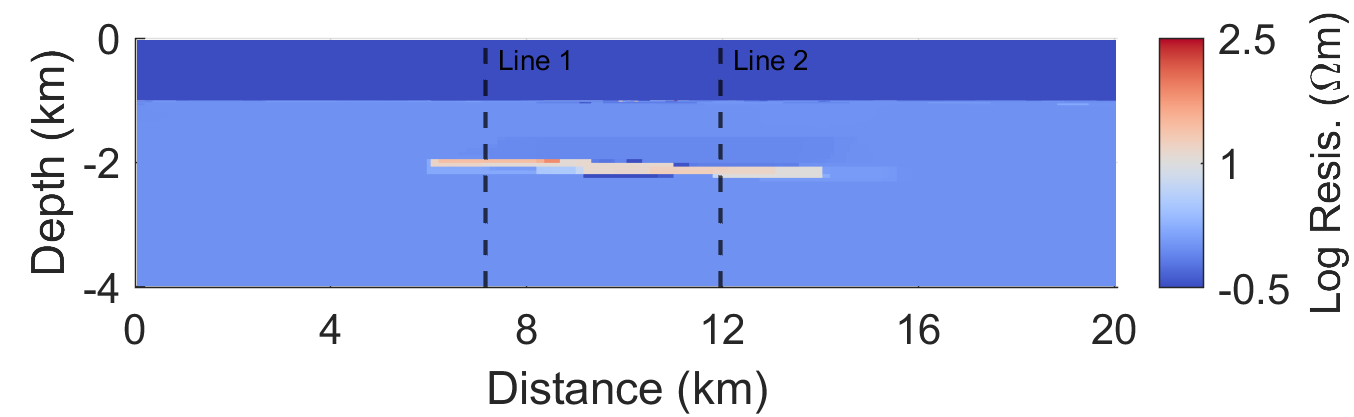}}\hfil
		\subcaptionbox{\label{fig10_k}}{\includegraphics[width=\mywidthB]{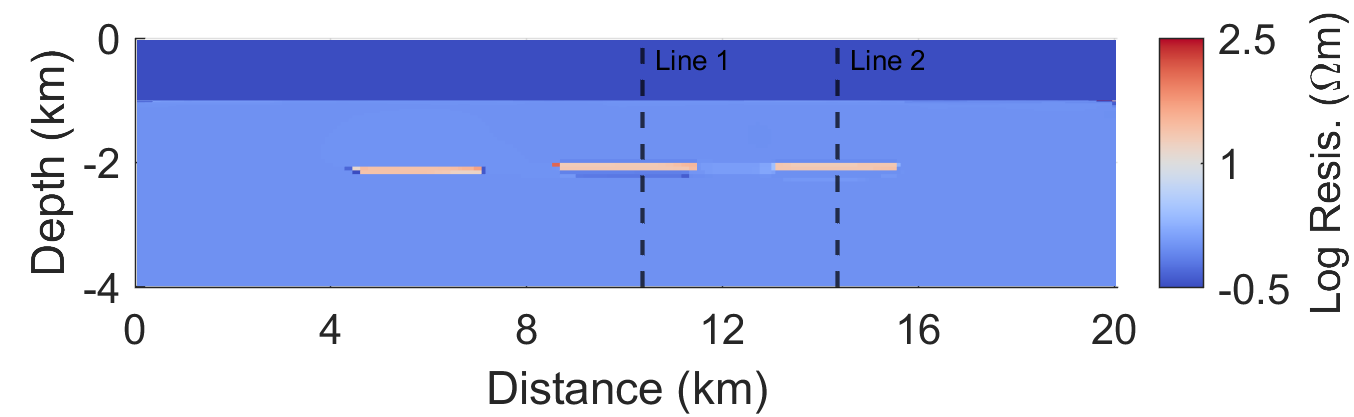}}\hfil
		\subcaptionbox{\label{fig10_l}}{\includegraphics[width=\mywidthB]{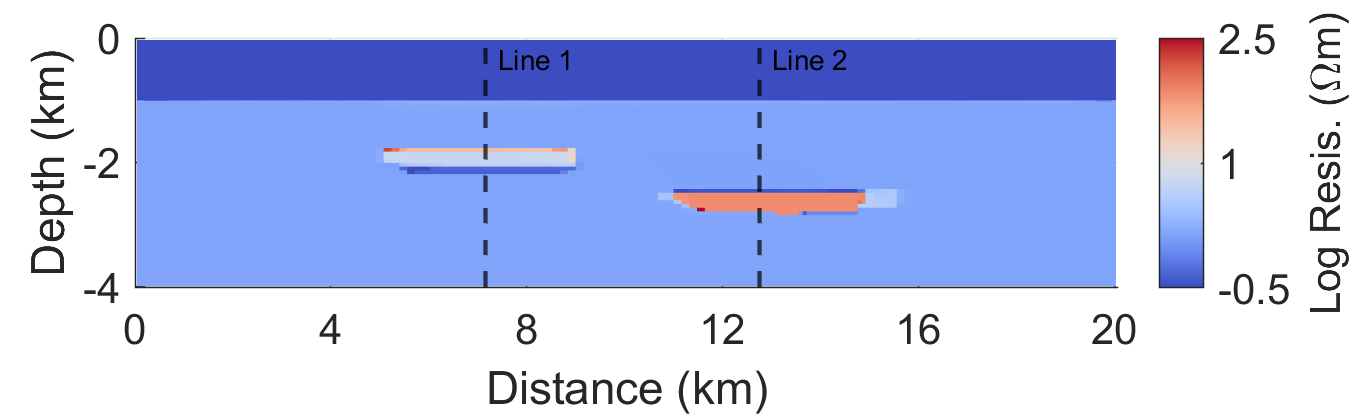}}
		\caption{The ground truth models and reconstructed models as introduced in Section \ref{Experiment2}. The four rows from top to bottom respectively represent the ground truth models, reconstructed models with $L_2$ regularization, W$L_2$ regularization and feature-based inversion.}
		\label{fig10}
	\end{minipage}
	
	\vspace{0.5cm}
	
	\begin{minipage}{\textwidth}
		\centering
		
		\subcaptionbox{\label{fig11_a}}{\includegraphics[width=\mywidthA]{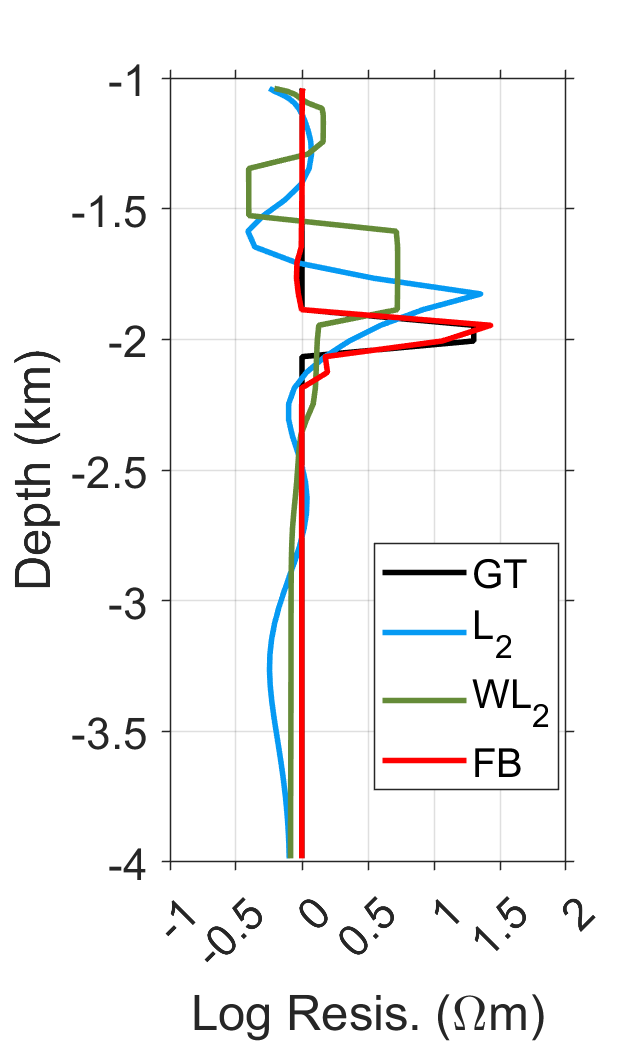}}\hfil
		\subcaptionbox{\label{fig11_b}}{\includegraphics[width=\mywidthA]{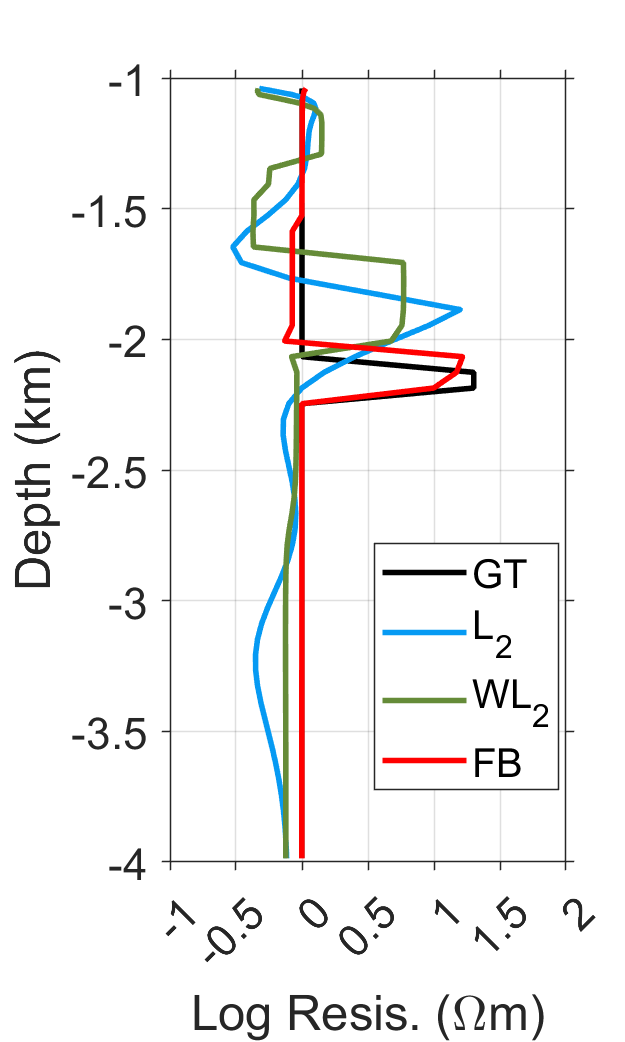}}\hfil
		\subcaptionbox{\label{fig11_c}}{\includegraphics[width=\mywidthA]{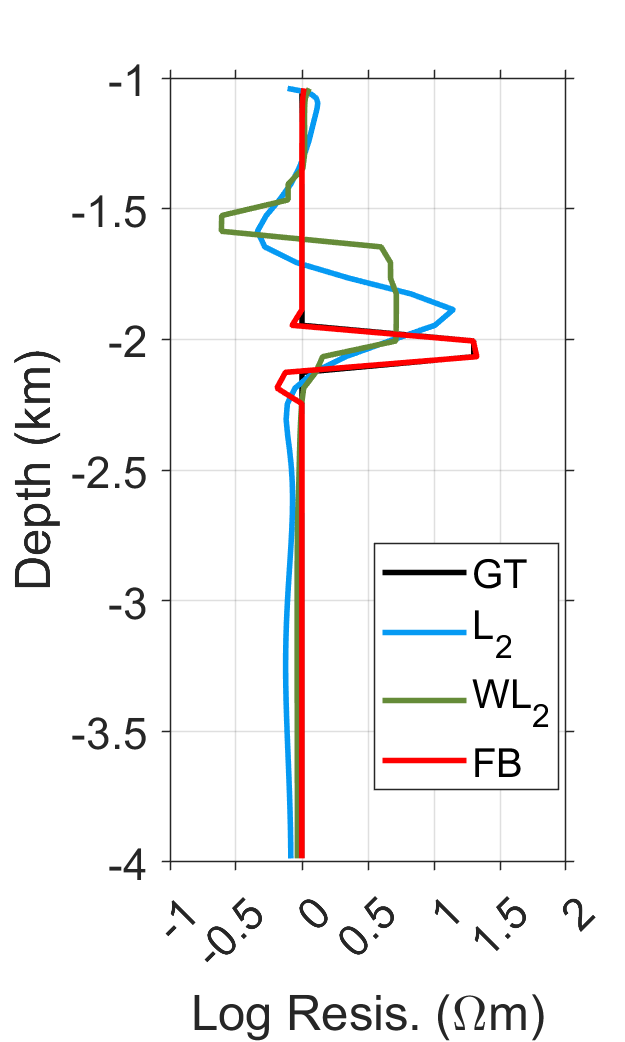}}\hfil
		\subcaptionbox{\label{fig11_d}}{\includegraphics[width=\mywidthA]{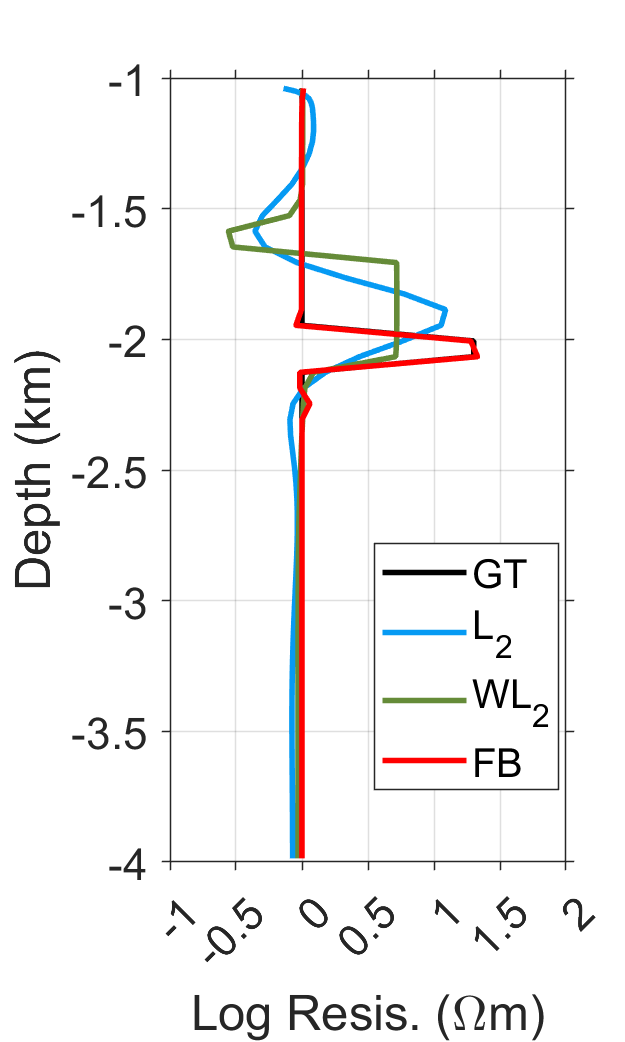}}\hfil
		\subcaptionbox{\label{fig11_e}}{\includegraphics[width=\mywidthA]{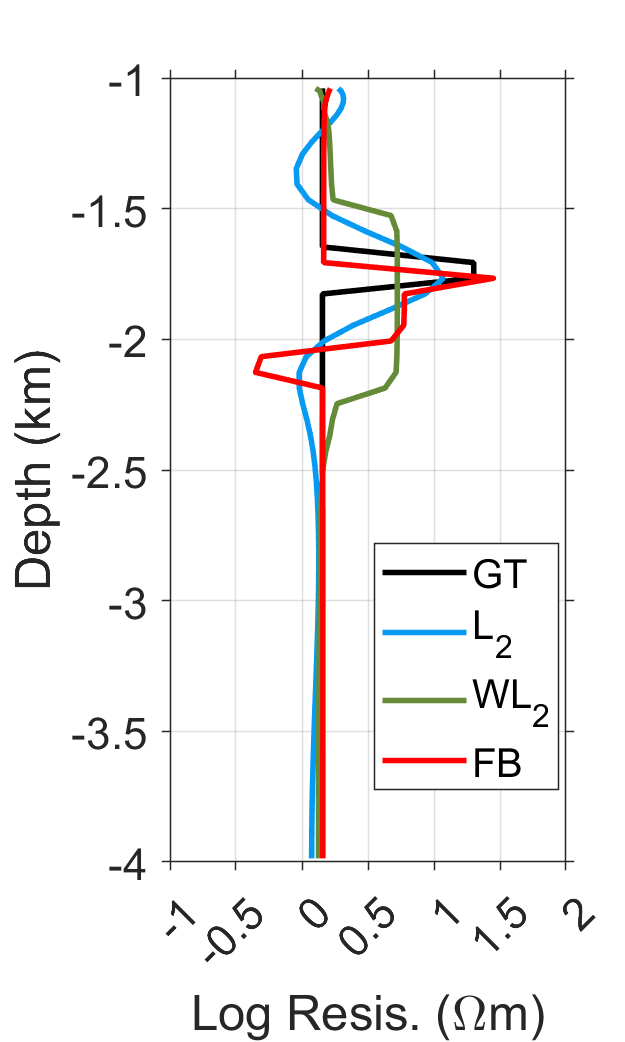}}\hfil
		\subcaptionbox{\label{fig11_f}}{\includegraphics[width=\mywidthA]{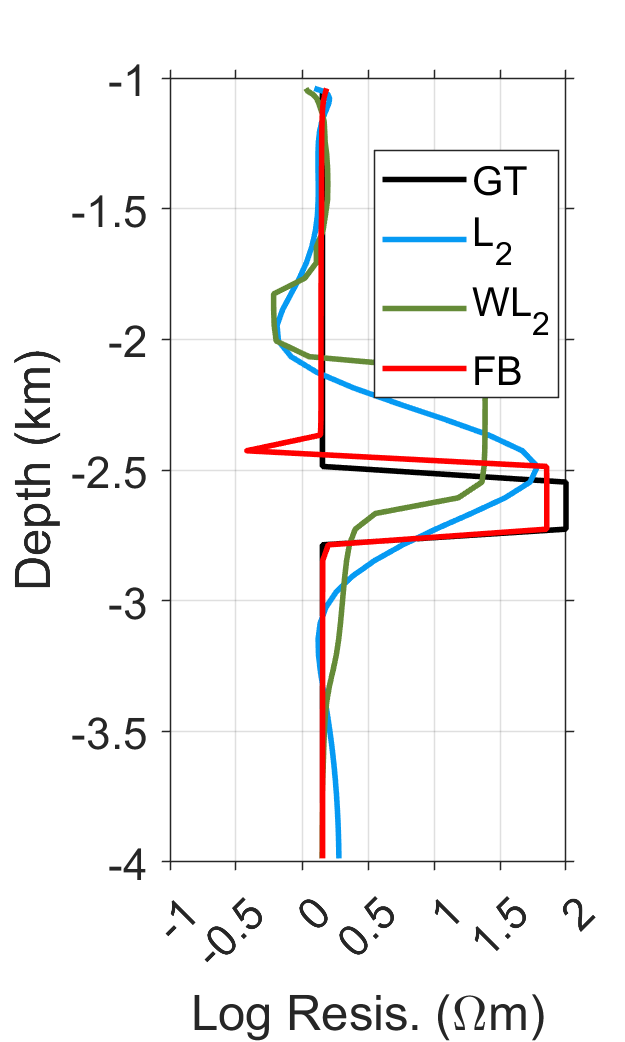}}
		
		\subcaptionbox{\label{fig11_g}}{\includegraphics[width=\mywidthB]{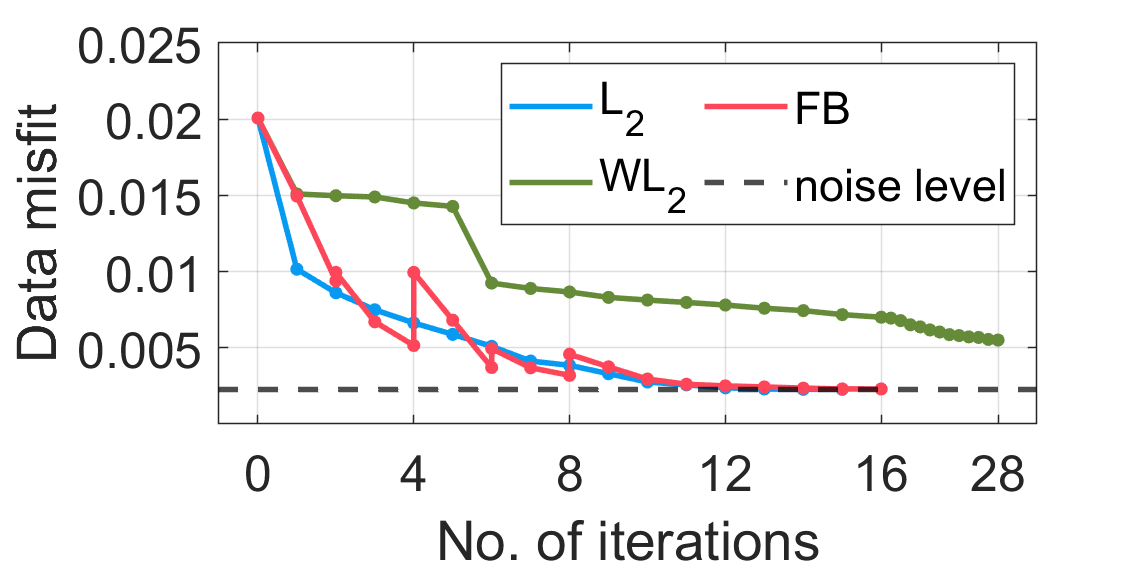}}\hfil
		\subcaptionbox{\label{fig11_h}}{\includegraphics[width=\mywidthB]{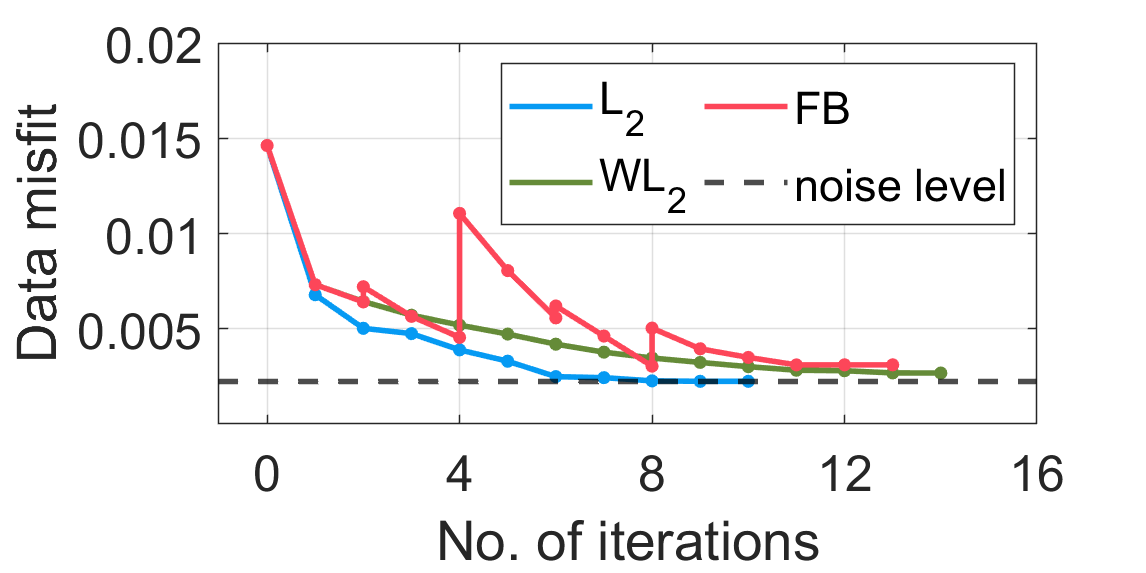}}\hfil
		\subcaptionbox{\label{fig11_i}}{\includegraphics[width=\mywidthB]{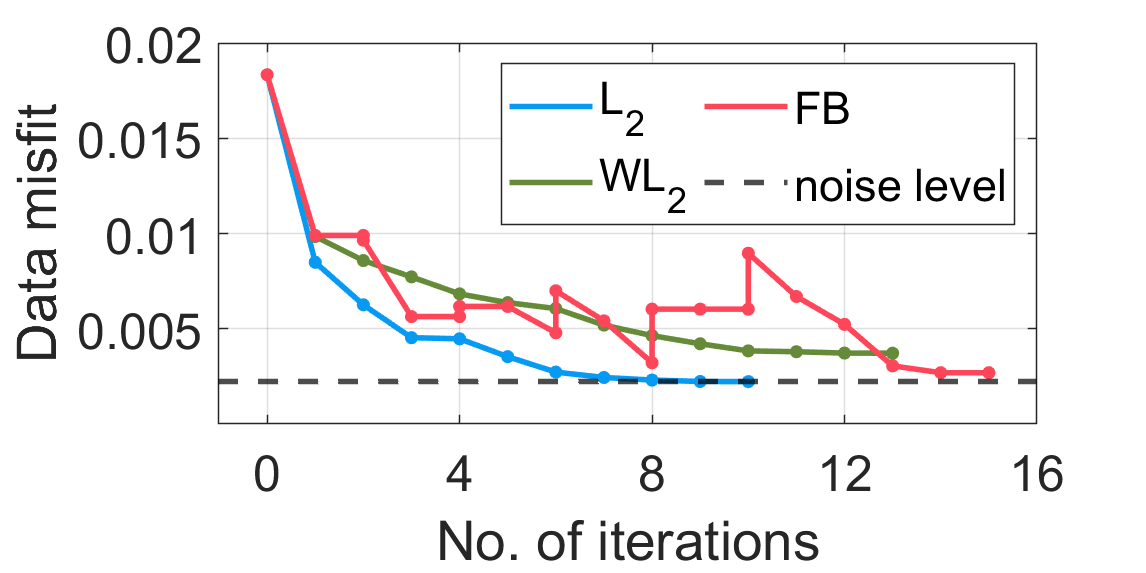}}
		
		\caption{\color{black}Figs. \ref{fig11_a} and \ref{fig11_b} respectively show the ground truth and recovered resistivity curves at $\mathtt{Line 1}$ and $\mathtt{Line 2}$ (the locations are denoted with black dash lines in Fig. \ref{fig10_a}),  and Fig. \ref{fig11_g} display the data misfit during the reconstruction of test model A. Similarly, Figs. \ref{fig11_c}, \ref{fig11_d} and \ref{fig11_h} show the resistivity curves corresponding to $\mathtt{Line 1}$ and $\mathtt{Line 2}$ in Fig. \ref{fig10_b}, and data misfit corresponding to test model B.  Figs. \ref{fig11_e}, \ref{fig11_f} and \ref{fig11_i} show the resistivity curves corresponding to $\mathtt{Line 1}$ and $\mathtt{Line 2}$ in Fig. \ref{fig10_c}, and data misfit corresponding to test model C.\color{black}}
		\label{fig11}
	\end{minipage}
\end{figure*}

Based on the survey geometry and transceiver configurations shown in Fig. \ref{model7}, we generate three ground truth models, respectively referred to as test model A, B and C, to simulate marine hydrocarbon exploration scenarios involving thin resistive layers with different shapes and geometries embedded in the seabed. Test model A is shown in both Fig. \ref{model7} and Fig. \ref{fig10_a}, and test model B and C are respectively shown in Figs. \ref{fig10_b} and \ref{fig10_c}. 
Test models A and B contain resistive layers 1 km below the seabed, and the resistivities of the seawater, sedimentary background and resistive layers are respectively 0.33 $\Omega$m, 1 $\Omega$m and 20 $\Omega$m. In test model A, the 8-km-length resistive layer is tilted by 2 degrees, while in test model B, three shorter resistive layers with lengths of 2.5 km, 3 km, and 2.5 km are arranged in parallel and separated by two 1.5-km gaps.
Test model C contains two 4-km-long resistive layers, one is 120 m thick, located 800 m below the seafloor and has a resistivity of 20 $\Omega$m, while the other is 240 m thick, located 1500 m below the sea floor and has a resistivity of 100 $\Omega$m. The background sedimentary layer has a resistivity of 1.43 $\Omega$m. 
In this experiment, the adopted frequencies are 0.25 and 0.75 Hz. Five-percent Gaussian noise is added to the simulated data in each case using (\ref{noise}):
\begin{equation}\label{noise}
	d_{\text{obs}, mnk}^{\text{noise}} = d_{\text{obs}, mnk}(1+\beta (n_{1, mnk}+in_{2, mnk})),
\end{equation}
where $\beta$ is the noise level, and $n_{1, mnk}$ and $n_{2, mnk}$ are independently sampled from standard normal distribution $\mathcal{N}(0, 1)$. $m,n$ and $k$ are respectively the index of transmitter, receiver and frequency.

\begin{figure*}[!t]
	\centering
	\includegraphics[width=1\linewidth]{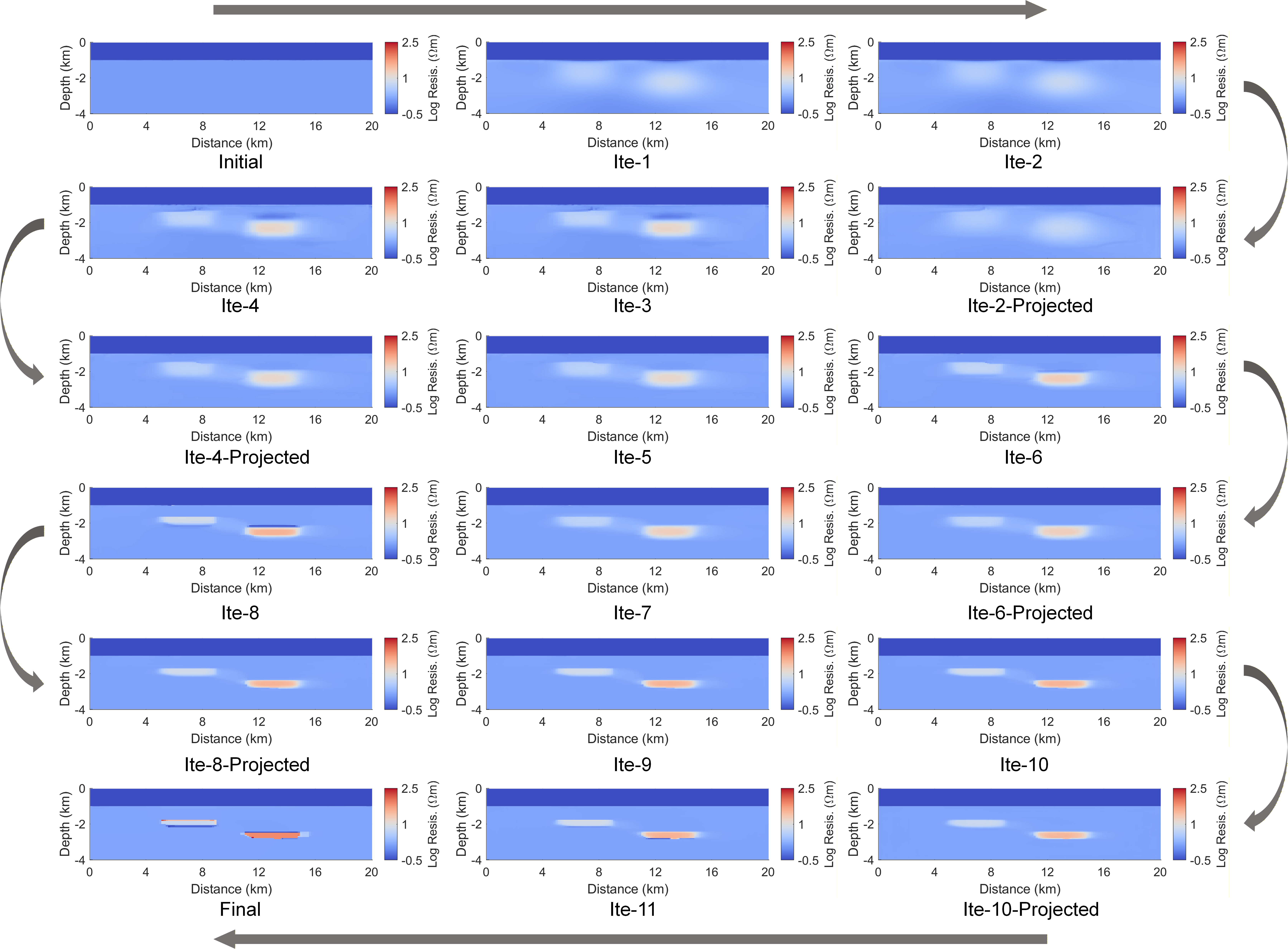}
	\caption{The model evolution \color{black} in the feature-based inversion of test model C as reported in Section. \ref{Experiment1}. \color{black}}
	\label{fig9}
\end{figure*}

We conducted inversions with conventional $L_2$ regularization, weighted $L_2$ regularization and the proposed feature-based inversion (abbreviated as $L_2$, W$L_2$ and FB hereafter). For the three test models, the initial models are  uniform backgrounds of 1 $\Omega$m, 1 $\Omega$m and 1.43 $\Omega$m, which are identical for all inversions. The reconstructed models obtained by $L_2$, W$L_2$ and FB inversion are respectively shown in the second, third and fourth rows of Fig. \ref{fig10}. 
For a clearer comparison, in each model we select two depth traces ($\mathtt{Line 1}$ and $\mathtt{Line 2}$, whose locations are shown by black dash lines in Figs. \ref{fig10_a}, \ref{fig10_b} and \ref{fig10_c}). The resistivity distributions of the ground truth models and reconstructions along the traces are presented in Figs. \ref{fig11_a} to \ref{fig11_f}, and data misfit curves are displayed in Figs. \ref{fig11_g} to \ref{fig11_i}. 

In both $L_2$ and W$L_2$ reconstructions, the resistive layers in all three test models can be identified with distinguishable boundaries, and their locations and sizes show no large deviations from the ground truth. However, in the conventional inversions, the recovered layers are  shallower than their actual depths, with deviations of about 300--400 m, and the layer thickness is overestimated. 
For example, in W$L_2$ reconstruction the resistive layer is about twice the real thickness. 
In W$L_2$ reconstructions, conductive anomalies can be observed across the three cases. For example, in Fig. \ref{fig10_g}, one of the conductive anomalies is close to the upper surface of the resistive layer, and the other is below the resistive layer with offset ranging from 8 to 12 km and depth at about 3 km. We note that similar artifact patterns in 2.5D mCSEM inversion have also been reported in previous studies (for instance, see Figure 7 in \cite{abubakar20082}), which reflect the non-uniqueness of the solution and hinder accurate reservoir evaluation.

\begin{figure*}[!t]
	\centering
	\includegraphics[width=1\linewidth]{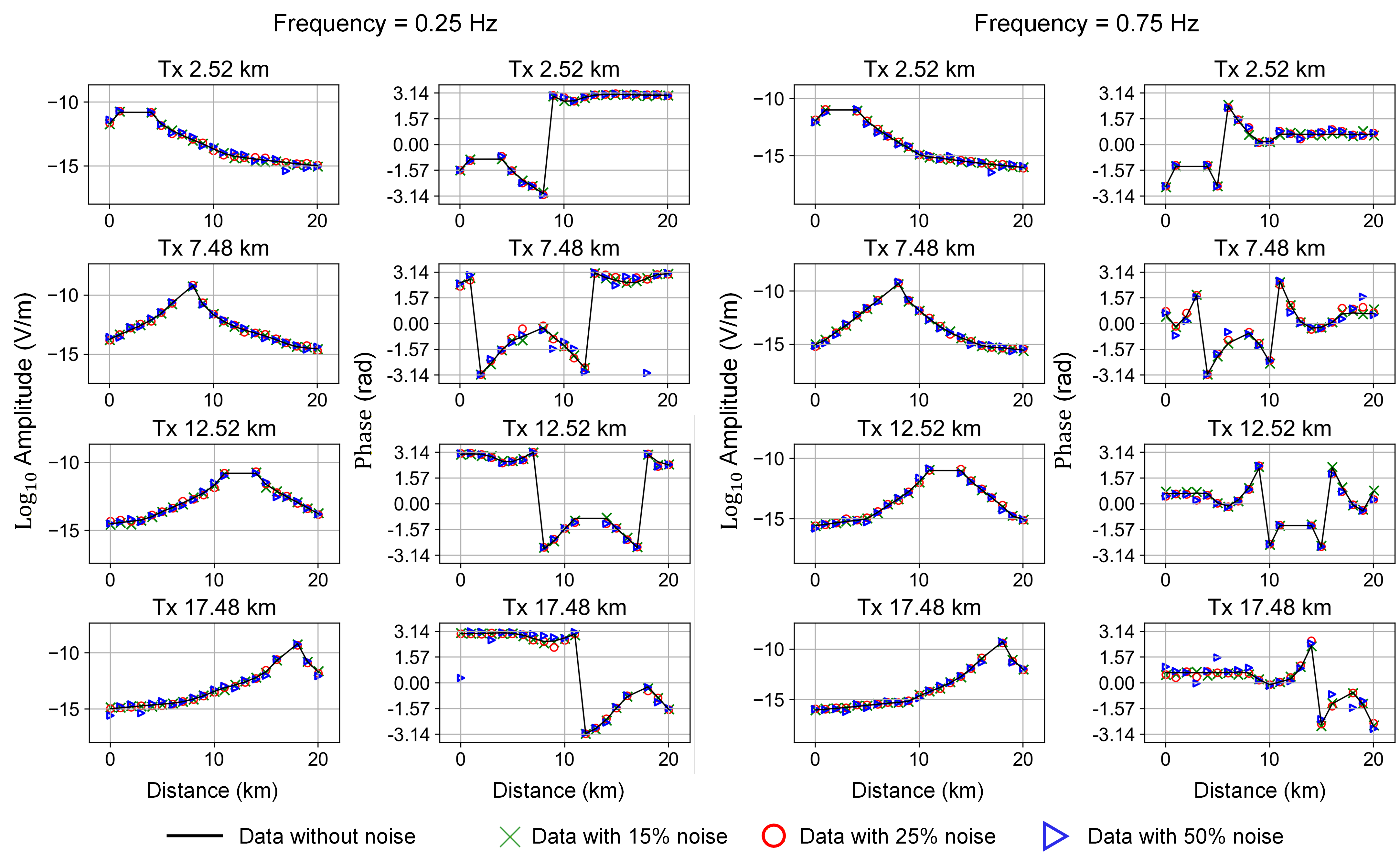}
	\caption{\color{black} The comparison between the noise-free simulated data, data with 15\% noise, 25\% noise and 50\% noise, corresponding to test model C. The four columns from left to right respectively show the $\log_{10}$ amplitude of 0.25 Hz data, phase of 0.25 Hz data, $\log_{10}$ amplitude of 0.75 Hz data, and phase of 0.75 Hz data. The four rows from top to bottom show the fields at all active receivers with the transmitter excited at 2.52 km, 7.48 km, 12.52 km, and 17.48 km horizontally. \color{black}}
	\label{test_model_C_data_vs_noise}
\end{figure*}

Comparing the inversion results,
clear improvements in FB inversions can be observed. Across Figs. \ref{fig11_a} to \ref{fig11_f}, the resistivity curves reconstructed by FB (plotted in red) are closer to the ground truth curves (plotted in black) than those obtained by $L_2$ (plotted in blue) and W$L_2$ (plotted in green). The artifacts observed in conventional inversions are largely suppressed, yielding more uniform backgrounds and well-resolved targets whose depths, thicknesses, and attribute values closely match the corresponding ground truth models. 
Across the three test models, both $L_2$ and FB inversions converge within 10--15 iterations with final data misfits close to the noise level, while the W$L_2$ inversion appears to be trapped in local minima for test models A and C. In the feature-based inversions, the \emph{projection steps} are iterations 2, 4, 6 and 8 for test model A and B,  and at iterations 2, 4, 6, 8 and 10 for test model C. 
The pairs of points with the same horizontal coordinates on the FB data misfit convergence curves in Figs. \ref{fig11_g}--\ref{fig11_i} respectively represent the data misfits before and after these projections. 

We also highlight the generalization ability of the proposed method: resistive layers with different shapes, such as inclined layers or layers composed of discontinuous segments, can be clearly delineated. We attribute this to the fact that the adopted VAE achieves a balance between reconstruction accuracy and generalization. Pre-training on a diverse set of natural images allows the neural network to learn general and rich prior knowledge for image feature extraction and reconstruction. In the fine-tuning stage, this prior knowledge is efficiently transferred to the distribution space of the conductivity model to be reconstructed.


\color{black} Some metrics, including data misfit (DM), intersection over union (IoU), mean square error (MSE) and structural similarity (SSIM), are adopted to assess the reconstructions. Their definitions are introduced in Appendix, and the results of test models A--C are reported in Table \ref{table1}. \color{black} 
$L_2$ reconstructions presents better MSE than W$L_2$, which may be explained by the observation that W$L_2$ inversion tends to introduce conductive anomaly with larger conductivities and thicker resistive layers compared to $L_2$ inversion. In contrast, W$L_2$ reconstructions outperform $L_2$ in SSIM, which may be explained by the observation that W$L_2$ yields a clearer boundary and more unique background compared to $L_2$ inversion. Both $L_2$ and W$L_2$ inversions have their own features, however, in all the experiments, the FB reconstructions outperform them in all the three model domain metrics.

To find out how the reconstruction is updated in the feature-based inversion, we visualize the \color{black} evolution of test model C in FB inversion in Fig. \ref{fig9}. \color{black} The models updated with Gauss--Newton optimization in the $i$-th iteration are indexed \emph{Ite-i}, and the projections of \emph{Ite-i} models to  $\mathcal{R}(\mathcal{D})$ are indexed \emph{Ite-i-Projected}. Comparing models before and after the projection, the non-uniform background and conductive anomalies are effectively suppressed and become more uniform, thereby alleviating non-uniqueness and avoiding convergence to local minima.
\color{black}

\color{black}
\begin{table}[h] 
	\caption{The comparison of data misfit (DM), intersection over union (IoU), mean square error (MSE) and structural similarity (SSIM) of test models A--C.}
	\centering
	\begin{tabular}{c|ccccc}
		\toprule
		&Methods & DM 	& IoU   & MSE    & SSIM  \\
		\midrule
		\multirow{3}{*}{Test model A} & $L_2$  & 0.227\% & 0.021 & 0.0234 & 0.619  \\
		&W$L_2$  & 0.551\% & 0 	 & 0.0432 & 0.628  \\
		&FB   	 & 0.230\% & \textbf{0.500} & \textbf{0.0079} & \textbf{0.907}  \\
		\hline
		\multirow{3}{*}{Test model B} & $L_2$  & 0.224\% & 0.146 & 0.0140 & 0.642  \\
		&W$L_2$  & 0.267\% & 0.200 & 0.0168 & 0.704  \\
		&FB   	 & 0.311\% & \textbf{0.675} & \textbf{0.0045} & \textbf{0.933}  \\
		\hline
		\multirow{3}{*}{Test model C}& $L_2$  & 0.222\% & 0.341 & 0.0191 & 0.840  \\
		&W$L_2$  & 0.372\% & 0.196 & 0.0523 & 0.853  \\
		&FB   	 & 0.269\% & \textbf{0.416} & \textbf{0.0110} & \textbf{0.957}  \\
		\bottomrule
	\end{tabular}
	\label{table1}
\end{table}
\color{black}

\begin{figure*}[!t]
	\centering
	
	\begin{minipage}{\textwidth}
	
	\raggedright
	\subcaptionbox{\label{fig12_a}}{\includegraphics[width=\mywidthB]{fig/model9_true.png}}\hfil 
	\subcaptionbox{\label{fig12_ab}}{\includegraphics[width=\mywidthB]{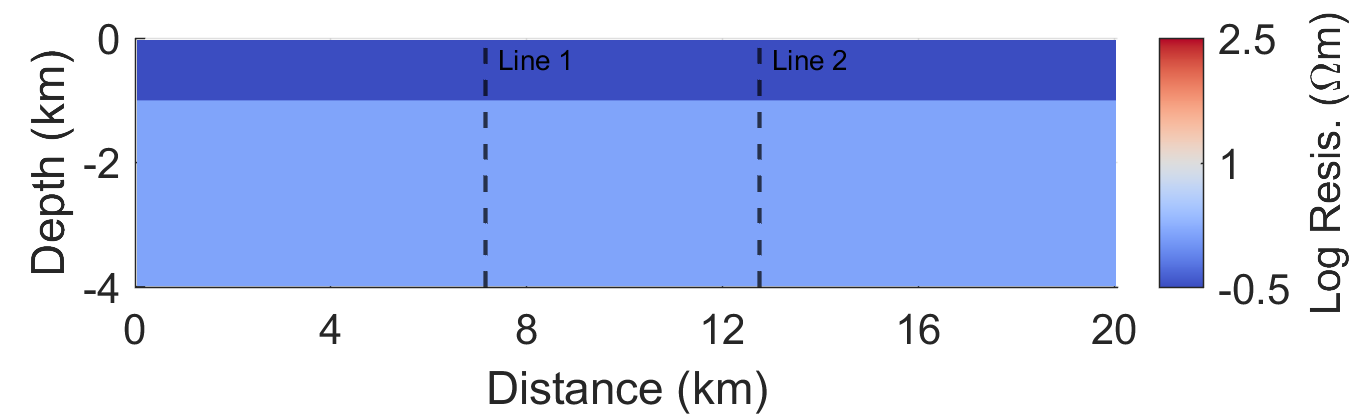}}\hfil 
	\hspace{\mywidthB}
	
	\centering
	\subcaptionbox{\label{fig12_b}}{\includegraphics[width=\mywidthB]{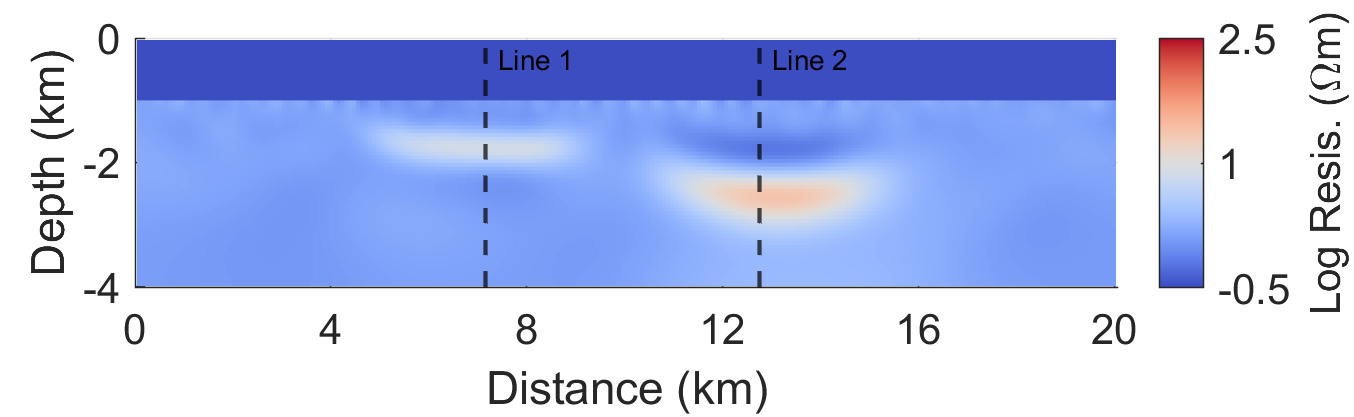}}\hfil
	\subcaptionbox{\label{fig12_c}}{\includegraphics[width=\mywidthB]{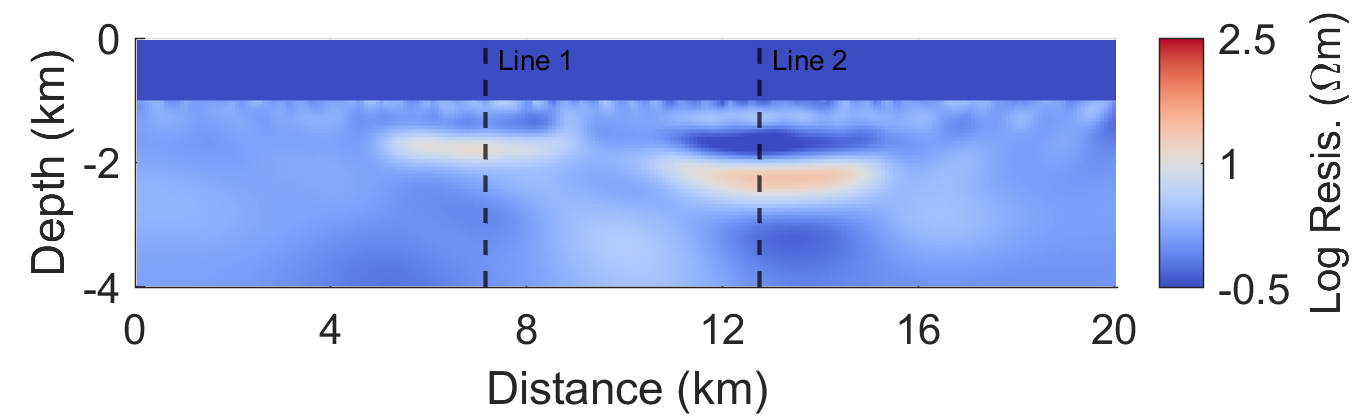}}\hfil
	\subcaptionbox{\label{fig12_d}}{\includegraphics[width=\mywidthB]{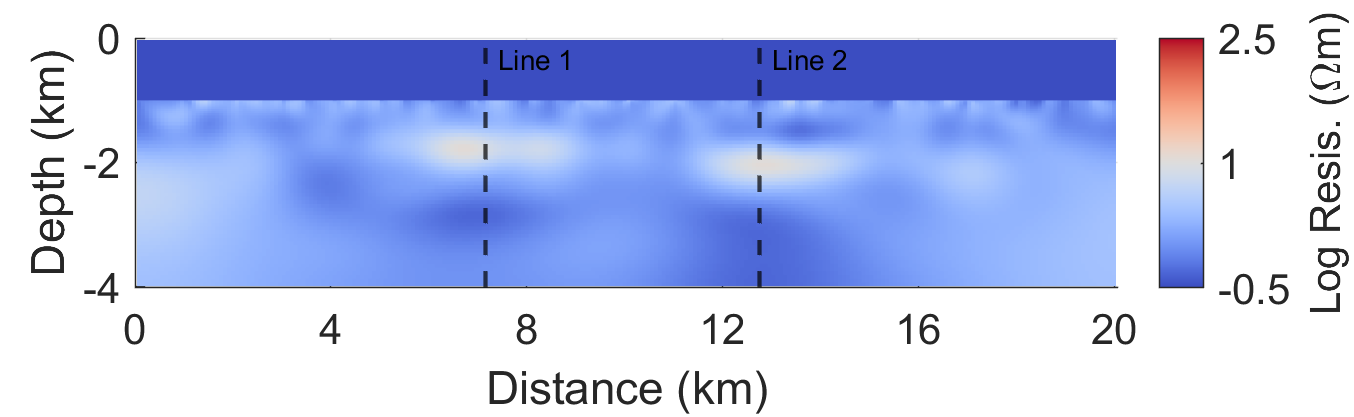}}
	
	\subcaptionbox{\label{fig12_e}}{\includegraphics[width=\mywidthB]{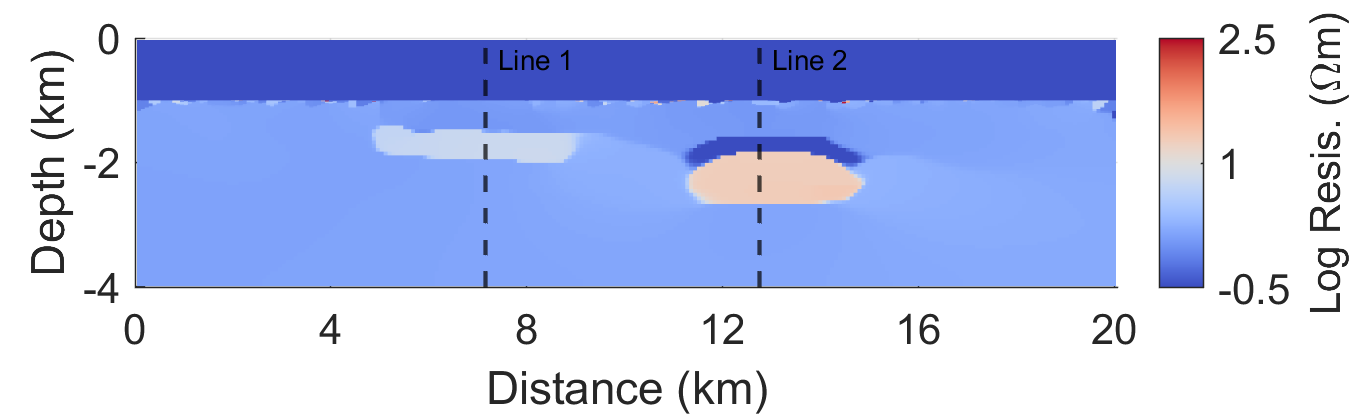}}\hfil
	\subcaptionbox{\label{fig12_f}}{\includegraphics[width=\mywidthB]{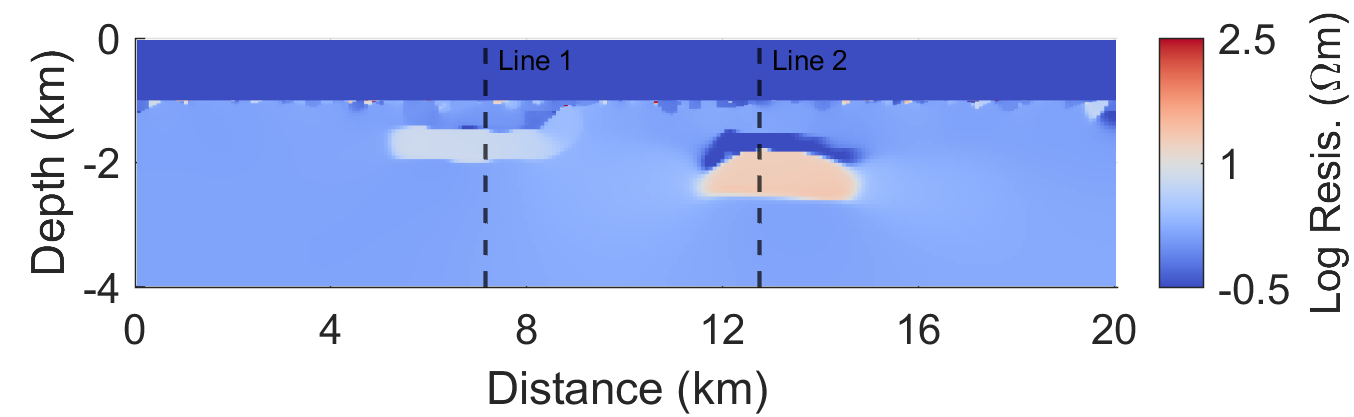}}\hfil
	\subcaptionbox{\label{fig12_g}}{\includegraphics[width=\mywidthB]{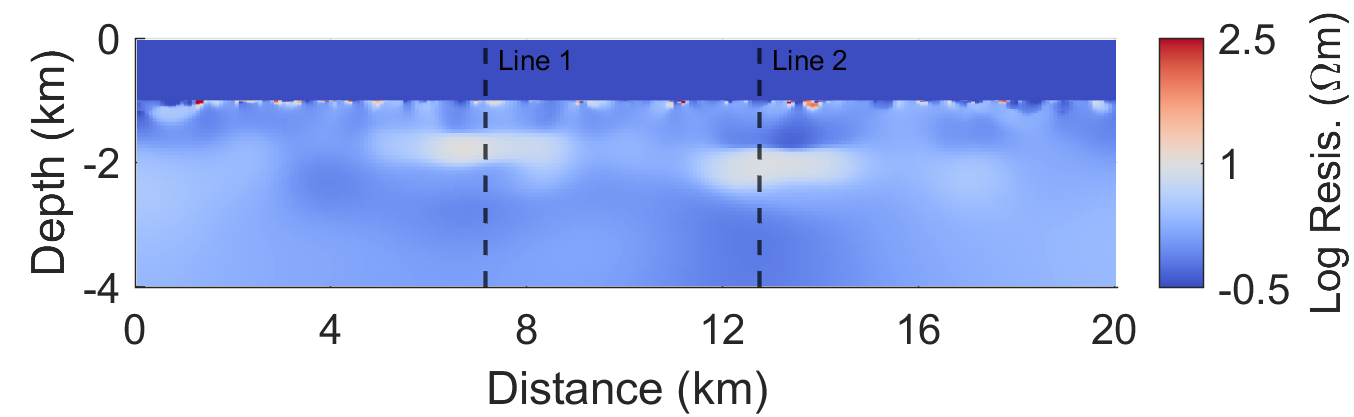}}
	
	\subcaptionbox{\label{fig12_h}}{\includegraphics[width=\mywidthB]{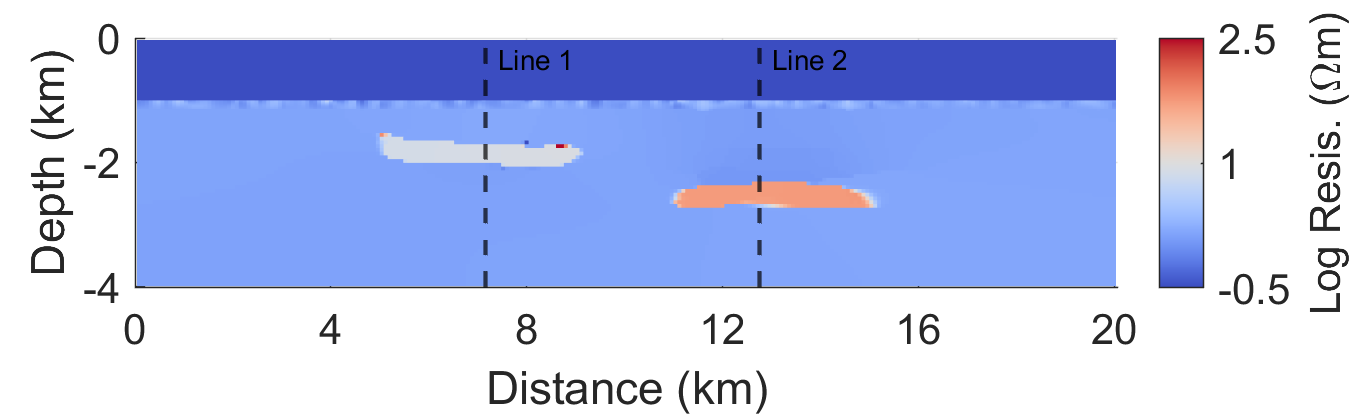}}\hfil
	\subcaptionbox{\label{fig12_i}}{\includegraphics[width=\mywidthB]{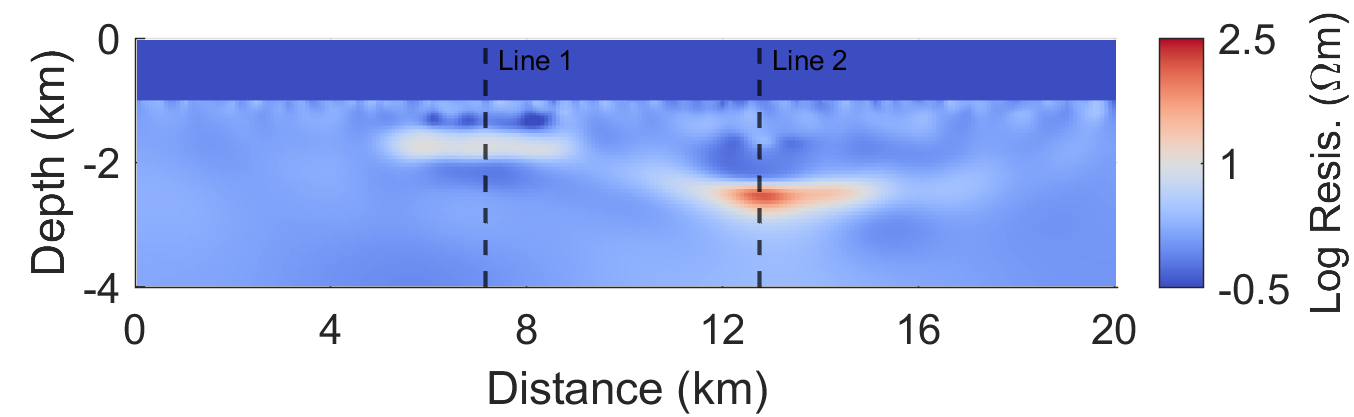}}\hfil
	\subcaptionbox{\label{fig12_j}}{\includegraphics[width=\mywidthB]{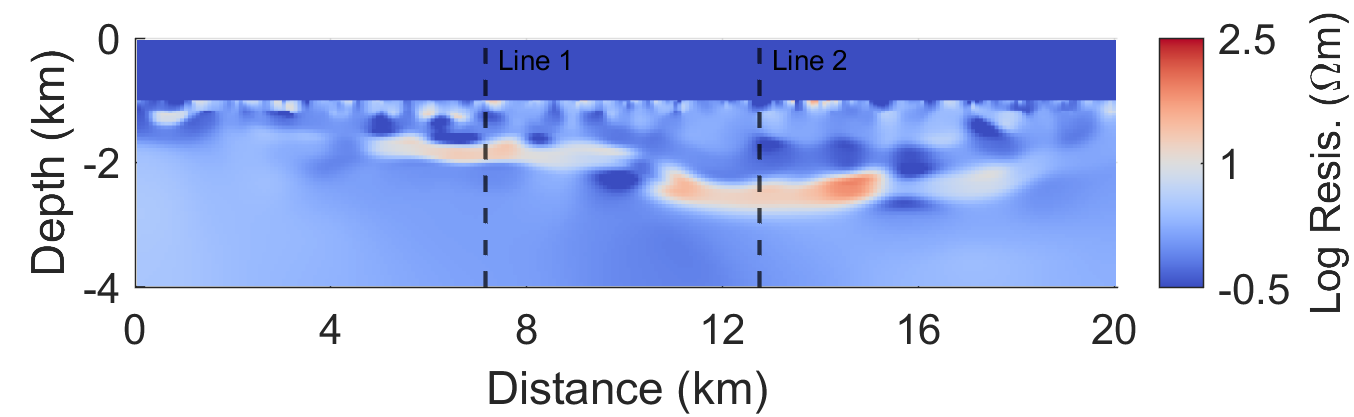}}
	\caption{\color{black}Fig. \ref{fig12_a}: Test model C. Fig. \ref{fig12_ab}: Initial model for the inversion of test model C data. Figs. \ref{fig12_b}--\ref{fig12_j}: The three rows from top to bottom represent inversions with $L_2$ regularization, W$L_2$ regularization and feature-based inversion, and the three columns from left to right show reconstructions corresponding to noise levels of 15\%, 25\% and 50\%.\color{black}}
	\label{fig12}
	\end{minipage}

	\vspace{0.5cm}

	\begin{minipage}{\textwidth}
	\centering
	
	\subcaptionbox{\label{fig12plus1_a}}{\includegraphics[width=\mywidthA]{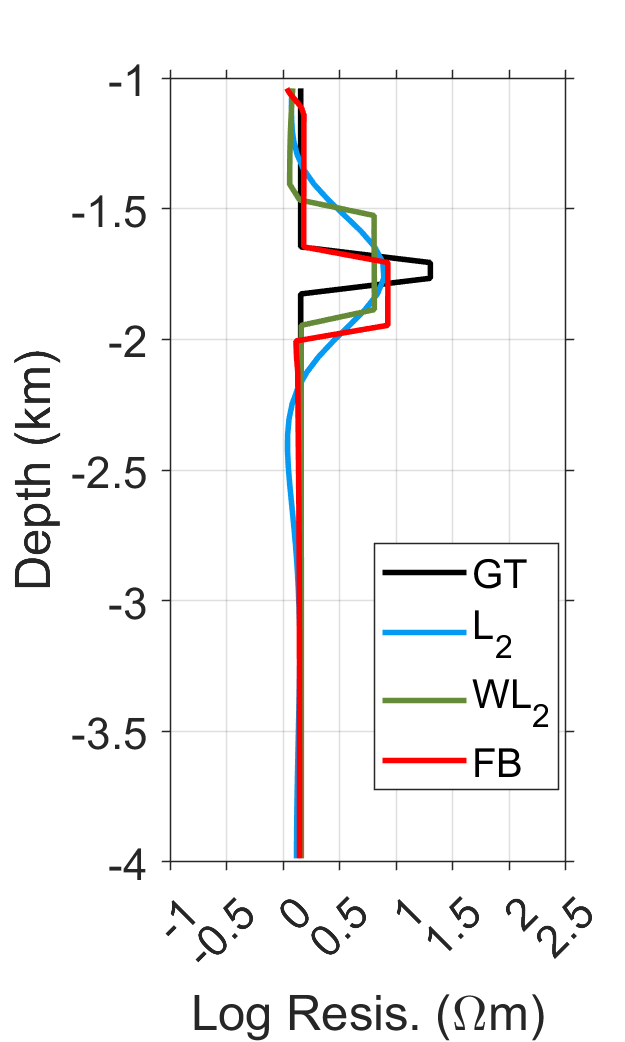}}\hfil
	\subcaptionbox{\label{fig12plus1_b}}{\includegraphics[width=\mywidthA]{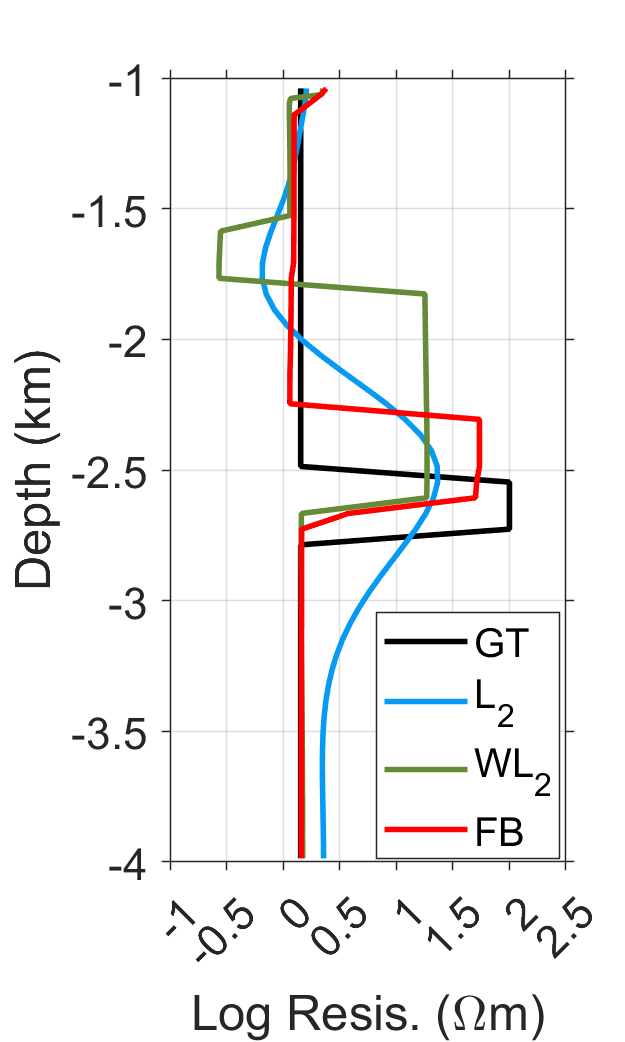}}\hfil
	\subcaptionbox{\label{fig12plus1_c}}{\includegraphics[width=\mywidthA]{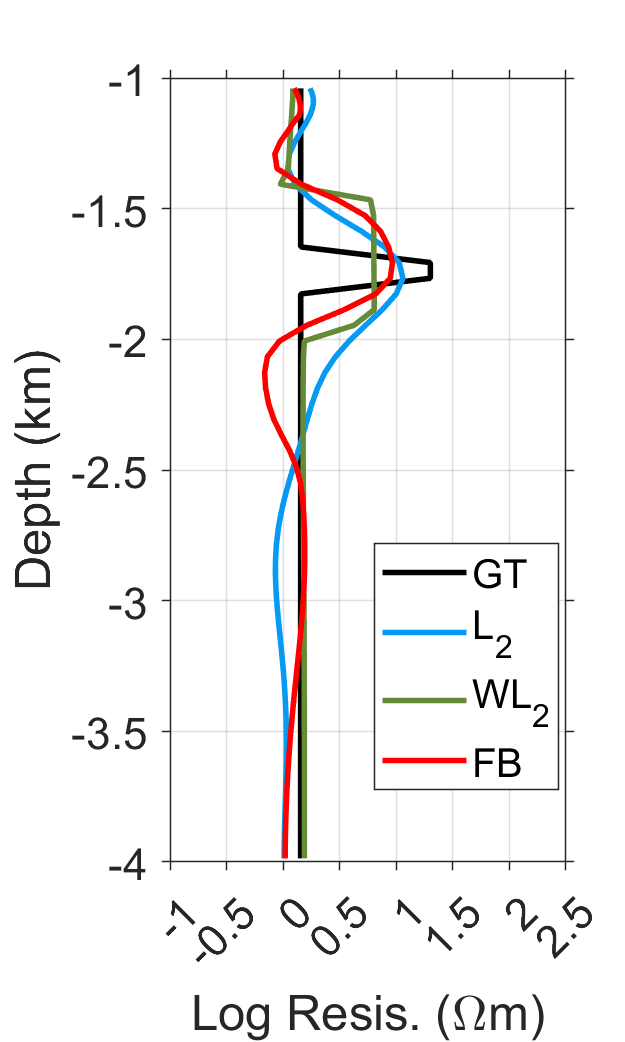}}\hfil
	\subcaptionbox{\label{fig12plus1_d}}{\includegraphics[width=\mywidthA]{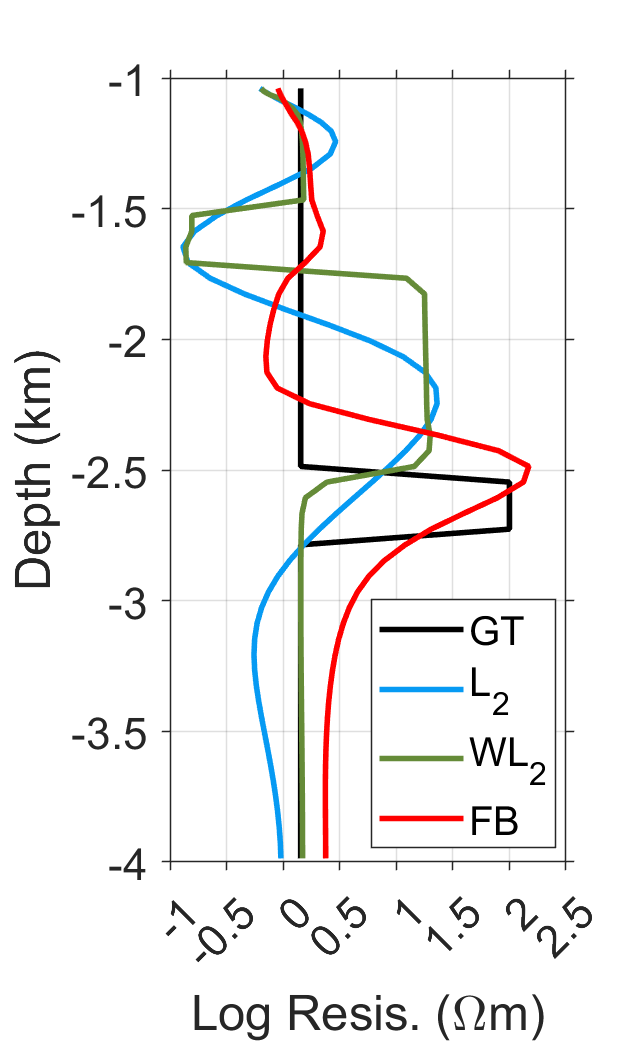}}\hfil
	\subcaptionbox{\label{fig12plus1_e}}{\includegraphics[width=\mywidthA]{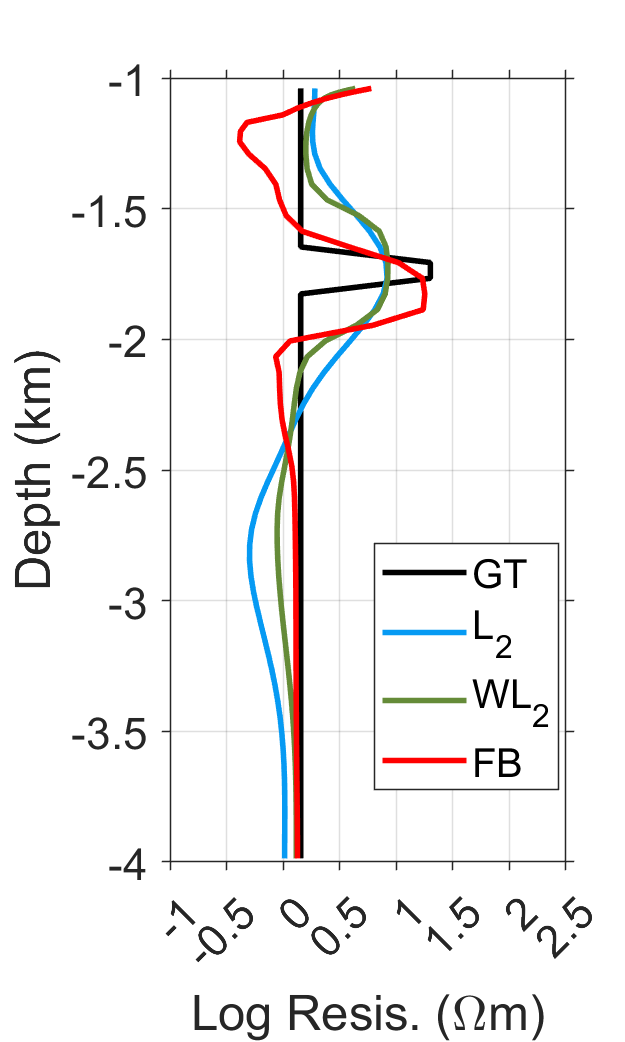}}\hfil
	\subcaptionbox{\label{fig12plus1_f}}{\includegraphics[width=\mywidthA]{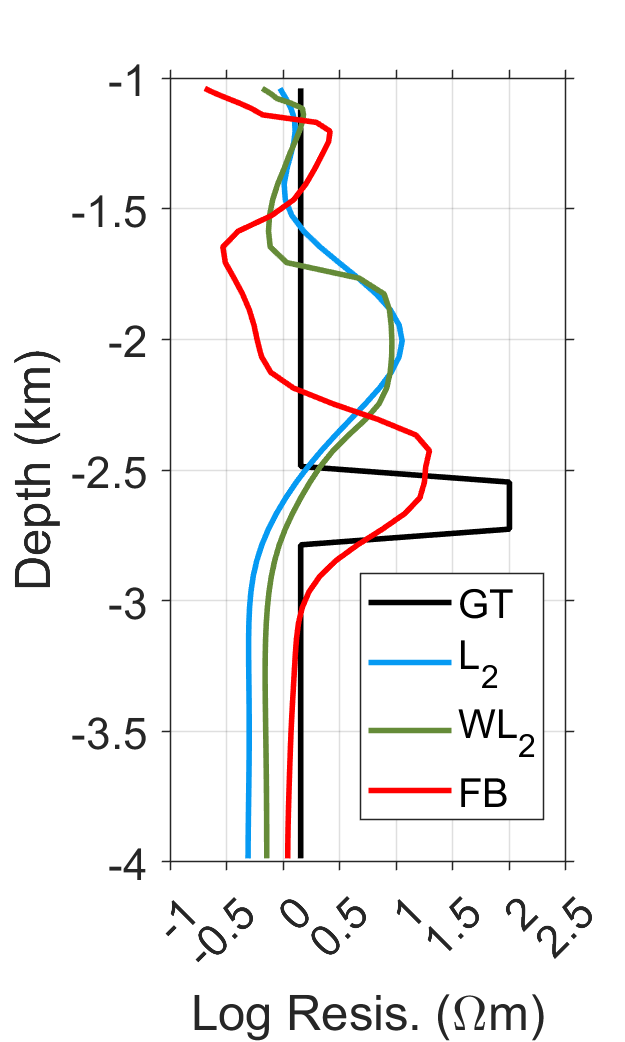}}
	
	\subcaptionbox{\label{fig12plus1_g}}{\includegraphics[width=\mywidthB]{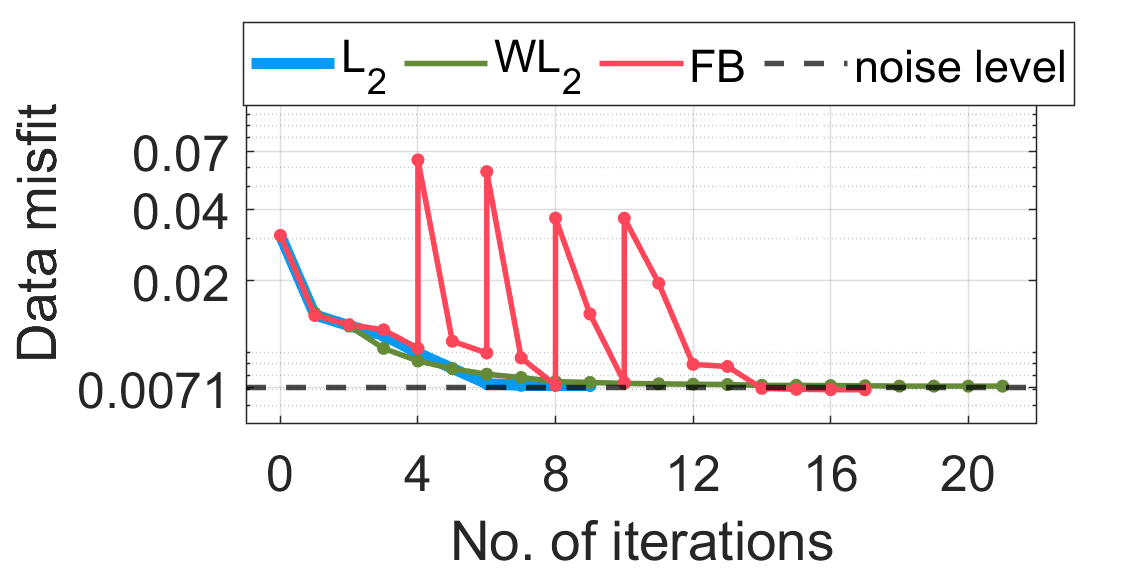}}\hfil
	\subcaptionbox{\label{fig12plus1_h}}{\includegraphics[width=\mywidthB]{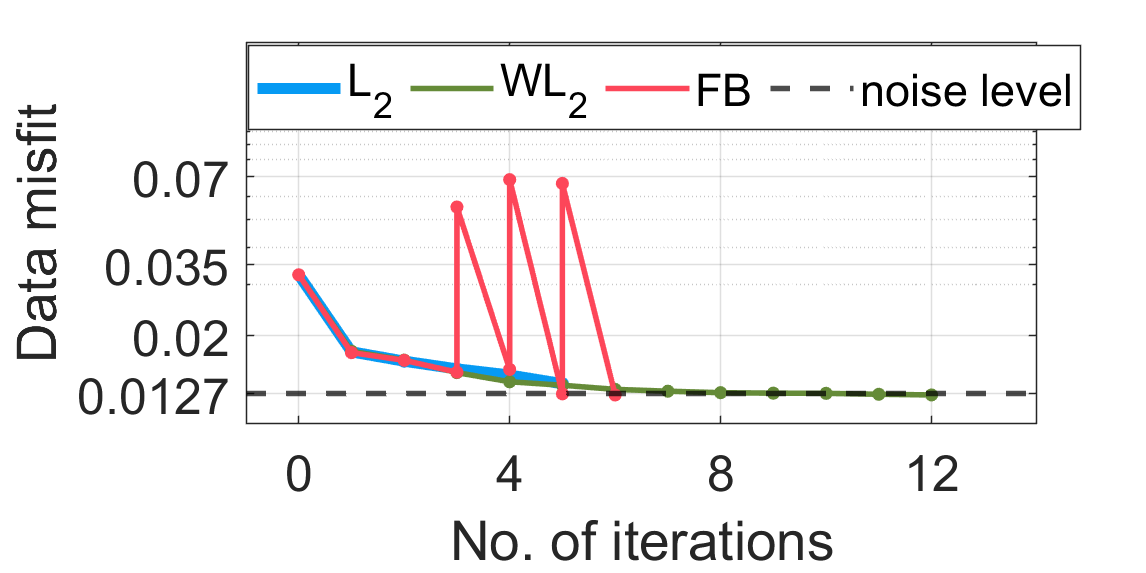}}\hfil
	\subcaptionbox{\label{fig12plus1_i}}{\includegraphics[width=\mywidthB]{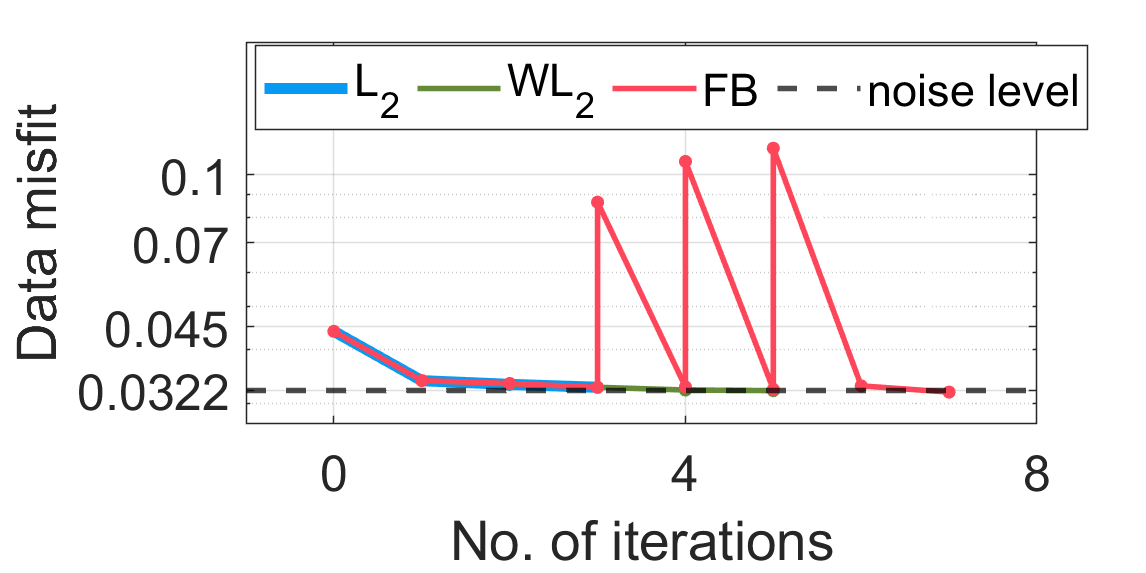}}
	
	\caption{\color{black}Figs. \ref{fig12plus1_a} and \ref{fig12plus1_b} respectively show the ground truth and recovered resistivity curves at $\mathtt{Line1}$ and $\mathtt{Line2}$ (the locations are denoted with black dash lines in Fig. \ref{fig12_a}), and Fig. \ref{fig12plus1_g} display the data misfit curves in the data inversion of test model C, under 15\% noise. Similarly, Figs. \ref{fig12plus1_c}, \ref{fig12plus1_d} and \ref{fig12plus1_h} show the resistivity curves corresponding to $\mathtt{Line1}$ and $\mathtt{Line2}$ in Fig. \ref{fig12_a} and data misfit curves under 25\% noise. Figs. \ref{fig12plus1_e}, \ref{fig12plus1_f} and \ref{fig12plus1_i} show the resistivity curves corresponding to $\mathtt{Line1}$ and $\mathtt{Line2}$ in Fig. \ref{fig12_a} and data misfit curves under 50\% noise.\color{black}}
	\label{fig12plus1}
	\end{minipage}
\end{figure*}

\begin{figure*}[!t]
	\centering
	\includegraphics[width=1\linewidth]{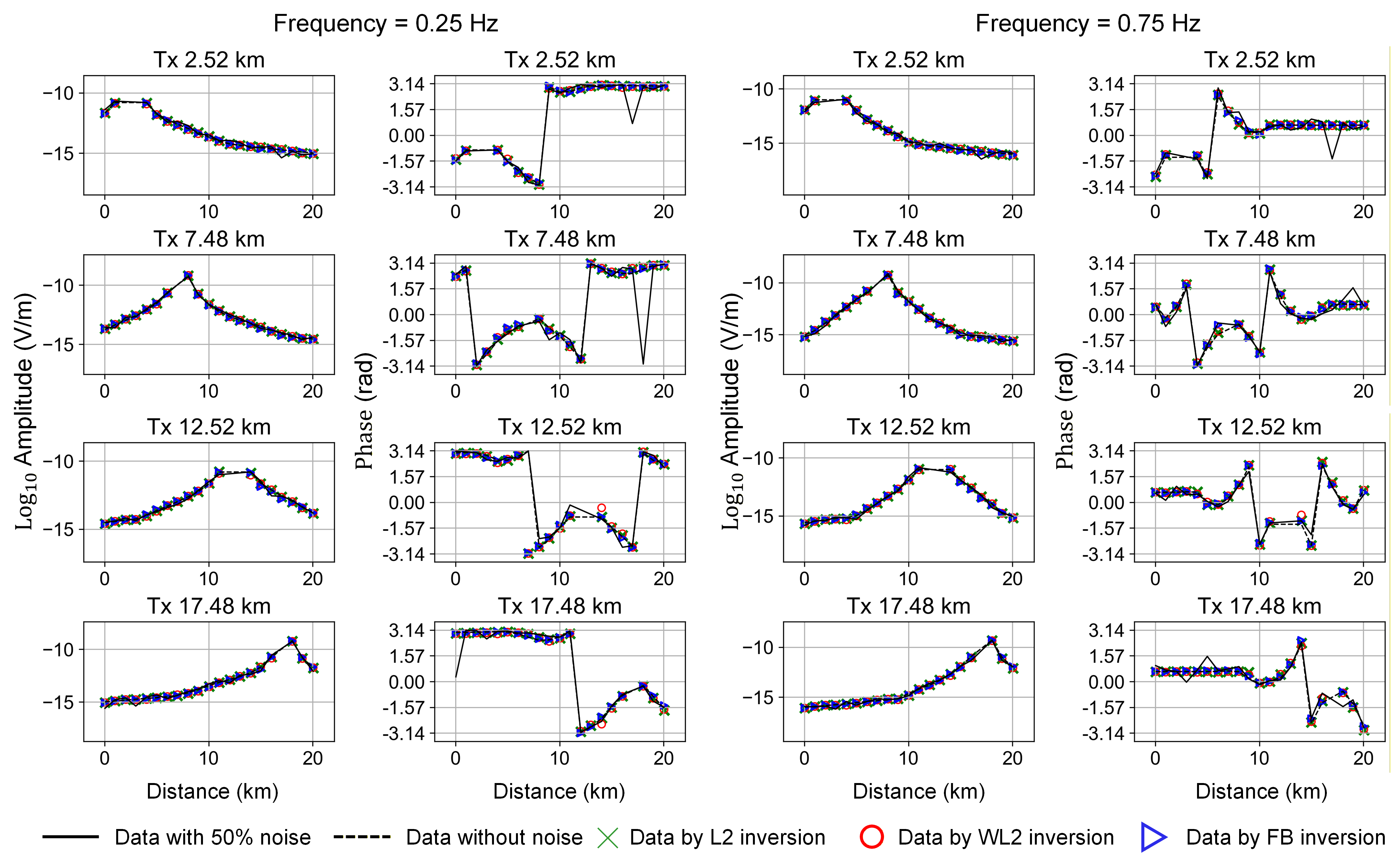}
	\caption{
		\color{black}
	 The comparison between the simulated data with 50\% noise, noise-free simulated data, and reconstructed data in $L_2$, weighted $L_2$ and FB inversions, corresponding to test model C. The four columns from left to right respectively show the $\log_{10}$ amplitude of 0.25 Hz data, phase of 0.25 Hz data, $\log_{10}$ amplitude of 0.75 Hz data, and phase of 0.75 Hz data. The four rows from top to bottom show the field measured at all the active receivers with transmitter excited at 2.52 km, 7.48 km, 12.52 km, and 17.48 km horizontally.
		\color{black}}
	\label{fig8plus}
\end{figure*}

\color{black}
\subsection{Simple Model Test with Noise}\label{Experiment2}
\color{black}
In this section, we investigate the impact of the noise level on the feature-based inversion. \color{black} Test model C is adopted here, and the configurations of the DoI, mesh, operating frequency, transmitters and receivers are the same as those in Section \ref{Experiment1}, while the noise level is changed to 15\%, 25\% and 50\%, respectively. Fig. \ref{test_model_C_data_vs_noise} shows a comparison between noise-free simulated data and data with different noise levels, excited by transmitters located at 2.52 km, 7.48 km, 12.52 km, and 17.48 km, corresponding to test model C. The DMs introduced by 15\%, 25\% and 50\% noise are respectively 0.0071, 0.0127 and 0.0322 (also calculated using (\ref{DM})). $L_2$, W$L_2$ and FB reconstructions using different levels of noise are shown in Fig. \ref{fig12}. \color{black}
The reconstructed resistivity traces as well as the data misfit curves are displayed in Fig. \ref{fig12plus1}. 

With the noise increasing from 5\% to 50\%, \color{black} all the three inversions \color{black} converge to the corresponding noise levels; however, the reconstruction quality degrades in all cases, which is expected. 
Reconstruction patterns similar to those reported in Section. \ref{Experiment1} are observed again: blurred resistivity layer boundaries for $L_2$ inversions, overestimated thickness and conductive anomalies above the resistive layer in W$L_2$ inversions, and the less uniform background for all the inversions. \color{black}
\color{black} However, across all noise levels, the FB reconstructions consistently perform the best among the three methods, with resistive layers  more clearly delineated. Even under the noise level of 50\%, the geometries of the resistive layers remain  identifiable. This may benefit reservoir interpretation when the measured data quality is limited. \color{black} 

\color{black} A clearer observation of the above findings can be obtained by comparing the ground truth and reconstructed resistive depth traces  in Figs. \ref{fig12plus1_a}--\ref{fig12plus1_f}. 
Although resistive anomalies can be captured, two major problems are observed in $L_2$ and W$L_2$ inversions: a vertical shift from the true depth of the resistive layer and an accompanying thin conductive layer above the resistive layer (both phenomena are more pronounced for the right layer, likely due to its greater burial depth and higher resistivity contrast than the left one). 
The FB inversion suppresses both phenomena, yielding a more accurate characterization of the ground truth DoI.

Fig. \ref{fig8plus} compares the data with 50\% noise, noise-free data, and reconstructed data obtained using different methods. In this figure, the four rows from top to bottom correspond to the fields at all the receivers excited by transmitters located at 2.52 km, 7.48 km, 12.52 km and 17.48 km, while the four columns from left to right denote the $\log_{10}$ amplitude for 0.25 Hz data, phase for 0.25 Hz data, $\log_{10}$ amplitude for 0.75 Hz data and phase for 0.75 Hz data. 
The final DMs for $L_2$, W$L_2$ and FB inversions are 0.0328, 0.0322 and 0.0319, respectively, all close to the noise DM of 0.0322. 
Rather than fitting the noise-contaminated data, the reconstructed data converge more closely to the clean data, as indicated by the dashed lines in Fig. \ref{fig8plus}, and data points with strongly deviating noise values are not fitted by any method.
This aligns well with our expectation of a sound inversion algorithm that fits the overall data trend while excluding extreme anomalous data points.

We note that this experiment further verifies the generalization capability of the proposed deep-learning-assisted FB inversion.
Because the neural parameters are decoupled from the forward problem and inference is performed only in the conductivity domain, the trained network can generalize across surveys with different measurement noise distributions, with a low risk of generalization failure. 
\color{black}

\begin{figure*}[tb]
	\centering
	\includegraphics[width=1.5\columnwidth]{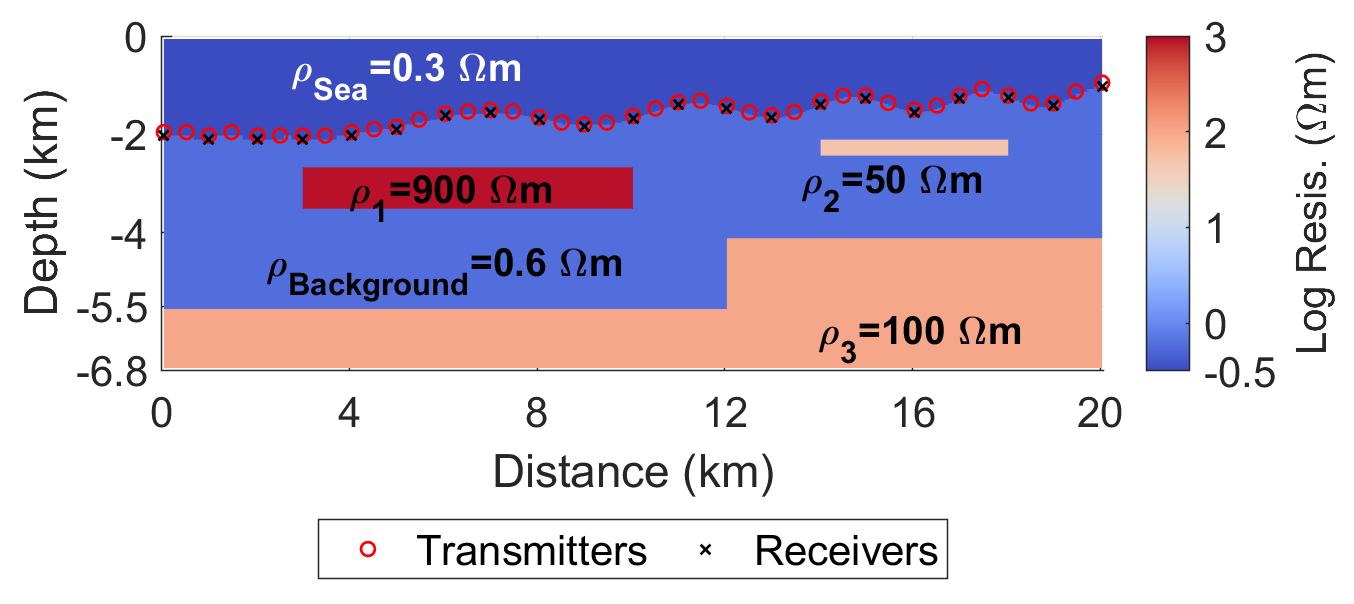}
	\caption{\color{black}A visualization of test model D. The seabed has a irregular topography, with twenty-one receivers and forty-one transmitters uniformly distributed horizontally and positioned at the seabed surface and 50 m above the seabed, respectively. The resistivities of the seawater, seabed background, and three resistive layers are labeled in the image. \color{black}}
	\label{testmodelD}
\end{figure*}

\begin{figure}[h]
	\centering
	\subcaptionbox{\label{fig6_1}}{\includegraphics[width=1\columnwidth]{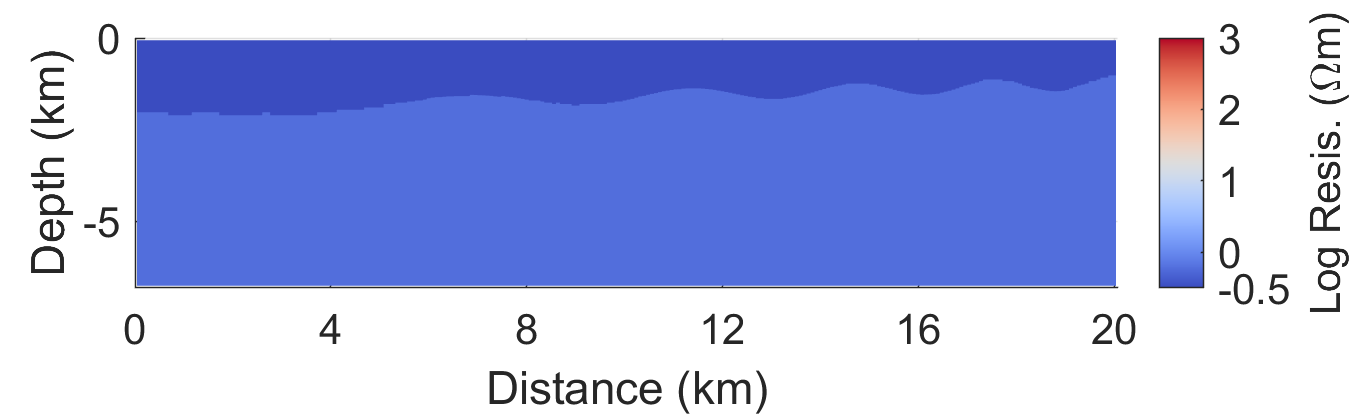}}\hfil
	\subcaptionbox{\label{fig6_2}}{\includegraphics[width=1\columnwidth]{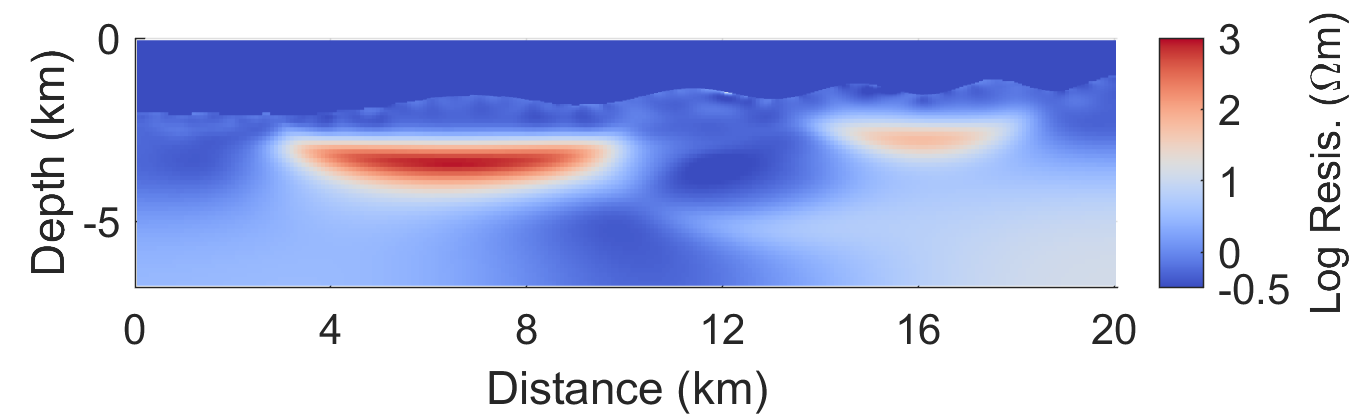}}\hfil
	\subcaptionbox{\label{fig6_3}}{\includegraphics[width=1\columnwidth]{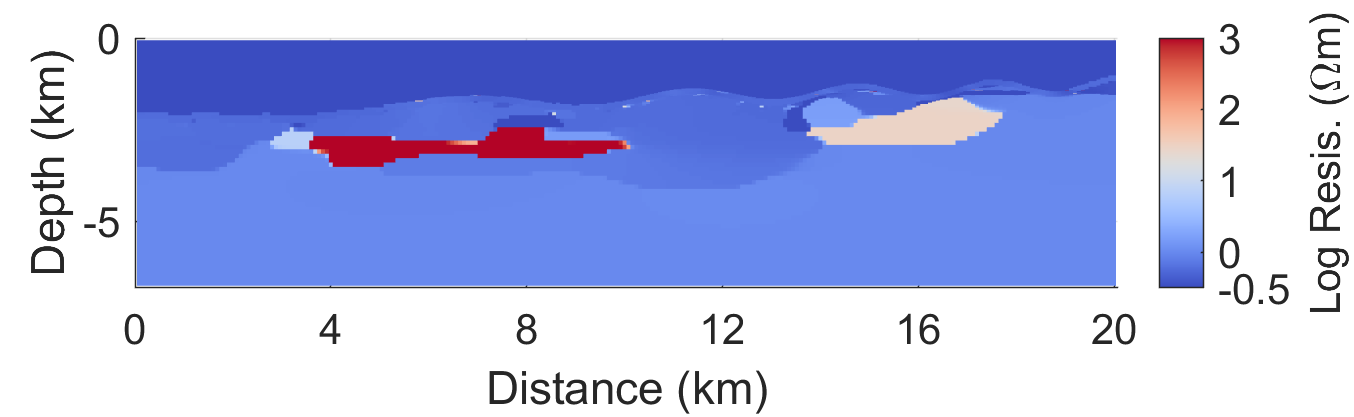}}\hfil
	\subcaptionbox{\label{fig6_4}}{\includegraphics[width=1\columnwidth]{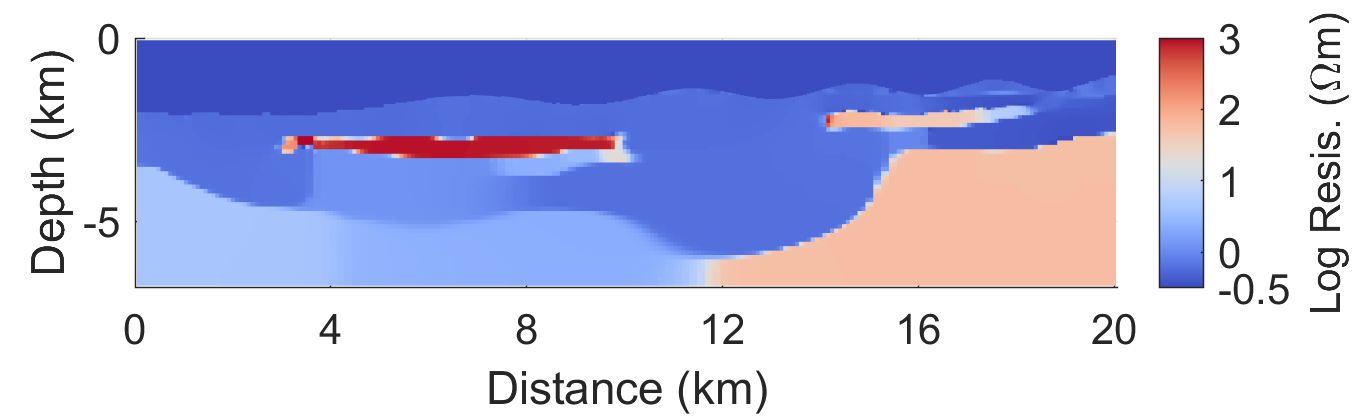}}
	\caption{\color{black} Fig. \ref{fig6_1}: The initial model in the inversions of test model D. Fig. \ref{fig6_2}: Model reconstructed by $L_2$ inversion. Fig. \ref{fig6_3}: Model reconstructed by W$L_2$ inversion. Fig. \ref{fig6_4}: Model reconstructed by feature-based inversion. \color{black}}
	\label{fig6}
\end{figure}

\begin{figure}[tb]
	\centering
	\includegraphics[width=0.95\columnwidth]{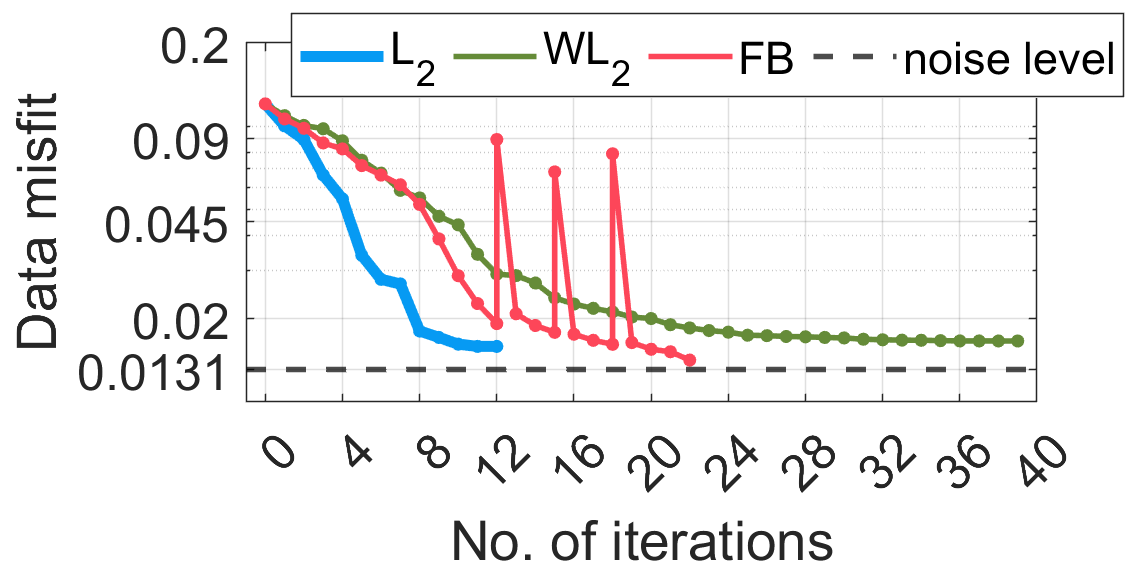}
	\caption{\color{black}Data misfit curves of the test model D inversions. \color{black}}
	\label{test_model_D_loss}
\end{figure}

\color{black}
\subsection{Complex Model Test with Noise}
\color{black}
\color{black}
We set up \color{black} a more complex model (test model D) following \cite{qin20243} (Fig. 13, in \emph{Marine Prospecting Study} section), as shown in Fig. \ref{testmodelD}. The domain of investigation is 20 km in width and 4.8--5.8 km in depth. The seabed has a varying upper surface, with depths ranging from 1 to 2 km below sea level. The resistivities of the seawater, sedimentary layer, and basement are 0.3, 0.6, and 100 $\Omega$m, respectively. Two resistive blocks are embedded within the sedimentary layer. The thick block has a resistivity of 900 $\Omega$m, is located between depths of 2.6 and 3.4 km, and extends horizontally for 7 km. The thin block has a resistivity of 50 $\Omega$m, is located between depths of 2.1 and 2.4 km, and extends horizontally for 4 km.  \color{black}

\color{black}
The adopted non-uniform mesh contains 251 grids in the horizontal direction and 146 grids in the vertical direction. Twenty-one receivers and forty-one transmitters are uniformly distributed horizontally, and positioned at the seabed surface and 50 m above the seabed, respectively. The locations of the transmitters and receivers are denoted by red circles and black crosses in Fig. \ref{testmodelD}, respectively. The operating frequencies are 0.1 and 1 Hz.  Gaussian noise with a level of 25\% is added to the simulated data, introducing a data misfit of 0.0131. 
\color{black}

\begin{figure}[tb]
	\centering
	
	\subfloat[]{\includegraphics[width=1\columnwidth]{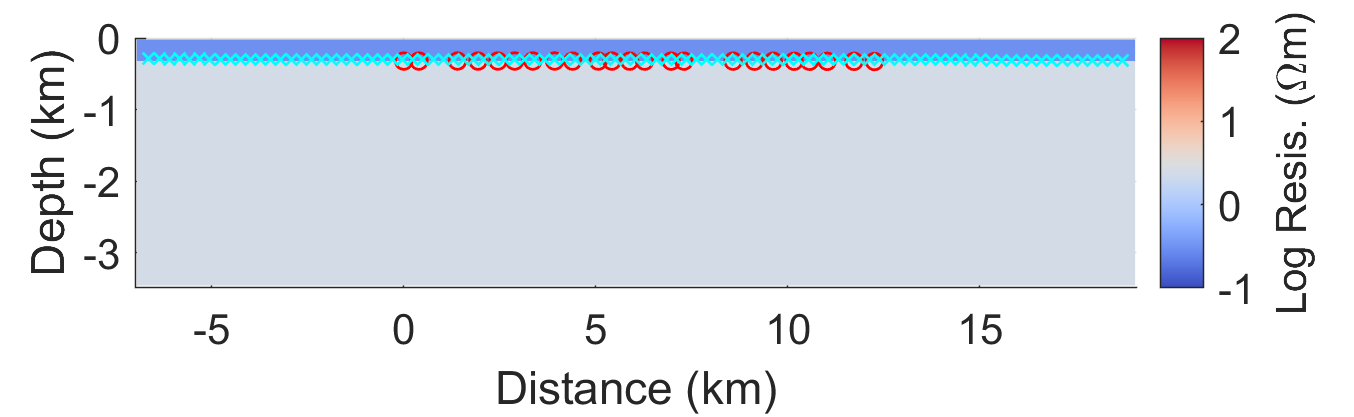}%
		\label{fig13_a}}
	\hfil 
	\subfloat[]{\includegraphics[width=1\columnwidth]{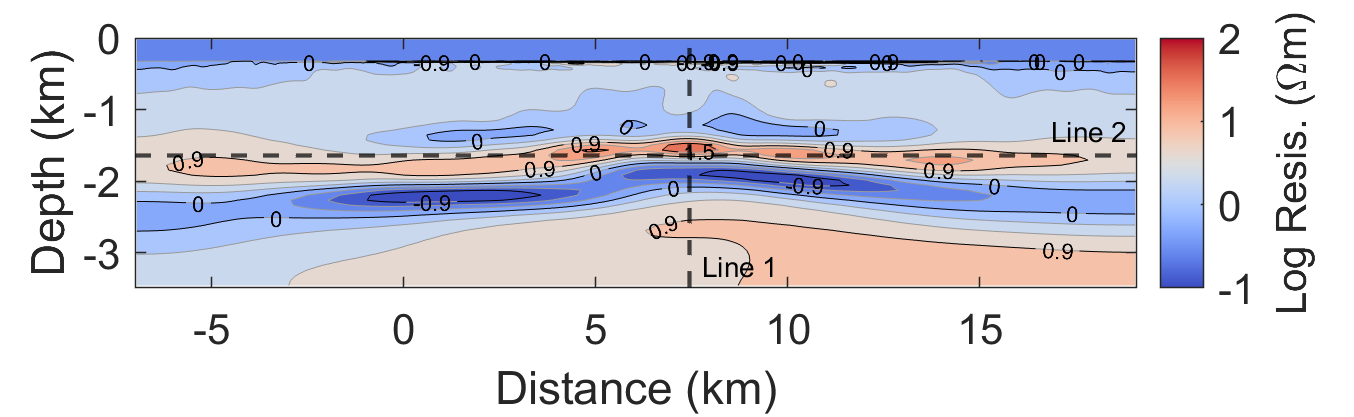}%
		\label{fig13_b}}
	\hfil
	\subfloat[]{\includegraphics[width=1\columnwidth]{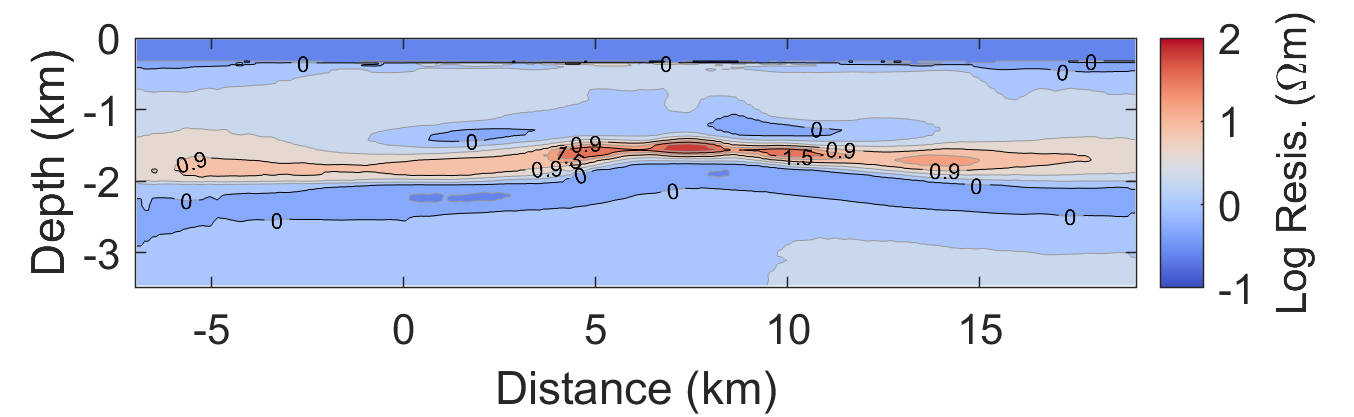}%
		\label{fig13_c}}
	\hfil
	\subfloat[]{\includegraphics[width=1\columnwidth]{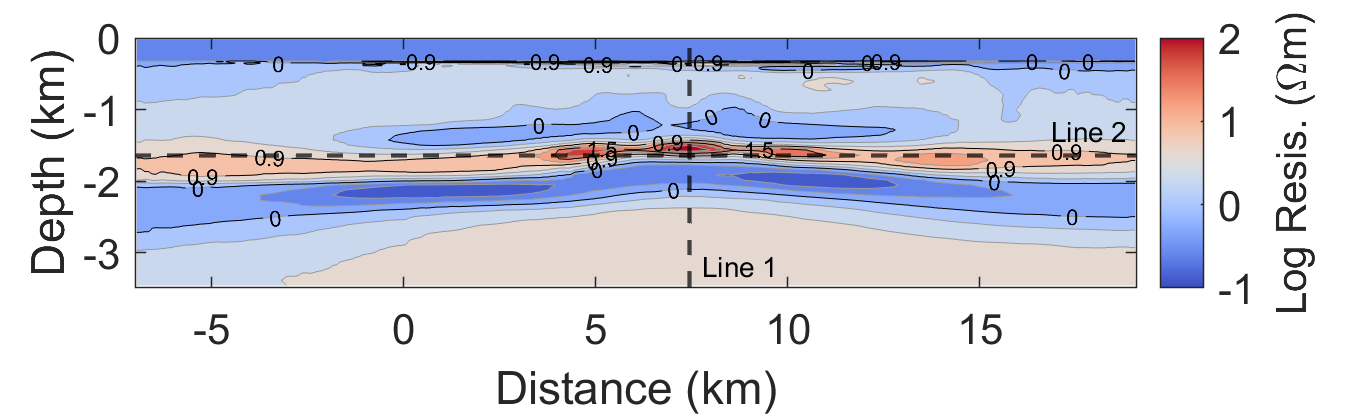}%
		\label{fig13_d}}
	\caption{\color{black} Fig. \ref{fig13_a}: The initial model in the inversions of Troll field data. Fig. \ref{fig13_b}: Model reconstructed by $L_2$ reconstruction. Fig. \ref{fig13_c}: Model projected to the VAE output range in FB inversion at iteration 14. Fig. \ref{fig13_d}: Model reconstructed by feature-based inversion. \color{black}}
	\label{fig13}
\end{figure}

\begin{figure}[tb]
	\centering
	\includegraphics[width=0.95\columnwidth]{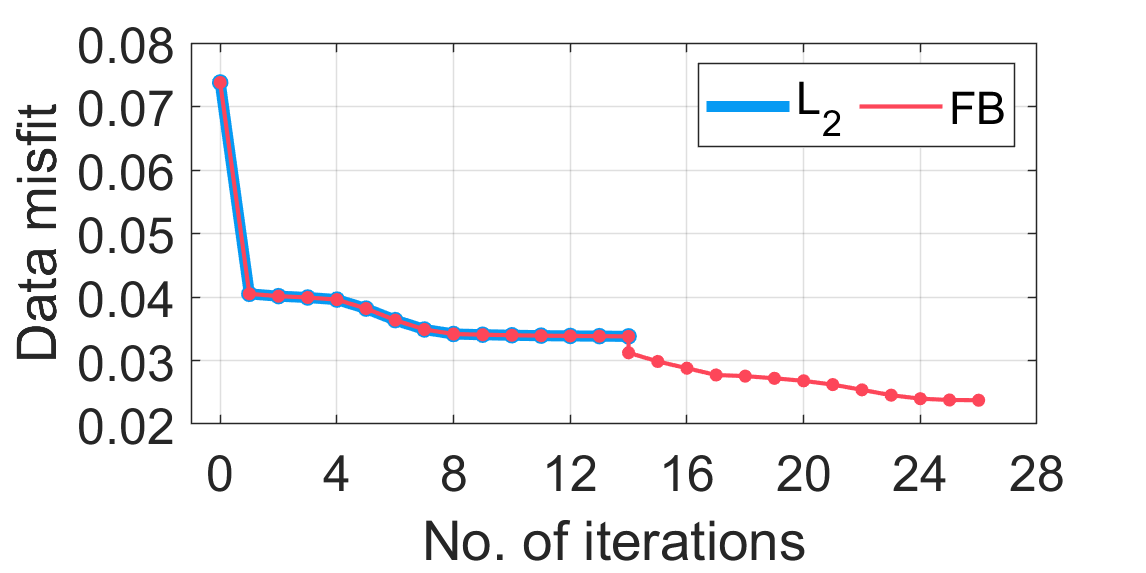}
	\caption{\color{black}Data misfit curves of the Troll field data inversions. \color{black}}
	\label{fig15_c}
\end{figure}

\begin{figure*}[tb]
	\centering
	
	\subfloat[]{\includegraphics[width=\mywidthC]{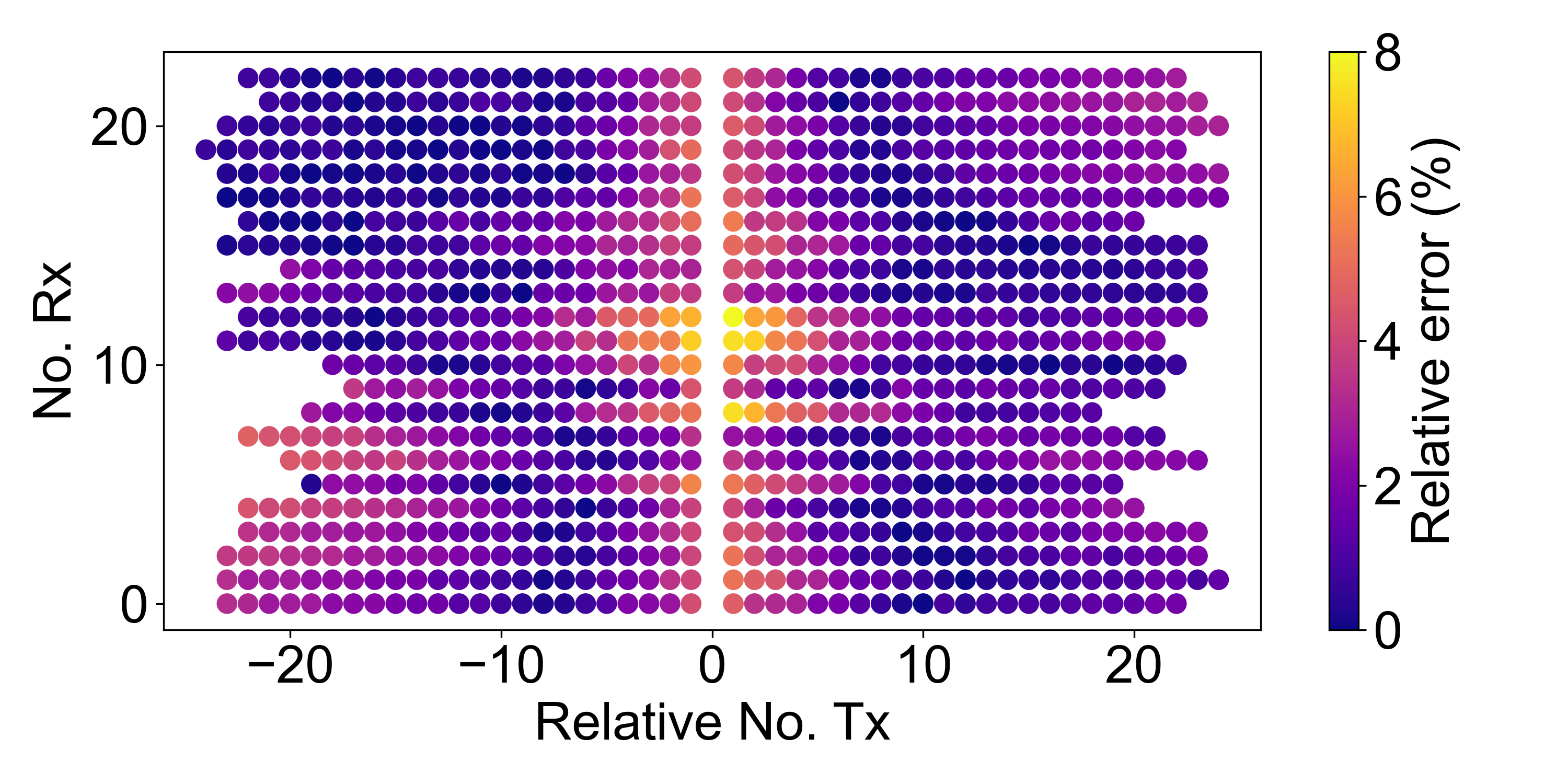}%
		\label{fig16_a}}
	\hfil 
	\subfloat[]{\includegraphics[width=\mywidthC]{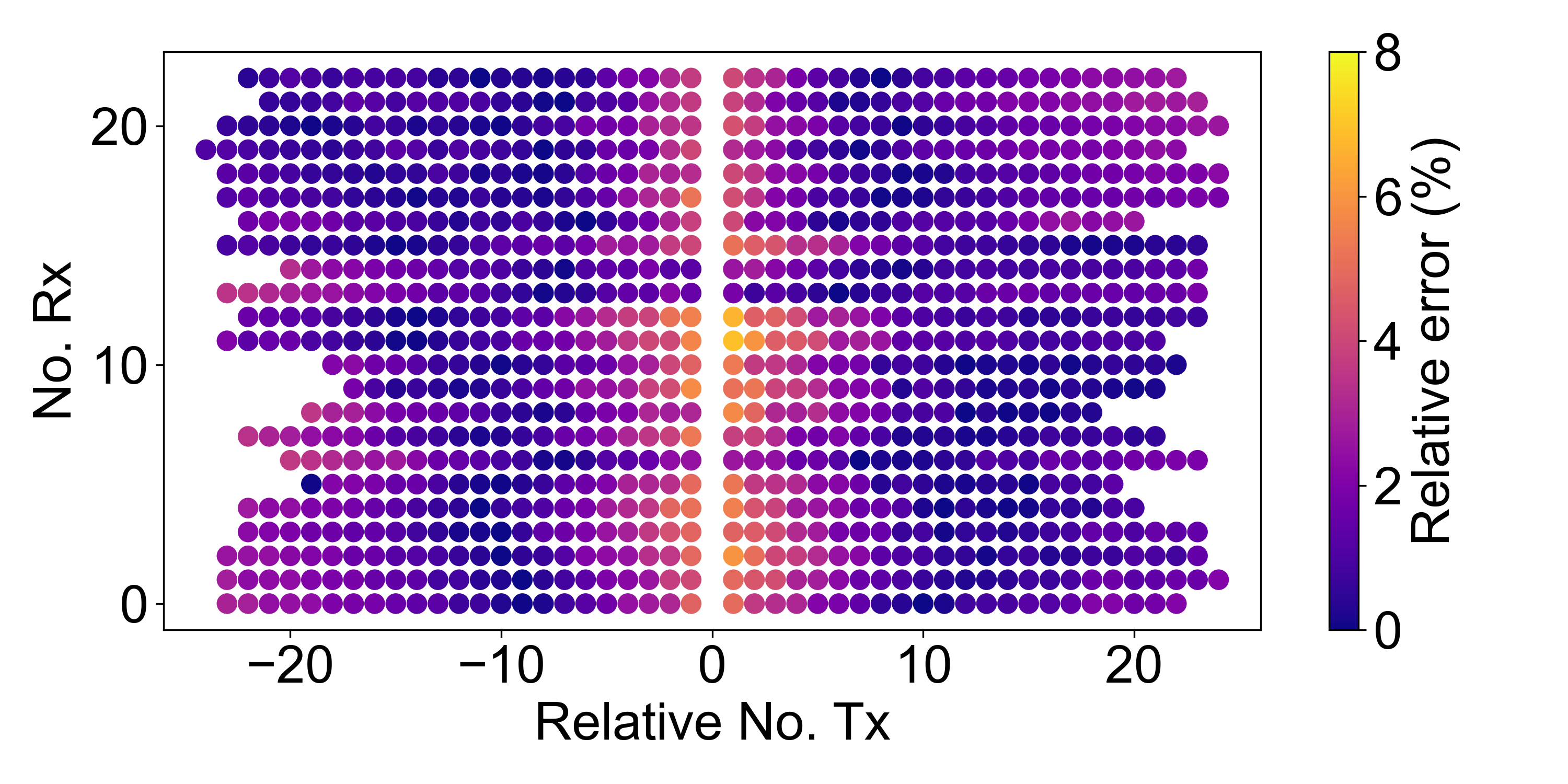}%
		\label{fig16_b}}
	
	\subfloat[]{\includegraphics[width=\mywidthC]{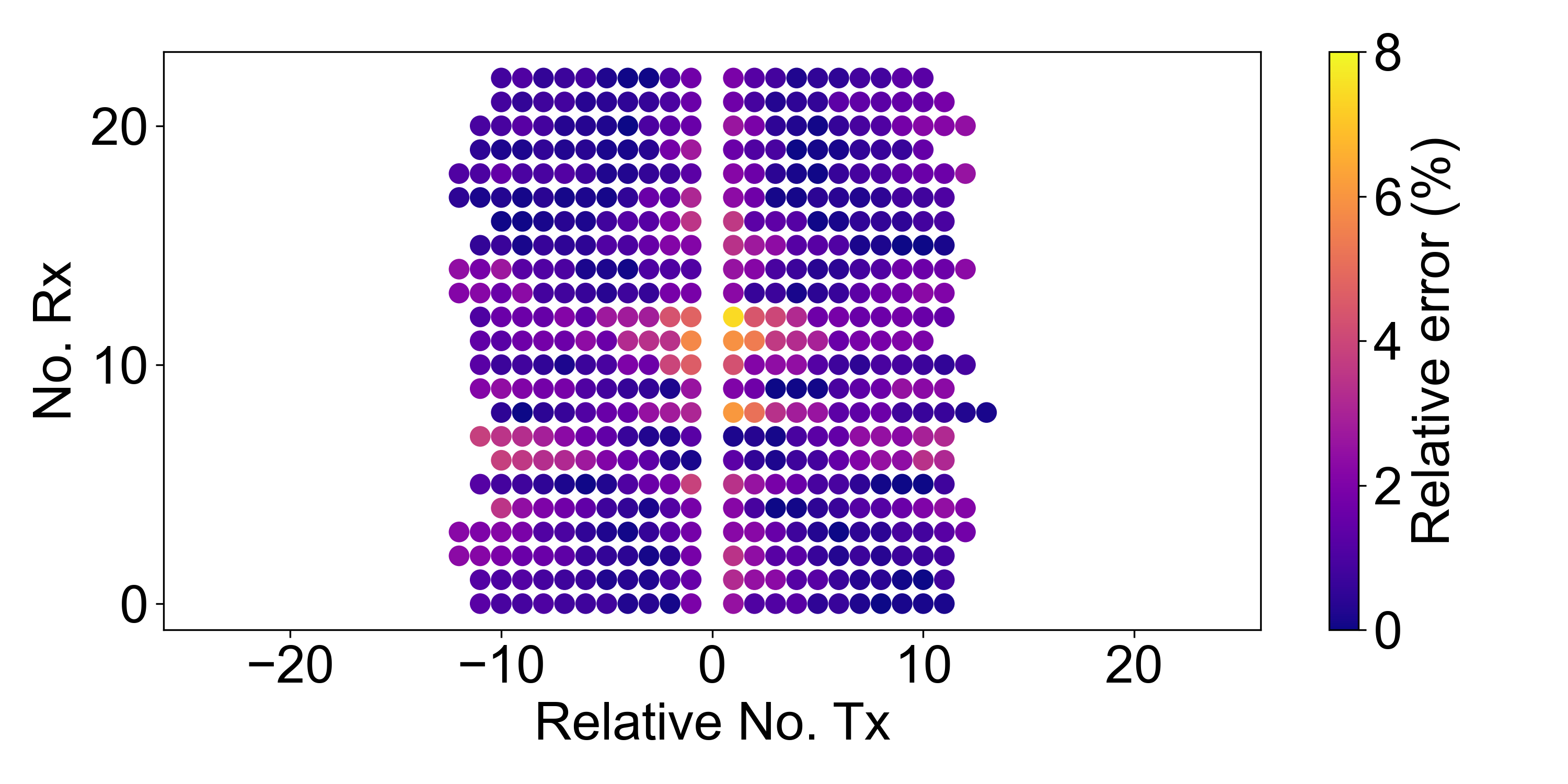}%
		\label{fig16_c}}
	\hfil
	\subfloat[]{\includegraphics[width=\mywidthC]{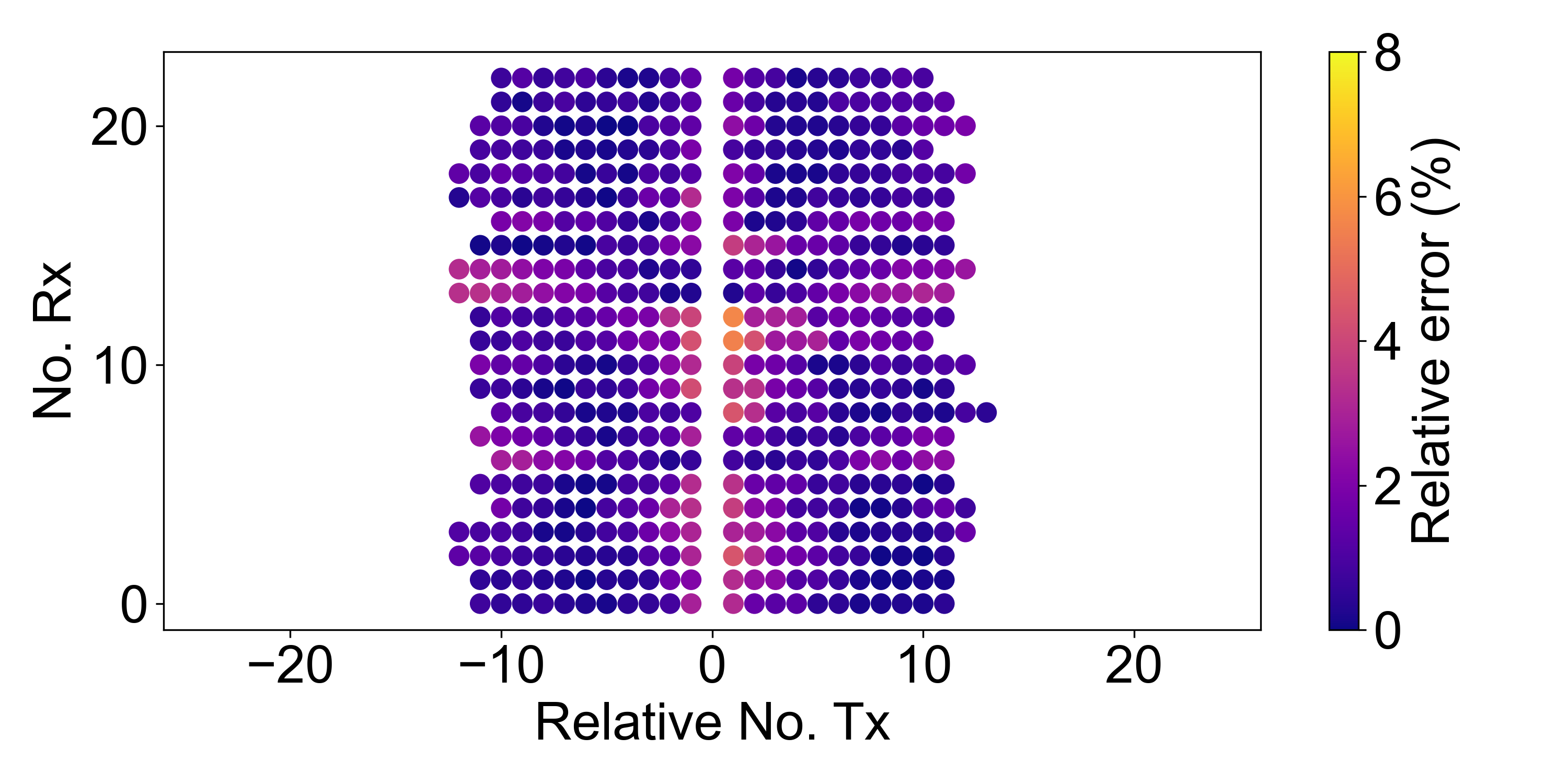}%
		\label{fig16_d}}
	
	\subfloat[]{\includegraphics[width=\mywidthC]{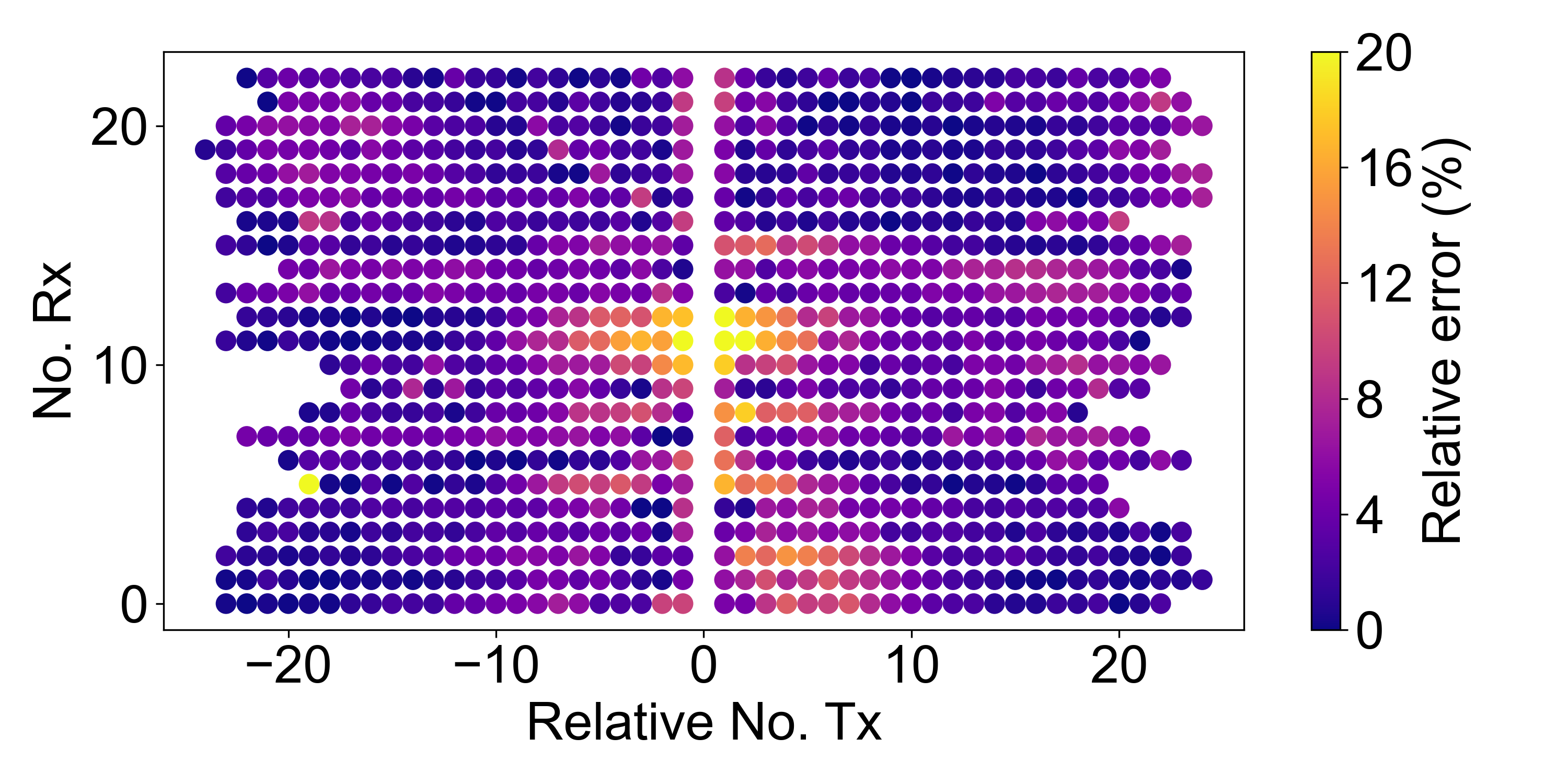}%
		\label{fig16_e}}
	\hfil
	\subfloat[]{\includegraphics[width=\mywidthC]{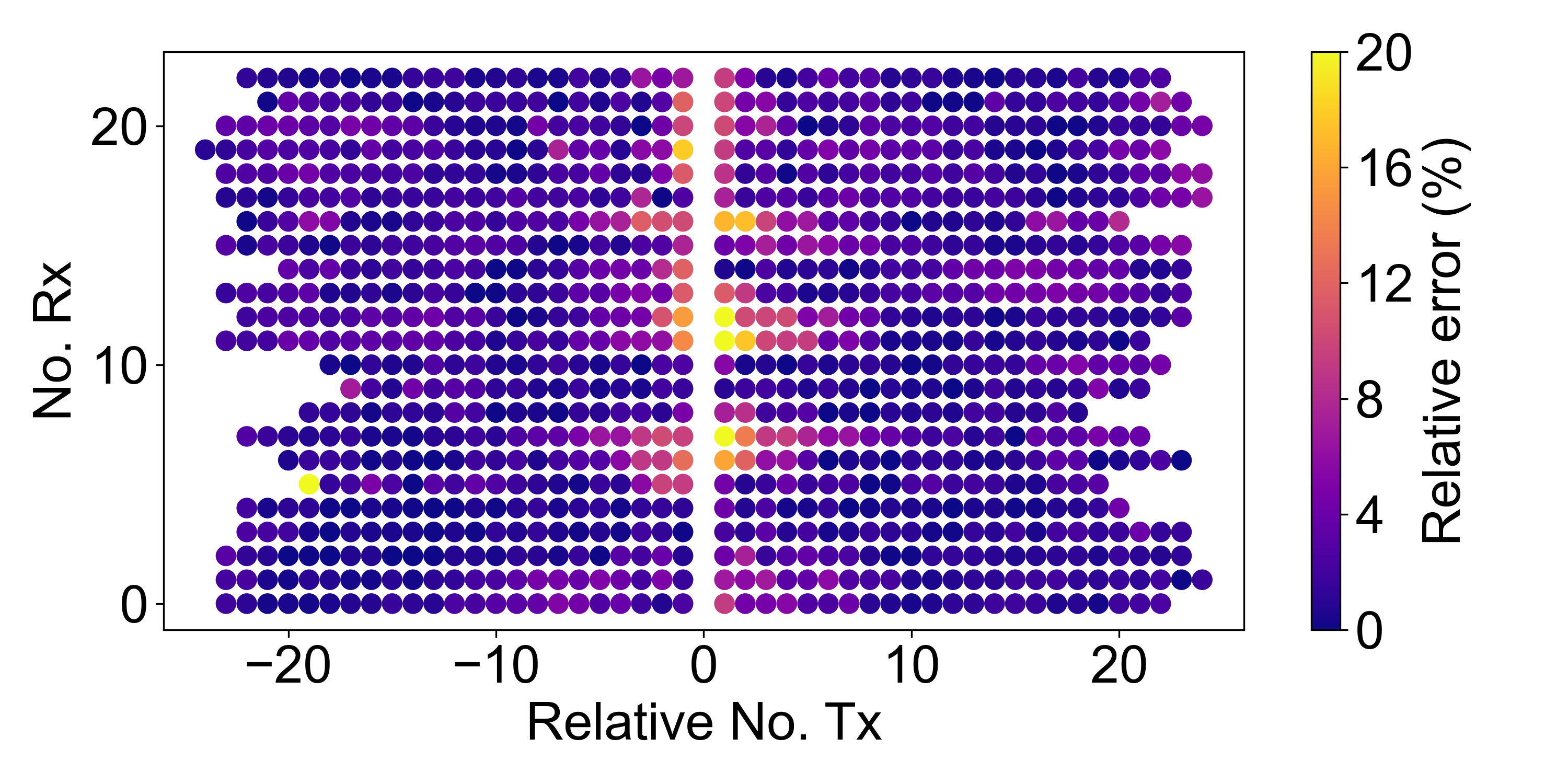}%
		\label{fig16_f}}
	
	\subfloat[]{\includegraphics[width=\mywidthC]{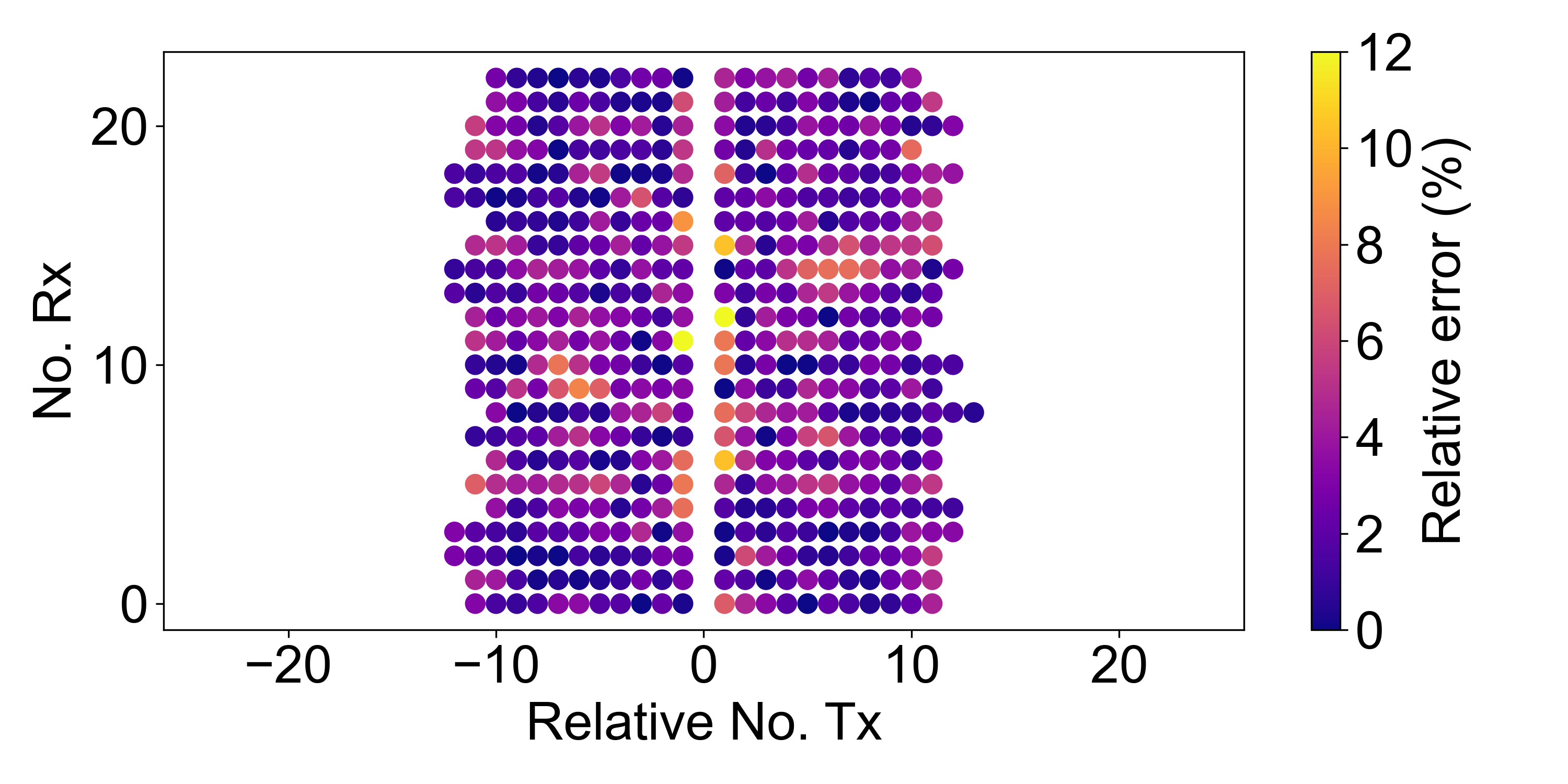}%
		\label{fig16_g}}
	\hfil
	\subfloat[]{\includegraphics[width=\mywidthC]{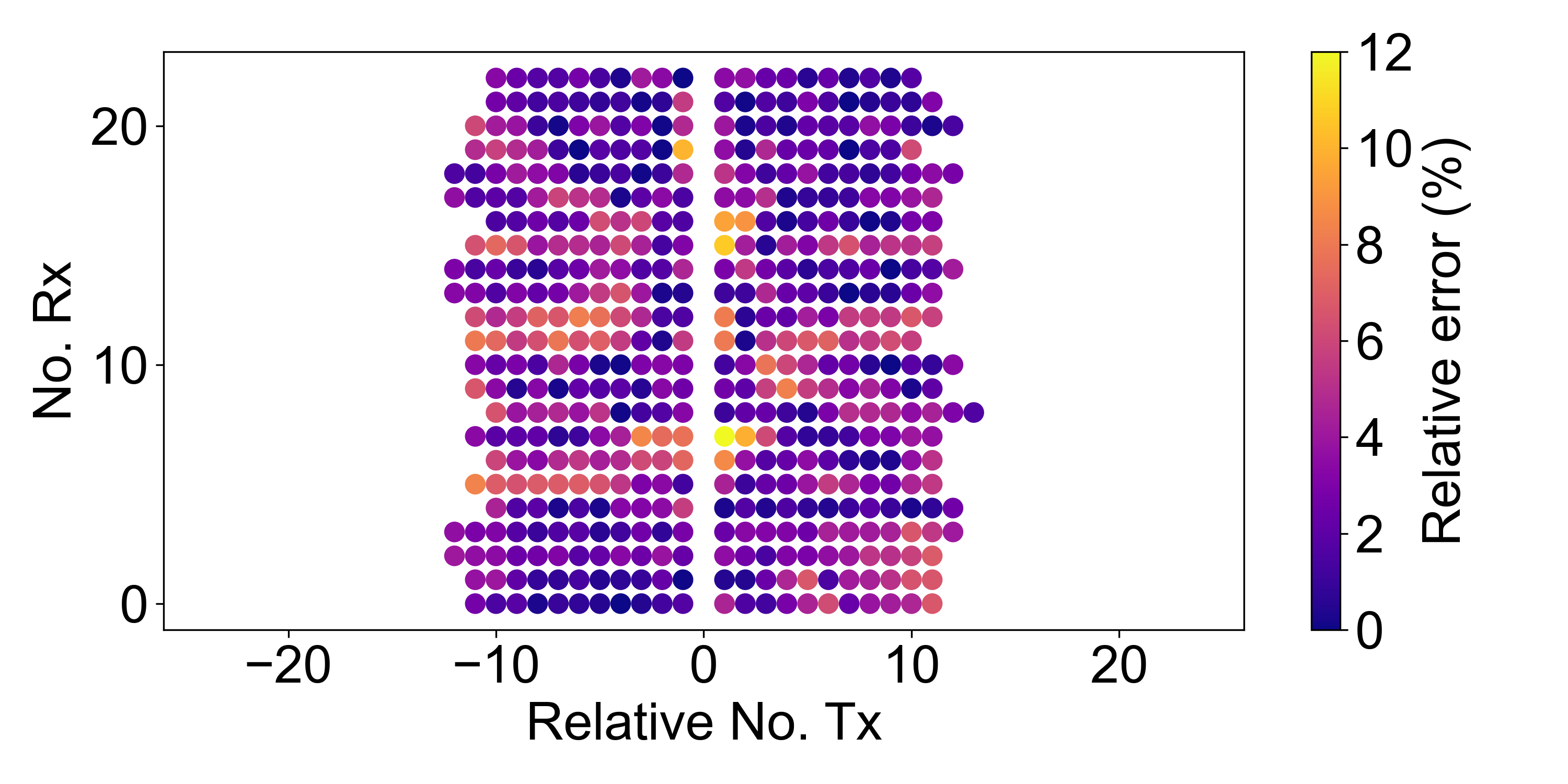}%
		\label{fig16_h}}
	
	

	\caption{\color{black}Visualization of the relative errors between the reconstructed data and measured data, for both normalized logarithm amplitude (NLA) and phase. The four rows respectively represent the NLA relative errors for 0.25 Hz, NLA relative errors for 0.75 Hz, phase relative errors for 0.25 Hz, and phase relative errors for 0.75 Hz. In the first column, the reconstructed data is recovered by $L_2$ inversion and in the second column, the reconstructed data is recovered by feature-based inversion.\color{black}}

	\label{fig16}
\end{figure*}

\begin{figure}[tb]
	\centering
	
	\subfloat[]{\includegraphics[width=0.45\columnwidth]{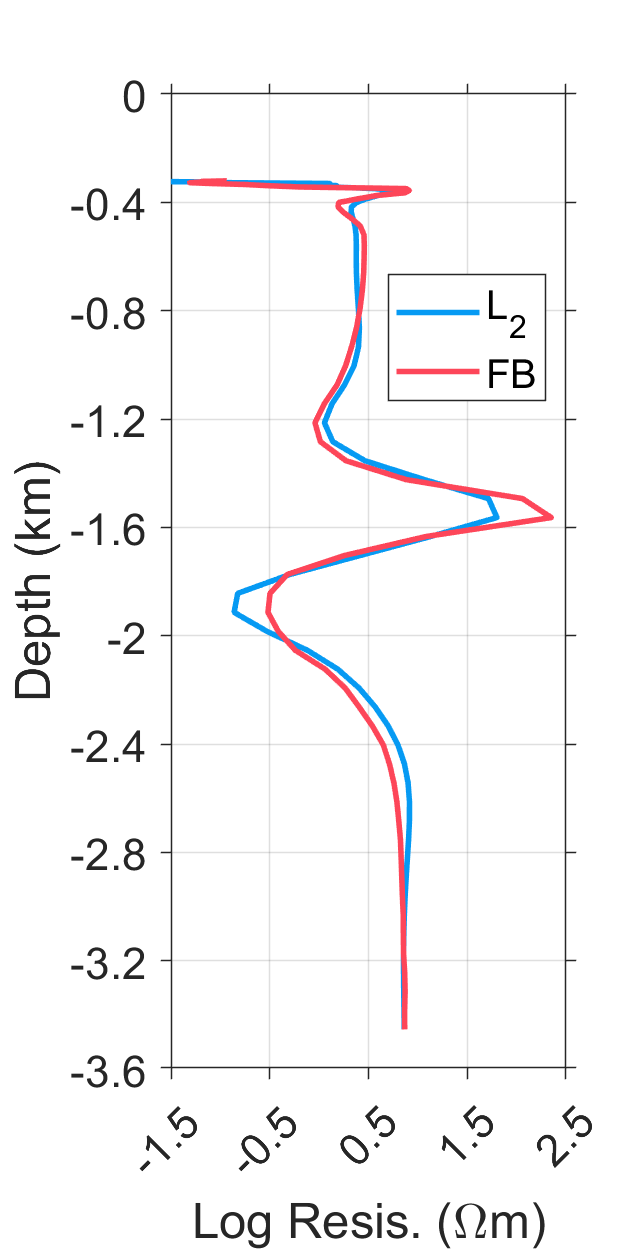}%
		\label{fig15_a}}
	\hfil 
	\subfloat[]{\includegraphics[width=0.75\columnwidth]{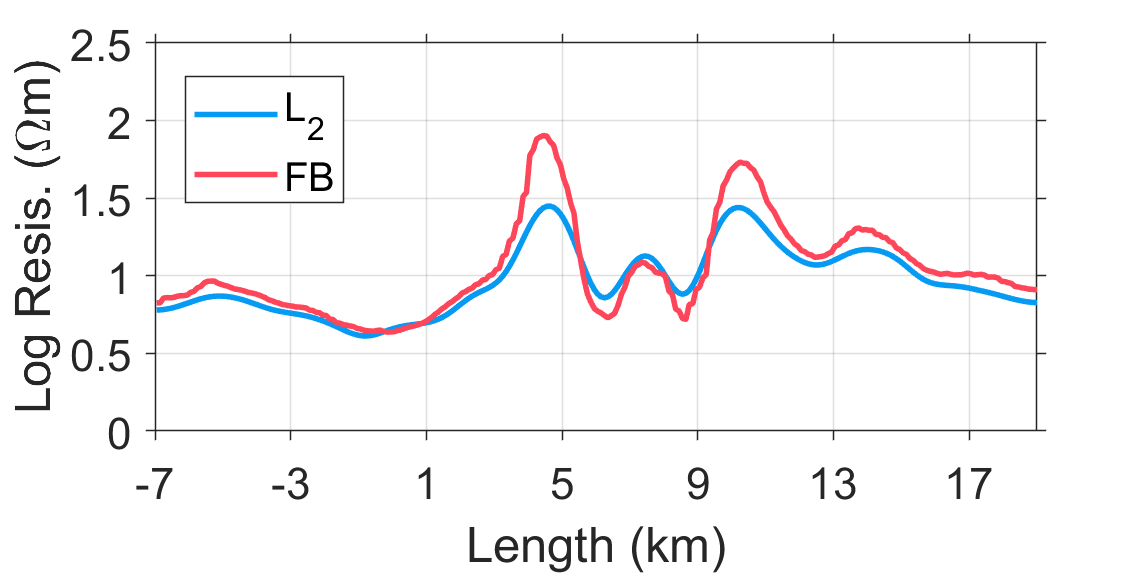}%
		\label{fig15_b}}
	\caption{Fig. \ref{fig15_a}: the vertical traces denoted by $\mathtt{Line 1}$. Fig. \ref{fig15_b}: the horizontal traces denoted by $\mathtt{Line 2}$.}
	\label{fig15}
\end{figure}

\color{black}
$L_2$, W$L_2$ and FB inversions are conducted to recover test model D. The initial model is the same for all the three inversions, with the sedimentary resistivity set to a uniform value of 0.6 $\Omega$m. Figs. \ref{fig6_1}--\ref{fig6_4} show the initial model and reconstructions recovered by $L_2$, W$L_2$ and FB inversions, respectively, and Fig. \ref{test_model_D_loss} shows the loss function curves of the inversions. 
For the FB inversion, because irregular DoI cannot be directly processed by the VAE, it is mapped to a regular rectangular region that fully encloses the DoI, and the pixels outside the DoI are filled with representative attribute values within the DoI. 
In our work, the regular region is set to the entire 20 km by 6.8 km region, as shown in Fig. \ref{testmodelD}, and the undefined pixels are filled with the resistivity value of the sedimentary background. The \emph{projection steps} are applied at iterations 12, 15 and 18. At the early stage of inversion, the conductivity distribution within the DoI in intermediate models is chaotic and differs from the ground truth; therefore, in our experience, the recovered model in FB inversion usually need to stabilize first before projection steps are applied. Because test model D has a more complex geometric pattern than test models A--C, more iterations are required before applying \emph{projection steps} to allow the recovered model to stabilize.

All the three inversions converge to similar final DMs, close to the noise level, while FB yields a slightly smaller final DM than two other inversions. The approximated locations of the two anomaly blocks can be identified in all the inversions. Again, consistent with the observations in the previous two sections, the $L_2$ inversion exhibits blurred boundaries. The right segment of the basement can be observed, but its boundary remains unclear and the recovered resistivity is lower than the true value. In W$L_2$ inversion, the shapes and boundaries of both blocks are better characterized, but the thickness of the right block is overestimated. In addition, the basement is not properly recovered in the W$L_2$ inversion. In the FB inversion, the boundaries of both blocks can be clearly identified, and the thickness of the right anomaly is closer to the ground truth than in the $L_2$ and W$L_2$ reconstructions, although the left anomaly still appears thinner than its true state. 
The existence and approximate geometry of the basement are also more clearly characterized than in the $L_2$ and W$L_2$ reconstructions, showing a left-lower–right-higher basement topography consistent with test model D. 
However, the left and right segments show different resistivities: the resistivity of the left segment is reconstructed lower than true value, while the resistivity of the right segment is reconstructed higher than true value. Overall, the FB inversion achieves better imaging results than the conventional inversions.

The generalization ability is further verified. In this experiment, the subsurface topography is irregular and differs from that in the training dataset and previous experiments, and the model is not included in the training dataset. In addition, the frequencies and spatial distributions of the transmitters and receivers differ from those in previous experiments. However, the FB inversion still produces reasonable results, further demonstrating its generalization capability.  
\color{black}

\section{Field Experiment}
To further verify the studied inversion, we invert the CSEM data collected in the TrollWest Gas Province (TWGP) in Norway. This is a single-profile survey. Twenty-four EM receivers are distributed on the sea floor. \color{black} The seawater depth along the line varies from approximately 300 to 360 m, and is commonly approximated as 320 m for modeling. \color{black} A 230 m long HED transmitter oriented along the survey line direction is towed by a ship and excite EM field with 0.25 Hz continuous square wave signal at different locations, spanning a survey line of over 25 km. In this survey, gas reservoir is buried in Jurassic sandstone, with the maximum thickness about 160 m, a depth about 1400 m and average resistivity around 200-500 $\Omega$m \cite{johansen2005subsurface}. The conductivity of water bearing Sognefjord Fm. sands and overburden sediments ranges from 0.5 to 2 S/m. More details about the survey can be found in \cite{johansen2005subsurface, hoversten2006integration}, and for existing inversion work on this survey, please refer to \cite{abubakar20082, abubakar2012model}.

In the inversion, we use EM data excited at 97 transmitter locations and measured at 23 receivers, as shown in Fig. \ref{fig13_a}. Data at 0.25 Hz and 0.75 Hz is reliably extracted. 
The depth of the sea water is set to 320 m, and the DoI to be reconstructed ranges from 0.32 km to 3.5 km in depth, and from -7 km to 19 km horizontally. 
The model is discretized into 140 grids vertically and 261 grids horizontally. The mesh used in the inversion is the same with the previous forward modeling. The conductivity of the sea water varies from 4 S/m to 3.3 S/m \cite{abubakar20082}. Since its hard to accurately model the excitation strength, we invert the normalized logarithm amplitude (NLA) computed from data $d$ with
\begin{equation}
	\text{NLA}_{mnk} = \frac{\text{log}_{10}|d_{mnk}|}{\sum_{m=1}^{M}\sum_{n=1}^{N}\text{log}_{10}|d_{mnk}|},
\end{equation}
where $m,n$ and $k$ are respectively indices of transmitter, receiver and frequency. 
We also invert the phase in addition to NLA.
At 0.25 Hz, each receiver corresponds to 45 valid data on average, while at 0.75 Hz, each receiver corresponds to 23 valid data on average.

We present $L_2$ and FB reconstructions. A uniform 0.4 S/m model (see Fig. \ref{fig13_a}) is used as \color{black} the initial model for both the $L_2$ and FB inversions. The $L_2$ inversion converges in 14 iterations, with a final DM of 3.38\%. Fig. \ref{fig13_b} shows the reconstruction obtained from the $L_2$ inversion. For the FB inversion, the regularization coefficients are the same with $L_2$ inversion, and the VAE parameters used here is the same with numerical experiments. One \emph{projection step} is configured at iteration 14, and after that the FB inversion takes another 12 steps to converge, resulting a final DM of 2.37\%. \color{black}  \color{black} The projected intermediate model at iteration 14 of FB inversion is shown in Fig. \ref{fig13_c}, and the final model by FB inversion is shown in Fig. \ref{fig13_d}. 
The DM curves of both inversions are presented in Fig. \ref{fig15_c}, and the relative errors between the measured data and data reconstructed by the two inversions are shown in Fig. \ref{fig16}. \color{black} 

For clarity, we select a vertical trace located at the offset of 7.45 km, and a horizontal trace located at the depth of 1.63 km (denoted by dashed lines in Fig. \ref{fig13_b} and \ref{fig13_d}), and plot the reconstructions in Fig. \ref{fig15_a} and \ref{fig15_b}.  We can observe that the overall conductivity pattern of TWGP can be recovered by the conventional inversion, that is, the resistive layer buried in the conductive background, with the location and size basically consistent with the prior knowledge. 
\color{black} From the $L_2$ inversion results, the reservoir is interpreted to be located at a depth of approximately 1400–1600 m, with a maximum thickness of about 210 m and a typical thickness of around 100 m. The lateral extent is estimated to be approximately 7.4 km, and the maximum resistivity reaches about 63 $\Omega$m. However, because the overall resistivity contrast is relatively weak, the reservoir boundaries are not sharply resolved. As a result, the estimates of reservoir length and thickness remain ambiguous and show noticeable deviations from prior geological expectations, where the average resistivity is known to be around 200-500 $\Omega$m and the maximum thickness about 160 m \cite{johansen2005subsurface, abubakar20082}. \color{black}  
Compared to conventional reconstructions, the features of the resistive layer are clearer in the projected model. From FB reconstruction, we can find that the depth of the reservoir is about 1400 to 1600 meters, with the maximum thickness about 160 m, a typical thickness about 100 m, length about 8 km and the maximum resistivity about 200 $\Omega$m. \color{black} The reservoir thickness and burial depth recovered by the feature-based inversion show improved agreement with available geological information \cite{johansen2005subsurface}. In addition, the estimated reservoir length is closer to previously reported inversion results \cite{abubakar2012model}, although the maximum recovered resistivity remains slightly lower than the expected values inferred from prior knowledge. \color{black} \color{black} 

We conclude that the FB inversion achieves improved inversion results compared to conventional inversions; however, in this field inversion case, the improvement is not very pronounced, likely because the conventional inversion already provides reasonably good results. We will investigate more representative inversion cases to better demonstrate the advantages of the feature-based inversion. \color{black}

\section{Conclusion}
In this work, we study the feature-based 2.5D \color{black} marine \color{black} controlled source electromagnetic (mCSEM) inversion using generative priors. Two-point-five dimensional modeling using finite difference method (FDM) is adopted to compute the response of horizontal electric dipole (HED) excitation in the 2D survey. Generative prior is incorporated in the inversion by reparametrizing the model with the variational autoencoder (VAE). In the inversion, the model is consecutively updated with Gauss-Newton method and projected to the range of the VAE decoder. Both numerical and field experiments verify that the proposed inversion is able to effectively incorporate \emph{a priori} knowledge and improve the reconstruction accuracy and possesses good generalization ability. 
Future improvements to this work may involve developing more efficient and robust optimization algorithms to eliminate the reliance on heuristics in setting the projection step, as well as optimizing the architecture and training of the VAE, \color{black} and extending the proposed method to land CSEM inversion\color{black}.

\color{black}
{\appendix[Metrics to evaluate inversion results]
	
	\setcounter{equation}{0} 
	\renewcommand{\theequation}{A\arabic{equation}}
	
	We introduce some metrics to measure the inversion quality, including data misfit (DM), intersection over union (IoU), mean square error (MSE) and structural similarity (SSIM) \cite{wang2004image}, as listed in (\ref{DM})-(\ref{SSIM}):
	\begin{equation}\label{DM}
		\mathrm{DM} = \frac{\|\mathbf{d}_\text{obs}-\mathbf{d}_\text{rec}\|_2}{\|\mathbf{d}_\text{obs}\|_2},
	\end{equation}
	and
	\begin{equation}\label{IoU}
		\mathrm{IoU}(\mathbf{I}_\text{rec},\mathbf{I}_\text{gnd})=\frac{|\mathbf{I}_\text{rec} \cap \mathbf{I}_\text{gnd}|}{|\mathbf{I}_\text{rec} \cup \mathbf{I}_\text{gnd}|},
	\end{equation}
	\begin{equation}\label{MSE}
		\mathrm{MSE} = \|\mathbf{m}_{\text{gnd}}-\mathbf{m}_{\text{rec}}\|^2_2,
	\end{equation}
	\begin{equation}\label{SSIM}
		\mathrm{SSIM} = \frac{(2\mu_x\mu_y+C_1)(2\sigma_{xy}+C_2)}{(\mu_x^2+\mu_y^2+C_1)(\sigma_x^2+\sigma_y^2+C_2)}.
	\end{equation}
	Here, $\mathbf{d}_\text{obs}$ and $\mathbf{d}_\text{rec}$ are observed and reconstructed data. $\mathbf{I}_\text{gnd}$ is the 0-1 identification function, in which 1 pixels indicate the target reservoir and 0 pixels indicate the background in the ground truth model. $\mathbf{I}_\text{rec}$ is also a 0-1 identification function, in which 1 pixels indicate where the value of the reconstructed conductivity is greater than half the peak value.  $\mathbf{m}_{\text{gnd}}$ and $\mathbf{m}_{\text{rec}}$ are respectively the ground truth and reconstructed model. $\mu_x, \mu_y, \sigma_x, \sigma_y$ and $\sigma_{xy}$ denote the mean, standard deviation and covariance between two images $x$ and $y$, and $C_1$ and $C_2$ are two positive constants.
}
\color{black}

\section*{Acknowledgments}
HZ would like to express his gratitude to Prof. Yoram Bresler from the University of Illinois at Urbana–Champaign for fruitful discussions on generative-prior-regularized inverse problems.

\bibliographystyle{IEEEtran}
\bibliography{CSEM-VAE} 



%
%
%
%

\vfill

\end{document}